\documentclass{aa}
\usepackage[utf8]{inputenc}
\usepackage[varg]{txfonts}
\usepackage{xcolor}
\usepackage{amssymb}
\usepackage{amsmath}
\usepackage{array}
\usepackage[colorlinks=true,citecolor=blue,linkcolor=blue,urlcolor=blue]{hyperref}

\bibpunct{(}{)}{;}{a}{}{,} 

\makeatletter
\renewcommand*\aa@pageof{, page \thepage{} of \pageref*{LastPage}}
\makeatother

\newcolumntype{L}[1]{>{\raggedright\let\newline\\\arraybackslash\hspace{0pt}}m{#1}}

\begin{document}

\title{The \emph{SRG}/eROSITA all-sky survey:\\The morphologies of clusters of galaxies I:\\A catalogue of morphological parameters}
\titlerunning{eRASS1 cluster morphologies I}
\authorrunning{J.~S.~Sanders et~al.}

\author{
  J.~S. Sanders \inst{1}
  \and Y.~E.~Bahar \inst{1}
  \and E.~Bulbul \inst{1}
  \and V.~Ghirardini \inst{1,2}
  \and A.~Liu \inst{1}
  \and N.~Clerc \inst{3}
  \and M.~E.~Ramos-Ceja \inst{1}
  \and T.~H. Reiprich \inst{4}
  \and F.~Balzer \inst{1}
  \and J.~Comparat \inst{1}
  \and M.~Kluge \inst{1}
  \and F.~Pacaud \inst{4}
  \and X.~Zhang \inst{1}
}

\institute{
  Max-Planck-Institut f\"ur extraterrestrische Physik,
  Gießenbachstraße 1, 85748 Garching, Germany
  \and INAF, Osservatorio di Astrofisica e Scienza dello Spazio, via Piero Gobetti 93/3, 40129 Bologna, Italy
  \and IRAP, CNRS, UPS, CNES, 14 Avenue Edouard Belin, 31400 Toulouse, France
  \and Argelander-Institut für Astronomie (AIfA), Universität Bonn, Auf dem Hügel 71, 53121 Bonn, Germany
}

\date{Received ---, Accepted ---}

\abstract{
  The first \emph{SRG}/eROSITA all-sky X-ray survey, eRASS1, resulted in a catalogue of over twelve thousand optically-confirmed galaxy groups and clusters in the western Galactic hemisphere.
  Using the eROSITA images of these objects, we measure and study their morphological properties, including their concentration, central density and slope, ellipticity, power ratios, photon asymmetry, centroid shift and Gini coefficient.
  We also introduce new forward-modelled parameters which take account of the instrument point spread function (PSF), which are slosh, which measures how asymmetric the surface brightness distribution is, and multipole magnitudes, which are analogues to power ratios.
  Using simulations, we find some non forward-modelled parameters are strongly biased due to PSF and data quality.
  For the same clusters, we find similar values of concentration and central density compared to results by ourselves using \emph{Chandra} and previous results from \emph{XMM-Newton}.
  The population as a whole has log concentrations which are typically around 0.3 dex larger than South Pole Telescope or \emph{Planck}-selected samples and the deeper eFEDS sample.
  The exposure time, detection likelihood threshold, extension likelihood threshold and number of counts affect the concentration distribution, but generally not enough to reduce the concentration to match the other samples.
  The concentration of clusters in the survey strongly affects whether they are detected as a function of redshift and luminosity.
  We introduce a combined disturbance score based on a Gaussian mixture model fit to several of the parameters.
  For brighter clusters, around $1/4$ of objects are classified as disturbed using this score, which may be due to our sensitivity to concentrated objects.
}

\maketitle

\section{Introduction}
The intracluster medium (ICM), the hot atmosphere which fills the potential well of clusters of galaxies is sensitive to many of the physical processes taking place when clusters form and internally within clusters.
This atmosphere is frequently assumed to consist of a spherically symmetric $\beta$-model, where the gas density profile follows $n(r) = n_0 \, [1+(r/r_\mathrm{c})^2]^{-3\beta/2}$ \citep{Cavaliere78}, where $n_0$ is an inner density, $r_\mathrm{c}$ is the core radius and $\beta$ controls the outer slope, where  $\beta=2/3$ is often assumed as a typical value.

The ICM is visible by its emission in the X-ray band, primarily through bremsstrahlung radiation, although X-ray emission lines are dominant for cool objects \citep[e.g.][]{bohringer10}.
Studying this emission, we know that many clusters show profiles which deviate from this $\beta$-model form and show shapes which are non-spherical.
A substantial fraction of clusters are cool core systems with steeply peaked surface brightness profiles where the ICM has very short mean radiative cooling times \citep[e.g.][]{Fabian12,McNamaraNulsen12}.
Without the injection of energy these systems would be rapidly radiatively cooling.
Feedback by active galactic nuclei (AGN) appears responsible for preventing this happening \citep[e.g.][]{Birzan04,Rafferty06}.

Galaxy clusters are also dynamic systems.
In the hierarchical formation of structure, galaxies, groups and clusters merge together to form larger systems \citep[e.g.][]{Voit05}.
These mergers can be minor or major.
In a minor merger, the passage of a subcluster through another cluster can cause a displacement of the hot gas from the potential well, which can introduce motions (or sloshing) in the ICM.
This is thought to produce the contact discontinuities, or cold fronts, seen frequently in clusters \citep[e.g.][]{MarkevitchCFShock07}, even in those which appear otherwise relaxed.

In more extreme major mergers the whole shape of the cluster can be affected, with the most prominent example being the Bullet cluster \citep{Clowe06}.
If a merger is strong enough it may completely destroy a cool core \citep[e.g.][]{Valdarnini21}.
Depending on the phase of a merger and the viewing angle, the system may be detected as two or more separate clusters, an elongated or non-circular single cluster or a circular cluster.

There are many different reasons why it is interesting to know whether a cluster or sample of clusters are disturbed.
If one would like to understand merger processes then identifying these systems is useful.
The environment of the cluster may also affect its halo properties \citep[e.g.][]{Wechsler02} and therefore its morphology, for example if it is within a supercluster.
Indeed, evidence for halo assembly bias in clusters has been detected \citep{Liu24} by eROSITA (extended ROentgen Survey with an Imaging Telescope Array).
Another important case is identifying clusters where there is a cool core, in order to study the impact of AGN feedback.
One common technique for measuring cluster masses is by the assumption of hydrostatic equilibrium, which may not apply if the object is substantially out of equilibrium \citep{Lau09,Biffi16}.
Those clusters with a higher degree of disturbance have a larger range of hydrostatic mass bias scatter \citep{Gianfagna23}.

In order to say whether a cluster is disturbed or not requires a quantitative measure.
There are several so-called morphological parameters which characterise the shape or peakiness of a cluster and therefore its level of disturbance.
We describe the parameters in our analysis in detail in Section \ref{sect:params}.
It is important to note that the quantities measured can be affected by either the signal to noise of the data being analysed or the angular size of the object relative to the point spread function (PSF) of the telescope.
The luminosity or redshift of the cluster can also affect what is measured.
Morphological parameters are are also affected by projection effects.
In the X-ray waveband we are also most sensitive to denser regions due to the bremsstrahlung emission process.
The parameters often cannot be blindly be compared between surveys.

eROSITA onboard \emph{SRG} (\emph{Spektrum Roentgen Gamma}) is an instrument optimised for surveying the sky in the X-ray waveband \citep{Predehl21}. The first eROSITA X-ray all sky survey is described in \cite{Merloni24}, in which is presented the catalogue from the western Galactic hemisphere.
The survey is characterised by a full coverage without gaps on the sky and a relatively uniform exposure except for the deeply observed ecliptic pole.
The $\sim 30$~arcsec PSF of the survey is also very uniform because each sky region passes through the field of view several times as the telescope scans over the sky.
A catalogue of clusters identified by the survey is presented in \cite{Bulbul24} (hereafter B24).
Extended sources from the main X-ray catalogue were optically confirmed \citep{Kluge24} to create a main eRASS1 catalogue of 12\,247 clusters spanning 13\,116 deg$^{2}$.
Despite optical confirmation, the catalogue contains contaminants, caused by falsely identified as extended X-ray sources, for example, due to bright point sources and background fluctuations \citep{Seppi22}, coincidental lying in an overdensity of red-sequence galaxies.
A contamination model estimates the purity of the main sample as 86\%.
Using a stronger cut on extent likelihood of the X-ray sources, a second purer catalogue of 5\,263 clusters spanning 12\,791 deg$^{2}$ was created.
This sample has an estimated purity of 95\% and is intended for cosmological analyses.
The integrated X-ray properties, including flux, temperature, total mass, gas mass, gas mass fraction and mass proxy $Y_\mathrm{X}$ were obtained for each object.
This sample of clusters is unmatched by other X-ray selected cluster samples.

In this paper we measure the morphological properties of the clusters detected in B24 in the main eRASS1 sample.
These parameters are published in the form of a catalogue.
This is the largest set of X-ray parameters collected for a cluster sample.
We also study the systematic effects on the parameters due to observational effects via simulations.
A combined disturbance parameter is also computed based on combining individual parameters.

Where not otherwise specified, $\log$ refers to $\log_{10}$, while $\ln$ is $\log_\mathrm{e}$.
We use the relative Solar abundances of \cite{Asplund09}.
A cosmology with $H_0=70$~km~s$^{-1}$~Mpc$^{-1}$, $\Omega_\mathrm{m}=0.3$ and $\Omega_\Lambda=0.7$ is assumed.

This paper is organised as follows.
In Section \ref{sect:params} we introduce the parameters we are measuring for the cluster, including new parameters measuring `slosh' and multipole magnitudes.
We describe how the parameters are obtained from the data in Section \ref{sect:analysis}.
In Section \ref{sect:sim} we describe the simulations made to understand the systematic uncertainties of the parameters.
Section \ref{sect:results} presents the results from the analysis of the eROSITA data.
A discussion of the various biases which affect the parameters is given in Section \ref{sect:biases}.
Values for different cluster subsets are compared to previous results in Section \ref{sect:other_samp}.
A combined disturbance parameter is described in Section \ref{sect:disturb}.

\section{Parameters}
\label{sect:params}

\begin{table*}
  \caption{
    Summary of morphological parameters.
  }
  \centering
  \begin{tabular}{llccl}
    \hline
    Name & Description & Type & Centre & $R_{500}$ \\ \hline
    $n_{\mathrm{s},0}$ & Log gas density at a radius of $0.02R_{500}$ relative to a scaled critical density & M & V & Y \\
    $n_{\mathrm{s},0}^*$ & Log gas density at a radius of $0.02R_{500}$ relative to a scaled critical density & M & P & Y \\
    $n_{50}$ & Log gas electron density at a radius of 50~kpc & M & V & N \\
    $n_{50}^*$ & Log gas electron density at a radius of 50~kpc & M & P & N \\
    $\alpha$ & Inner density slope at $0.04R_{500}$ radius & M & V & Y \\
    $\alpha^*$ & Inner density slope at $0.04R_{500}$ radius & M & P & Y \\
    $\alpha_{50}$ & Inner density slope at 50~kpc radius & M & V & N \\
    $\alpha_{50}^*$ & Inner density slope at 50~kpc radius & M & P & N \\
    $c_{500}$ & Log ratio of integrated model surface brightness in apertures of $0.1R_{500}$ and $R_{500}$ & M & V & Y \\
    $c_{500}^*$ & Log ratio of integrated model surface brightness in apertures of $0.1R_{500}$ and $R_{500}$ & M & P & Y \\
    $c_{80-800}$ & Log ratio of integrated model surface brightness in apertures of 80 and 800 kpc & M & V & N \\
    $c_{80-800}^*$ & Log ratio of integrated model surface brightness in apertures of 80 and 800 kpc & M & P & N \\
    $F$ & Ratio of offset between best fitting cluster centre and peak and $R_{500}$ & M/I & -- & Y \\
    $P_{10}$, $P_{20}$, $P_{30}$, $P_{40}$ & Log power ratio with orders from 1 to 4 & I & F & Y \\
    $P_{10}^*$, $P_{20}^*$ & Log power ratio with orders from 1 to 2 & I & P & Y \\
    $G$ & Gini coefficient (0--1) & I & F & Y \\
    $A_\mathrm{phot}$ & Photon asymmetry & I & F & Y \\
    $A_\mathrm{phot}^*$ & Photon asymmetry & I & P & Y \\
    $w$ & Centroid shift & I & F & Y \\
    $\epsilon$ & Ellipticity, the ratio of minor to major axis (0--1) & M & M & N \\
    $H$ & Slosh, the degree of sloshing factor (0--1) & M & M & N \\
    $M_1$, $M_2$, $M_3$, $M_4$ & Multipole magnitudes (0--1) for orders 1 to 4 & M & M& N \\
    \hline
    $D_\mathrm{shape}$ & Combined disturbance score (0--1), based on $\epsilon$, $H$, $M_1$ to $M_4$ and $F$ & M/I & -- &  Y \\
    $D_\mathrm{comb}$ & Combined disturbance score (0--1), also including $c_{80-800}$ & M/I & -- &  Y \\
    \hline
  \end{tabular}
  \tablefoot{
    The Type column lists whether the parameter is the output of models (M) or computed from images (I).
    I values can also depend on the model fits (e.g. cluster centre, background rate, model surface brightness profile for masking).
    The type of cluster centre is given, including symmetric model position, where the value can vary during the analysis (V), when it is fixed to the best-fit location (F), the cluster peak (P) or comes from the model being fitted, which can vary during the analysis (M).
    The column $R_{500}$ specifies whether the parameter depends on the cluster radius.
    The disturbance indicators combine other parameters and are discussed in Section \ref{sect:disturb}.
  }
  \label{tab:params}
\end{table*}

\subsection{Introduction}
In our analysis we measure a variety of parameters for each cluster.
These include parameters used in previous studies in clusters and new parameters developed by us during the course of our analysis.
The parameters are listed in Table \ref{tab:params} and described below.

\subsection{Central density}
A high central gas density, and therefore likely a short mean radiative cooling time, is often associated with the presence of a cool core \citep[e.g.][]{Fabian12}.
Therefore central gas density is often used as a morphological parameter \citep[e.g.][]{Ghirardini22}.
Cool cores are often located in regular (or `relaxed') clusters, although there not a one-to-one match of regular clusters and those with a cool core \citep[e.g.][]{Hudson10,Lovisari17}.

If a cool core evolves with redshift with the rest of the cluster, it makes sense to measure it a fixed fraction of $R_{500}$.
In addition, as the critical density of the universe evolves it can make sense to measure the density relative to the average within a cluster, if it contains the universal baryon fraction.
Therefore we compute the scaled density from the electron density profile $n_\mathrm{e}(r)$ as
\begin{equation}
  n_{\mathrm{s},0} = \log \frac{n_\mathrm{e}(0.02 R_{500})}{\mathrm{cm}^{-3}} - \log \left( 500 \, \rho_\mathrm{crit} \frac{f_\mathrm{B}}{\mu_\mathrm{e} m_\mathrm{u}} \frac{1}{\mathrm{cm}^{-3}} \right),
  \label{eqn:ns}
\end{equation}
where $\rho_\mathrm{crit}$ is the critical density at the redshift, $f_\mathrm{B}$ is the baryon fraction (fixed to $0.175$ from \citealt{Spergel07}), $\mu_\mathrm{e} m_\mathrm{u}$ is the mass per electron for an ionised plasma (here $1.175 \times 1.661 \times 10^{-24}$~g).
The choice of $0.02 R_{500}$ is chosen for consistency with \cite{Ghirardini22}, although the density in that work is not relative to the critical density.
Due to the wide dynamic range, we report our density parameters in log space.

However, if a cool core evolves independently from the rest of the cluster, it might be better to measure a physical electron density at a fixed physical radius.
Therefore, we also measure a parameter for the log electron density at a radius of 50~kpc,
\begin{equation}
  n_{\mathrm{50}} = \log \left[ n_\mathrm{e}(50 \, \mathrm{kpc}) / \mathrm{cm}^{-3} \right].
\end{equation}

The density and uncertainties are computed from a Markov chain Monte Carlo (MCMC) analysis using \texttt{MBProj2D} (B24) of each cluster (Section \ref{sect:analysis}), with the \texttt{emcee} sampler \citep{ForemanMackey12}.
We take a random sample of entries from the MCMC chain, compute the log density from each one, then compute the median and $1\sigma$ percentiles as the parameter.
The assumed density functional form can have some impact on the measured central densities, especially for low mass or high redshift objects where the PSF size is larger than the radius the density is measured at.
However, the wide prior on the central density slope we used reduces the bias due to the parametrization and other priors.

For several of the parameters we examine, the quantity can be dependent on the position of the cluster centre (for discussion of this see Section \ref{sect:pos_bias}).
This is particularly important for quantities affected by the shape of the central density profile.
Therefore for these parameters we compute their value with the central position at two positions: the best-fitting cluster centre (fitting the whole cluster) and using the peak of emission.
The peak values are denoted with a ${}^*$, while the best-fit centre values are not marked in this way.
The central density parameters are $n_{\mathrm{s},0}^*$, $n_{\mathrm{s},0}$, $n_{50}^*$ and $n_{50}$.

\subsection{Inner density slope or cuspiness}
Similarly, to the central density, the cuspiness parameter \citep{Vikhlinin07}, $\alpha$, is another indicator of the presence of a cool core and therefore a cluster with a regular morphology.
Cool core clusters have steeply rising density profiles, while clusters with a more unrelaxed X-ray appearance have a flatter central density profile.
The slope is defined at some radius, $r$, by
\begin{equation}
  \alpha(r) = - \frac{\mathrm{d} \log n_\mathrm{e}}{\mathrm{d} \log r}.
\end{equation}
Following \cite{Vikhlinin07}, the parameter $\alpha$ is the slope at $0.04R_{500}$.
We also calculate a slope at fixed physical radius, $\alpha_\mathrm{50}$ at 50~kpc radius.
Similarly to the central density, we also compute values computed using the peak cluster position instead of the best fit position, $\alpha^*$ and $\alpha_{50}^*$.

\subsection{Concentration}
The concentration is defined as the ratio of integrated surface brightness within two apertures \citep[e.g.][]{Santos08}.
This is again a parameter which is sensitive to the presence of a cool core.
We use two different sets of apertures.
The first are fixed physical apertures of 80 and 800 kpc, where
\begin{equation}
  c_{80-800} = \log \frac{ I_X(80 \: \mathrm{kpc}) }{ I_X(800 \: \mathrm{kpc})},
\end{equation}
where $I_X(r)$ is the integrated surface brightness within radius $r$.
This definition has the advantage of not being affected by the determination of the cluster $R_{500}$.
In addition, if a cool cores do not evolve in time, as found previously in relaxed systems \citep{McDonald17}, then a fixed aperture could be more appropriate.

We can also measure the concentration using a fraction of the $R_{500}$ for the system,
\begin{equation}
  c_{500} = \log \frac{ I_X(0.1 R_{500}) }{ I_X(R_{500})}.
\end{equation}
The fixed aperture version uses a larger set of apertures than some other definitions (e.g. $40-400$~kpc), but $80-800$~kpc is better suited to the $\sim 30$~arcsec eROSITA PSF at moderate redshifts (for example, at $z=0.3$, a radius of $40$~kpc is only 9~arcsec).
$c_{80-800}$ and $c_{500}$ will be equivalent for $R_{500}=800$~kpc, which is close to the median value of $R_{500}=740$~kpc.

We note that here we take the log of the values, unlike in other works.
With this definition, flatter surface brightness profiles have more negative values (a completely flat profile would give a value of $-2$), while steeply peaked profiles tend towards a value of zero.

If we simply took the surface brightness on the sky, the concentration would be affected by the PSF and background.
To avoid this issue, the concentration is computed using our cluster model, before convolution with the PSF and the addition of background, rather than the observed surface brightness.
We use the MCMC chains to compute the parameter and its uncertainties.
For more compact clusters, the effect of the PSF is to increase the size of the uncertainties, rather than bias the results to less concentration.
However, these results are dependent on how realistic the density parametrization and its priors are.
Disturbed objects may not be well fitted by our parametrization or our priors may be inaccurate in these cases, potentially giving rise to a bias.

As the concentration also depends on the central position, we also calculate the peak position concentrations, $c_{80-800}^*$ and $c_{500}^*$.

\subsection{Fit-peak offset}
\label{sect:fit_peak_offset}
As discussed above, the preferred cluster position from the symmetric model analysis and the peak X-ray emission may be offset from each other.
This offset is itself a morphological parameter, as perfectly regular clusters would have their X-ray peaks at their centres.
We define the fit-peak offset, $F$, to be
\begin{equation}
  F = \frac{ \theta \: D_{A} }{ R_{500} },
\end{equation}
where $\theta$ is the angular separation between the fit and peak position in radians, $D_{A}$ is the angular diameter distance to the cluster and $R_{500}$ is the cluster radius.

There are several ways to define the X-ray peak of a cluster.
In the lower count regime of eROSITA, one of the most reliable ones is to smooth the X-ray image and find the position of the maximum pixel.
We smooth the exposure-corrected and background-subtracted image using a Gaussian with $\sigma=24$~arcsec, after replacing source regions of neighbouring sources with a realisation of the best fitting model of the main source and background (see Section \ref{sect:image_params}).
We search for the maximum within a radius of $R_{500}$ from the best-fit cluster centre and compute $F$ from its offset.
Uncertainties are calculated by bootstrap resampling of the counts in the input image before smoothing, then calculating the $1\sigma$ range of $F$.

\subsection{Power ratio}
Power ratios are a method for the multipole decomposition of the surface brightness distribution \citep{BuoteTsai95}.
The power decomposed into a particular multipole is measured relative to a zero-th order value.
The values are measured within a particular aperture, for which we use $R_{500}$.
If the higher order multipoles for a cluster have relatively high power compared to the zero-th order, then this indicates an object with a non-spherical morphology.

The ratio for order $m$ is defined as
\begin{equation}
  P_{m0} = \frac{P_m}{P_0},
\end{equation}
where the zero order value is
\begin{equation}
  P_0 = \left(a_0^2 \ln R^{2}_{500} \right)^2
\end{equation}
and for higher orders
\begin{equation}
  P_m = \frac{1}{2 m^2 R^{2m}_{500}} \left( a^2_m + b^2_m \right),
\end{equation}
where for polar coordinate $x=(r, \phi)$ given the surface brightness $S(x)$
\begin{equation}
  a_m = \int_{r<R_{500}} S(x) \, r^m \cos m\phi \, \mathrm{d}^2 x
\end{equation}
and
\begin{equation}
  b_m = \int_{r<R_{500}} S(x) \, r^m \sin m\phi \, \mathrm{d}^2 x.
\end{equation}
In our results we quote the power ratio logarithm, $\log P_{m0}$, as they cover a wide dynamical range, for orders $m=1$ to $4$.

We note that the surface brightness has been background-subtracted by the best fitting background rate from the spherical \texttt{MBProj2D} analysis (Section \ref{sect:analysis}).
As the method does not take account of contaminating sources within $R_{500}$ these regions are filled with a Poisson realisation of the best fitting spherical \texttt{MBProj2D} model.
The details of this filling procedure are given in Section \ref{sect:image_params}.
Statistical uncertainties were computed using bootstrap resampling of the pixels within $R_{500}$ and the quoted value is the median from this analysis.

We note that the power ratios are computed directly from the X-ray image and therefore do not take account of the effect of the PSF.
Blurring by the PSF will affect objects with a smaller angular size, reducing the power measured, although both terms of the ratio will be affected to different degrees.
In addition, the non-Gaussian Poisson noise will affect the determined power in the low count regime (see Section \ref{sect:noise_distance}).
Previous studies have shown that there is bias in the parameters due to the number of photons \citep[e.g.][]{Weissmann13}.
The ratio $P_{10}$ gives the relative strength of the dipole to the monopole, while $P_{20}$, $P_{30}$ and $P_{40}$ can be associated with quadrupole, hexapole and octopole moments, respectively.
$P_{20}$, the quadrupole moment, is also strongly correlated with the ellipticity of an object.
The lower order power ratios are affected by the choice of the cluster centre.
We therefore also compute $P^*_{10}$ and $P^*_{20}$ measured around the cluster peak.

\subsection{Gini coefficient}
The Gini coefficient is a well-known economic indicator of income inequality.
Nevertheless, it can be applied to astronomical images \citep{Abraham03}.
It is primarily an indicator of how peaked the surface brightness distribution is.
High Gini coefficients should be correlated with high concentrations.
Following the definition used by \cite{Lotz04}, we measure the X-ray variation using
\begin{equation}
  G = \frac{1}{|\bar{K}| \, n(n-1)} \sum_{i} (2i-n-1) \, |K_i|,
\end{equation}
where the $n$ pixels in the input aperture have sorted surface brightness values $K_i$ and the mean value is $\bar{K}$.

As eROSITA images have a low number of counts per pixel (typically 0 or 1), the Gini coefficient is insensitive to morphology when measured from raw images because the pixels cannot be ordered.
Therefore we measure the Gini coefficient from a Gaussian-smoothed background-subtracted exposure-corrected image, using a fixed smoothing scale of $\sigma=6$ pixels (24 arcsec).
Similarly to how we measure the Power ratios, we measure using an aperture of $R_{500}$.
In addition, we fill masked areas and other source regions in the image like for the Power ratio analysis (Section \ref{sect:image_params}).
We use bootstrap resampling of the pixels in the aperture to obtain the median value and its $1\sigma$ uncertainties.
We note that the Gini coefficient does not take account of the PSF of the instrument, as it works purely from an image.
The result of blurring by the PSF will be to reduce the differences between pixels, leading to lower values of $G$ for compact or distant objects.
As the Gini coefficient depends on the smoothing of the image and the instrument PSF, it cannot be compared directly with previous surveys.

\subsection{Photon asymmetry}
The photon asymmetry parameter ($A_\mathrm{phot}$; \citealt{Nurgaliev13}) measures the degree of symmetry in the source image.
Simulations show this parameter is insensitive to the redshift of the cluster, if the instrument has a small PSF relative to the examined regions.
We note that eROSITA has a significant PSF on these scales, so the photon asymmetry is not redshift independent for our clusters.

It is computed by first using Watson's test to compare the photons in an annulus with a radially-symmetric cluster, using
\begin{equation}
  U_N^2 [F_N, G] = N \min_{\phi_0} \int (F_N - G)^2 \, \mathrm{d}G,
\end{equation}
where the total number of counts is given by $N$, $F_N$ is the cumulative distribution of the angle of all the photons in an annulus, $G$ is the expected distribution and $\phi_0$ is a starting angle for the distribution.
If there are $C$ counts in the annulus, then the distance between these distributions is
\begin{equation}
  \hat{d}_{N,C} = \frac{N}{C^2} \left( U_N^2 - \frac{1}{12} \right).
\end{equation}
The photon asymmetry is then computed using
\begin{equation}
  A_\mathrm{phot} = 100 \sum^{X}_{k=1} C_k \hat{d}_{N_{k},C_{k}} / \sum_{k=1}^{X} C_k,
\end{equation}
where we are computing it over $X=4$ annuli, with edges of $0.05$, $0.12$, $0.2$, $0.3$ and $1.0 \, R_{500}$.
The uncertainties are obtained by repeating the calculation using a bootstrap resampled set of the photons.

The effect of PSF blurring should be to reduce the measured asymmetry for those clusters with small angular extent.
Photon asymmetry is also sensitive to the choice of cluster centre.
A version of this parameter with the peak position is also included as $A_\mathrm{phot}^*$.

Although photon asymmetry is assumed to be insensitive to redshift, we checked this using simple symmetric models.
We made a $\beta$ model cluster ($r_c=0.17R_{500}$, $\beta=2/3$) image with no PSF broadening or background and with the annuli are properly resolved in the image ($R_{500}=120$~arcsec, with $0.25$~arcsec pixels).
If the model is scaled to have $N$ counts within $R_{500}$, and then a Poisson realisation is made, we find $A_\mathrm{phot} \approx 20\,N^{-1}$.
The reference implementation\footnote{\url{https://github.com/ndaniyar/aphot}} gives consistent results with our code.
Therefore, for faint symmetric clusters of a few tens of counts, $A_\mathrm{phot}$ can be relatively large (0.8 for 20 counts, 0.4 for 50 counts and 0.2 for 100 counts) and therefore not redshift independent.
This simple model does not account for the effect of background nor for the 4 arcsec integer positions we use here.
However, more realistic simulations of the intrinsic distribution can be found in Section \ref{sect:sim}.

\subsection{Centroid shift}
The centroid shift parameter is a commonly applied parameter which measures the variance of the centroid of the emission with increasing apertures, although different authors use different definitions \citep{Mohr95,Poole06,OHare06,BohringerPratt10}.
We define it as
\begin{equation}
  w = \frac{1}{R_{500}} \left( \frac{1}{N-1} \sum_{i=1}^{N} \left| \mathbf{C}_i - \mathbf{C}_\mathrm{peak} \right|^2 \right) ^{1/2},
\end{equation}
where $\mathbf{C}_i$ is the centroid within aperture $i$ and $\mathbf{C}_\mathrm{peak}$ is the peak position.

In our analysis we compute the centroid shift using the shifts from $N=10$ linearly radially-increasing apertures within $R_{500}$ about the peak position.
The peak position is measured from a Gaussian-smoothed image of the cluster (smoothed using $\sigma=24$~arcsec), with masked and fitted sources replaced by model realisations (Section \ref{sect:image_params}).
The value and its uncertainties are generated from the median and $1\sigma$ percentiles of $w$ values measured from Poisson realisations of the smoothed image.

We note that this quantity is calculated directly from the image and therefore does not take account of the instrument PSF.
Blurring of the image means that for apertures smaller or around the PSF size, offsets from the peak will be reduced, leading to lower values of $w$.
This is most important for compact objects or those at higher redshifts.
Similarly to other quantities, we compute $\log w$ as its dynamic range is large.

\begin{figure}
  \includegraphics[width=\columnwidth]{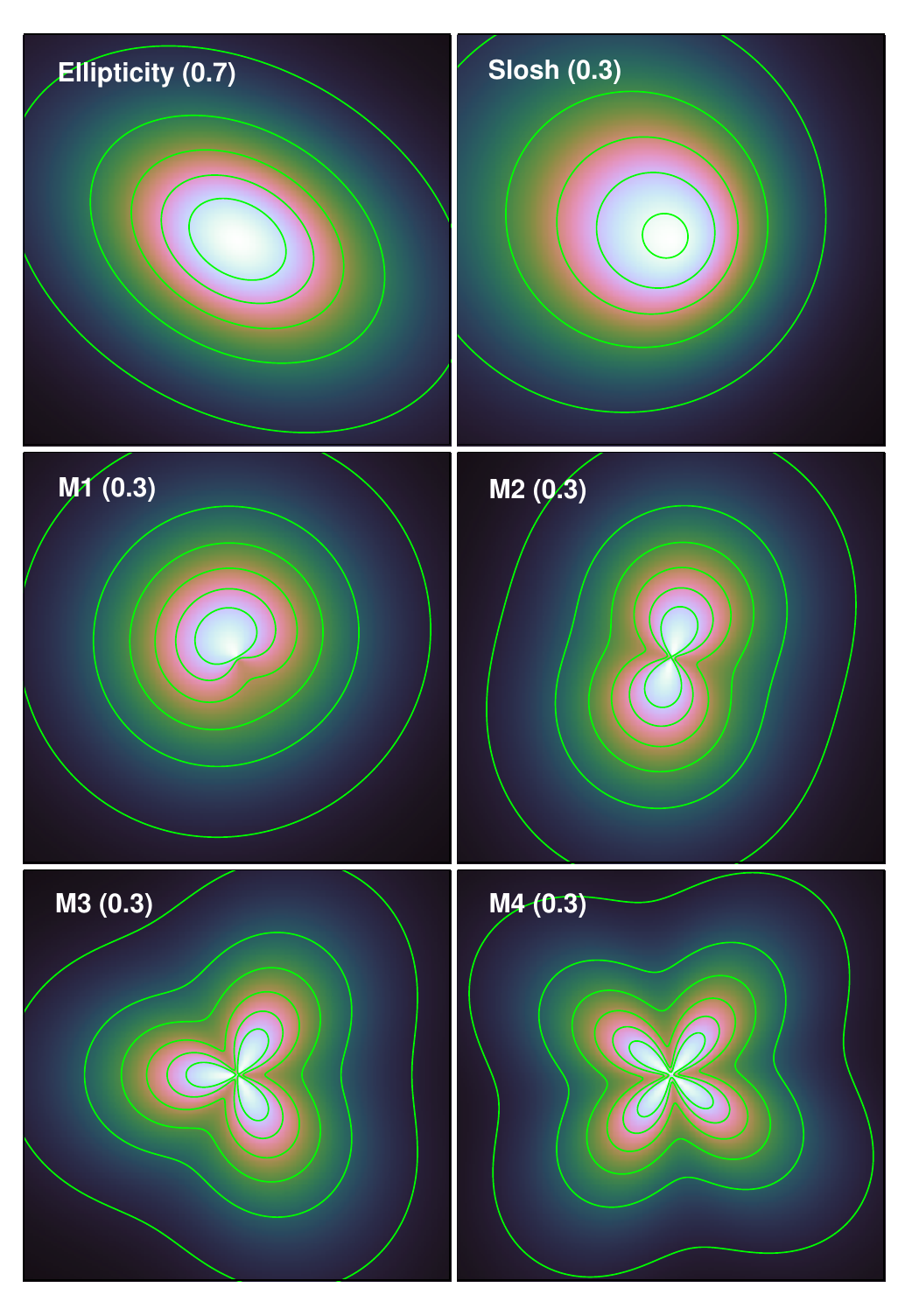}
  \caption{
    Example models exhibiting the effect of the different shape parameters.
    Shown are an elliptical model (top left), a slosh model (top right), an  $M_{1}$ model (centre left), an $M_{2}$ model (centre right), an $M_{3}$ model (bottom left) and an $M_{4}$ model (bottom right).
    For the models we use an angle of $\theta=30 \deg$.
    The numeric values show the magnitude of the respective parameter.
  }
  \label{fig:shape_example}
\end{figure}

\subsection{Ellipticity}
The ellipticity parameter ($\epsilon=b/a$) follows the definition in \cite{Ghirardini22}, of the ratio of the minor ($b$) to the major axis ($a$).
A value of $1$ is circular and $0$ is the extreme elliptical case.
The ellipticity of the X-ray emission of clusters may be sensitive to the physics of the ICM \citep{Lau12}, in addition to being affected by mergers.
Ellipticity is implemented within \texttt{MBProj2D} to produce an elliptical surface brightness distribution on the sky.
An example model with ellipticity is shown in Fig.~\ref{fig:shape_example}.
It models the surface brightness on the sky in Cartesian coordinates where the projected surface brightness is
\begin{equation}
  S'(x, y) = S(\{\epsilon[x \cos \theta_0 - y\sin \theta_0]^2 + [x\sin \theta_0 + y\cos \theta_0]^2/\epsilon\}^{1/2}),
\end{equation}
where $S(r)$ is the original radial surface brightness and $\theta_0$ is the inclination angle of ellipse (between $0$ and $180\deg$).
The radial surface brightness is parametrized in the same way as for the standard profile analysis.
As is standard when modelling clusters with \texttt{MBProj2D}, observational effects such as PSF and background are also included in the total model and uncertainties are calculated from the MCMC chain.

We note that the ellipticity parameter can describe clusters which are likely physically unrealistic (where $\epsilon \lesssim 0.2$).
Informative, rather than flat, priors could be useful for a future analysis.
The eccentricity parameter might be better than ellipticity, as its parameter space is biased towards more circular objects.

\subsection{Slosh}
Slosh ($H$) is a new parameter we design to model how asymmetric the surface brightness is about the cluster.
It has the appearance of the gas sloshing in the potential well of the cluster, as is thought to be the origin of cold fronts \citep[see][]{MarkevitchCFShock07}.
Like ellipticity, we model this within \texttt{MBProj2D} as a modification to the projected surface model brightness.
Similarly, an example model with slosh is shown in Fig.~\ref{fig:shape_example}.

The projected surface brightness on the sky, $S'$, at a radius $r$ and angle $\theta$, is
\begin{equation}
  S'(r,\theta) = A(H) \: S(r[1+H \cos(\theta+\theta_0)]), 
\end{equation}
where the slosh factor, $H$, lies between 0 and 1, $S(r)$ is the original symmetric distribution and $\theta_0$ is the sloshing angle (between $0$ and $360\deg$).
As $H$ tends towards 0, the distribution becomes the original symmetric one, while values of 1 give an extreme asymmetric distribution.
Here, $A(H)$ is a factor to ensure that the total brightness of the cluster model is not modified by changes in $H$.
By considering the relative area of a circle with the slosh transformation applied, compared to its original area, one finds that
\begin{equation}
  A(H) = (1-H^2)^{3/2}.
\end{equation}
Within \texttt{MBProj2D} the surface brightness is computed more easily in Cartesian coordinates ($x,y$) around the cluster centre, where
\begin{equation}
  S'(x, y) = A(H) \: S([x^2+y^2]^{1/2} + H[x\cos \theta_0 - y\sin \theta_0]).
\end{equation}

If there is significant disturbance of the cluster, the position of the centre is likely to be different from the outer regions or one obtained assuming the cluster is symmetric.
As described in Section \ref{sect:analysis}, the centre of the main cluster is left free in the analysis.

\subsection{Multipole magnitude}
Forward modelling of parameters has some advantages over values measured directly from images.
For example, we can include the PSF of the source and the background within the modelling and use MCMC to get a better idea of the parameter uncertainties.
Here we create forward models of the cluster which include azimuthal variations similar to power ratios.
The surface brightness of a symmetric model profile $S(r)$ is modified with angle to produce the following distribution,
\begin{equation}
  S'(r, \theta) = [1 + M_m \sin(m \theta + \theta_0)] \: S(r),
\end{equation}
where $m$ is a multipole index, $\theta_0$ is an inclination angle which ranges from 0 to $360/m \deg$, and $M_m$ is the multipole magnitude which varies between 0 and 1.
Images of models with these multipole variations are shown in Fig.~\ref{fig:shape_example}.
For \texttt{MBProj2D} we compute the model image in Cartesian coordinates before comparison with the data.
Firstly, the pixel sky coordinates ($x,y$) are rotated on the sky in reverse around the cluster centre by $\theta_0$ to compute rotated coordinates ($x_r,y_r$).
Using $\sin \theta = y_r / r$ and $\cos \theta = x_r / r$, we can compute $\sin (m \theta)$ using the standard multiple-angle formulae.
When $r=0$, we use $\theta=0$ above.

In our analysis we compute the morphological parameters $M_1$ to $M_4$ by fitting the multipole model to the data.
As can be seen by the images, the parameter $M_1$ has a shape which is somewhat similar to the shape of a cluster with the slosh parameter ($H$) applied.
This leads to a correlation between these parameters, particularly for objects with lower data quality.
Parameter $M_2$ also looks similar to an elliptical cluster ($\epsilon$) and so these parameters are also correlated.

\subsection{Test of new parameters}
\label{sect:chandra_descr}
We tested our new morphological parameters on images of 83 galaxy clusters detected from South Pole Telescope (SPT) surveys \citep{Bleem15} using the Sunyaev-Zel'dovich (SZ) technique and observed by \emph{Chandra} in the X-ray waveband \citep{McDonald13SPT,Sanders18}.
Section \ref{sect:spt_data} describes these data and their analysis.
These clusters show a wide variety of different morphologies and their images have high spatial resolution and good data quality.
The data were analysed seven times, one for each of the morphological parameters $\epsilon$, $H$, $M_1$, $M_2$, $M_3$ and $M_4$, and also for a spherical cluster.

\begin{figure}
  \includegraphics[width=\columnwidth]{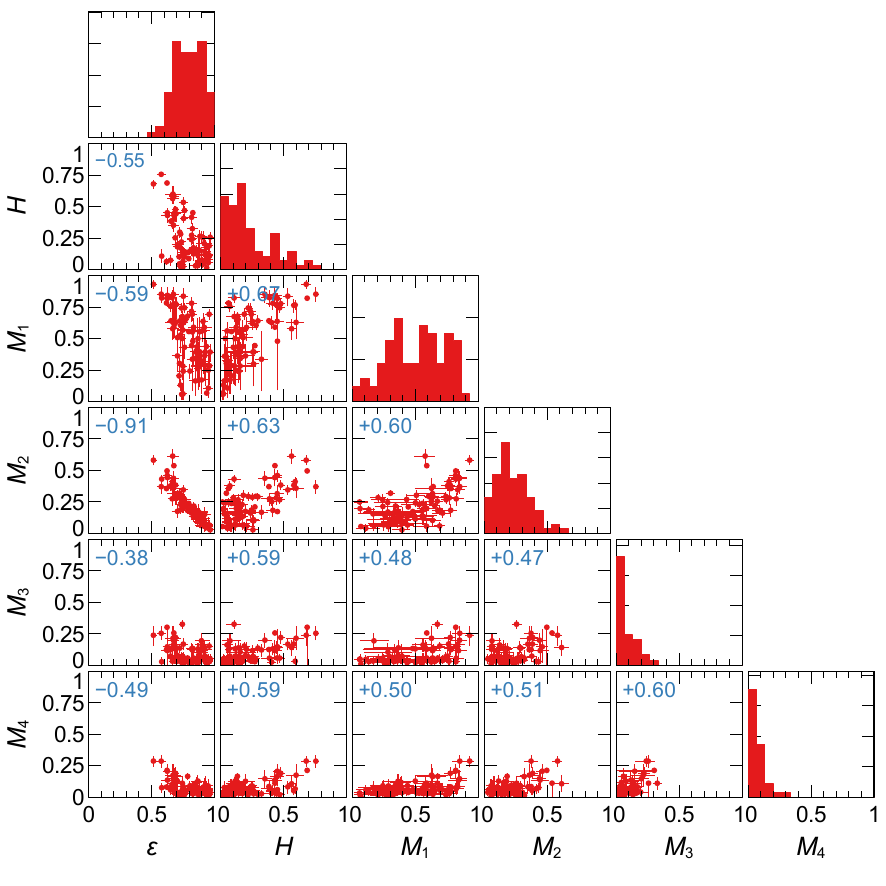}
  \caption{
    Corner plot of the median \texttt{MBProj2D} shape parameters for \emph{Chandra} observations of clusters in the SPT cluster sample.
    The quantities are plotted against each other, while the rightmost panels show the probability density of the values.
    The Pearson correlation coefficient is shown in each panel.
  }
  \label{fig:spt_shape}
\end{figure}

Fig.~\ref{fig:spt_shape} shows a corner plot of the parameters for the clusters in the sample.
The plot shows that there is significant variation in these parameters between the clusters.
Some parameters are clearly correlated with each other, including $\epsilon$ and $M_2$, and $M_1$ and $H$.

\begin{figure}
  \includegraphics[width=\columnwidth]{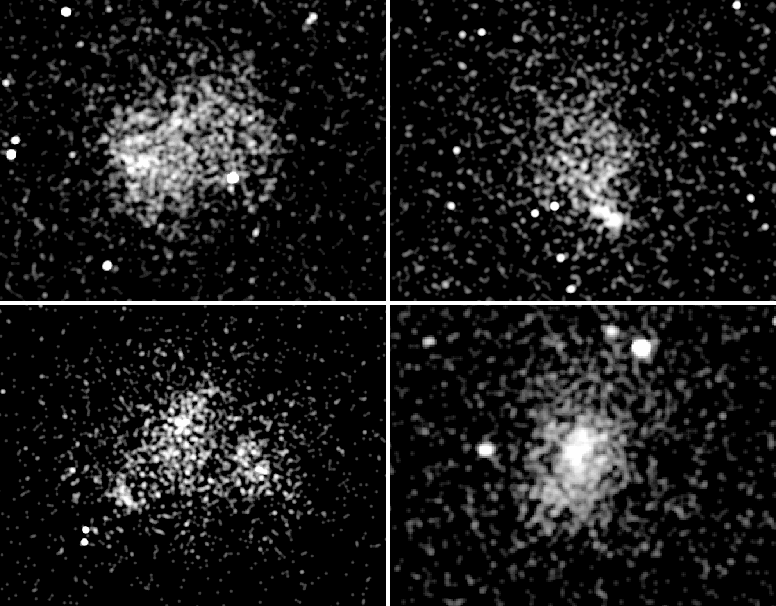}
  \caption{
    Example \emph{Chandra} images from the SPT sample shown in Fig.~\ref{fig:spt_shape}.
    (Top left) SPT-CLJ0014-4952 [$\epsilon=0.85$, $H=0.50$, $M_1=0.36$, $M_2=0.26$, $M_3=0.06$, $M_4=0.03$]. (Top right) SPT-CLJ0307-6225 [$\epsilon=0.58$, $H=0.75$, $M_1=0.88$, $M_2=0.37$, $M_3=0.27$, $M_4=0.28$]. (Bottom left) SPT-CLJ0304-4401 [$\epsilon=0.75$, $H=0.14$, $M_1=0.64$, $M_2=0.26$, $M_3=0.33$, $M_4=0.11$]. (Bottom right) SPT-CLJ2342-5411 [$\epsilon=0.64$, $H=0.08$, $M_1=0.71$, $M_2=0.38$, $M_3=0.07$, $M_4=0.07$].
  }
  \label{fig:spt_examples}
\end{figure}

If we select clusters which are indicated as having a strong non-circular geometry by these parameters, we find these are the most disturbed.
In Fig.~\ref{fig:spt_examples} we see SPT-CLJ0014-4952 has a strong slosh and $M_2$ signal.
SPT-CLJ0307-6225 has pronounced non-circular $\epsilon$, $H$ and $M_1$.
SPT-CLJ0304-4401 has a large $M_1$, $M_2$ and $M_3$ signal because it has a triangular morphology.
SPT-CLJ2342-5411 is elliptical, with large $\epsilon$ and $M_1$ values.
The modelling appears to perform well by picking up the disturbances that the human eye can see in the images.
We further compare the \emph{Chandra} and eRASS1 cluster results in Section \ref{sect:other_samp}.

\section{Data analysis}
\label{sect:analysis}
\subsection{Introduction}
The inputs of our analysis are created using the same pipeline as in B24, although we only use the monochromatic X-ray images in a band of 0.2-2.3 keV.
We only use the data from the cameras without a light leak (i.e. the so-called TM8 data which combines data from telescope modules, TMs, 1, 2, 3, 4 and 6).
These images and exposure maps were produced using the standard \texttt{eSASS} analysis software \citep{Brunner22}, version \texttt{eSASSusers\_211214\_0\_4}, which is equivalent to the publicly available version \texttt{eSASS4DR1}.
We assume that the clusters lie at the best redshift from the cluster catalogue.
For those parameters which assume a value of $R_{500}$, this value also comes from the cluster catalogue.
The best fit positions we use, for consistency, comes from our \texttt{MBProj2D} analysis assuming a radially symmetric model, and are not necessarily the same as in B24.

\subsection{\texttt{MBProj2D} analyses}
Several of the parameters come from the result of modelling the cluster with \texttt{MBProj2D}.
For these analyses, we follow a common procedure which we describe here.
As we are only fitting monochromatic images, we assume isothermal temperature profiles fixed to the median temperature from the MCMC chain obtained by B24 and use the Galactic absorbing column density from the catalogue.

\begin{table*}
  \caption{Priors on the \texttt{MBProj2D} analyses with the different models.}
  \centering
  \begin{tabular}{lll}
    \hline
    Description & Parameter & Prior \\ \hline
    \multicolumn{3}{c}{Standard (circular and other models)} \\ \hline
    Central density & $\ln (n_{0} / \mathrm{cm}^{-3})$ & $\mathcal{U}(-14, 5)$ \\
    Outer slope & $\beta$ & $\mathcal{TN}(2/3, 1/3, 0.2, \infty)$ \\
    Inner slope & $\alpha$ & $\mathcal{TN}(0, 5, -10, 10)$ \\
    Core radius & $\ln (r_\mathrm{c}/\mathrm{kpc})$ & $\mathcal{TN}(\ln 400, 3, \ln 5, \ln 5000)$\\
    Cluster R.A. shift$^*$ & $\delta x / \mathrm{arcsec}$ & $\mathcal{N}(0, 30)$ \\
    Cluster Dec. shift$^*$ & $\delta y / \mathrm{arcsec}$ & $\mathcal{N}(0, 30)$ \\
    Background & $\ln (B/\mathrm{ct}/\mathrm{s}/\mathrm{arcsec}^{2})$ & $\mathcal{U}(-18.4,-8.4)$ \\
    Point source normalisation$^{**}$ & $\ln(N/\mathrm{photon}/\mathrm{keV}/\mathrm{cm}^2/\mathrm{s})$ & $\mathcal{U}(-17,-12)$\\
    \hline
    \multicolumn{3}{c}{Elliptical} \\ \hline
    Rotation angle${}^\dagger$ & $\theta_0 /{}^{\circ}$ & $\mathcal{U}^s(\theta_\mathrm{fit}-90, \theta_\mathrm{fit}+90)$ \\
    Ellipticity & $\epsilon$ & $\mathcal{U}^s(0, 0.999)$ \\
    \hline
    \multicolumn{3}{c}{Slosh} \\ \hline
    Rotation angle${}^\dagger$ & $\theta_0 /{}^{\circ}$ & $\mathcal{U}^s(\theta_\mathrm{fit}-180, \theta_\mathrm{fit}+180)$ \\
    Slosh & $H$ & $\mathcal{U}^s(0.001, 1)$ \\
    \hline
    \multicolumn{3}{c}{Multipole 1} \\ \hline
    Rotation angle${}^\dagger$ & $\theta_0 /{}^{\circ}$ & $\mathcal{U}^s(\theta_\mathrm{fit}-180, \theta_\mathrm{fit}+180)$ \\
    Multipole magnitude & $M_1$ & $\mathcal{U}^s(0, 1)$ \\
    \hline
    \multicolumn{3}{c}{Multipole 2} \\ \hline
    Rotation angle${}^\dagger$ & $\theta_0 /{}^{\circ}$ & $\mathcal{U}^s(\theta_\mathrm{fit}-90, \theta_\mathrm{fit}+90)$ \\
    Multipole magnitude & $M_2$ & $\mathcal{U}^s(0, 1)$ \\
    \hline
    \multicolumn{3}{c}{Multipole 3} \\ \hline
    Rotation angle${}^\dagger$ & $\theta_0 /{}^{\circ}$ & $\mathcal{U}^s(\theta_\mathrm{fit}-60, \theta_\mathrm{fit}+60)$ \\
    Multipole magnitude & $M_3$ & $\mathcal{U}^s(0, 1)$ \\
    \hline
    \multicolumn{3}{c}{Multipole 4} \\ \hline
    Rotation angle${}^\dagger$ & $\theta_0 /{}^{\circ}$ & $\mathcal{U}^s(\theta_\mathrm{fit}-45, \theta_\mathrm{fit}+45)$ \\
    Multipole magnitude & $M_4$ & $\mathcal{U}^s(0, 1)$ \\
    \hline
  \end{tabular}
  \tablefoot{
    Here $\mathcal{U}(a,b)$ is a uniform prior between $a$ and $b$.
    $\mathcal{N}(\mu,\sigma)$ is a normal distribution centred on $\mu$ with width $\sigma$.
    $\mathcal{TN}(\mu,\sigma,a,b)$ is a truncated normal distribution with the peak at $\mu$, with width $\sigma$, truncated below $a$ or above $b$.
    $\mathcal{U}^s(a,b)$ is a uniform distribution with softened edges, where the log prior is $1-e^{(a-x)/0.01}$ if $x<a$, $1-e^{(x-b)/0.01}$ if $x>b$, and zero otherwise.
    In this case, during model evaluation, the parameter is forced to the range of $a$ to $b$.
    $^*$ Only the main cluster position is allowed to vary from the source detection position during the MCMC run. Other cluster positions are once fitted for, then fixed to these positions.
    If fitting using the peak position, these parameters are frozen.
    $^{**}$ Normalisations are defined at 1~keV assuming a powerlaw photon index of $1.7$.
    $^\dagger$ The best fitting value of $\theta_0$ is used as $\theta_\mathrm{fit}$ to reduce boundary issues when sampling.
    In the output catalogue, the value of $\theta_0$ is reduced to the range about $\theta_\mathrm{fit}=0$.
  }
  \label{tab:priors}
\end{table*}

The cluster profiles in B24 use a fairly strong prior on the inner density slope model parameter, $\alpha$.
This leads to some bias on the central slope measurement and concentration.
Therefore we reanalysed the cluster images using a density model with a weak prior on $\alpha$ (Table \ref{tab:priors}).
As discussed by \cite{Kaefer19}, clusters typically follow a $\beta$ model with $\beta=2/3$ in the outskirts, provided the core is otherwise modelled.
The data quality in the eRASS1 survey also limits how sophisticated a model can be fitted to the data.
Therefore we use a simplified form for the density profile of \cite{Vikhlinin09}, compared to B24.
We allow $\beta$, the standard $\beta$-model outer slope, to vary over a more restricted range around $\beta=2/3$ using a prior.
The assumption of $\beta=2/3$ could bias results in objects which deviate from that (e.g. flatter profiles in groups, see for example \citealt{Johnson09}), although those parameters which depend on the density profile are sensitive to the central density.
We do not allow the outermost slope in the parametrization to vary, by fixing $\epsilon$ to 0 and freezing $r_s$.
The second $\beta$ component is also not included in the model.
The fitted profile therefore is therefore a $\beta$ model with an inner powerlaw core and has the form
\begin{equation}
  n_e^2 = n_0^2 \frac{(r/r_c)^\alpha}{[1+(r/r_c)^2]^{3\beta-\alpha/2}},
\end{equation}
where $n_0$ is the central density and $r_c$ is the core radius.
We note that due to the definition being density-squared, $\alpha$ here is twice the inner density slope and is not the same as the cluster density inner slope morphological parameter, $\alpha$.

The input images use the same masking and sizes as B24.
If a point source was fitted for in B24, we fit for it here.
Unlike B24, we only use monochromatic 0.2-2.3 keV band input images and use a fixed 4~arcsec bin size to ensure the best resolution.
As our PSF, we took the standard eROSITA calibration 1.486 keV survey-averaged PSF, but averaged it in radius to make it symmetric.
Signal outside 4~arcmin radius was also removed to match the definition of the eROSITA ancillary response matrices.

As in B24, we assume the sky background is flat across the fields examined.
As most of the background in this energy band is due to X-ray sources, rather than noise intrinsic to the detector, we multiplied this background by the vignetted exposure map.
Faint point sources with $\mathtt{ML\_RATE}<0.4$ in the 0.2-2.3 keV band were masked out in the fit, while brighter ones were fitted as part of the model.
If there were other clusters within the fitted image, these were included separately in the model.
The position of the main cluster was allowed to vary in the MCMC analysis, while point source positions and positions of other clusters were fixed at their best-fitting values.

\texttt{MBProj2D} was run with a standard spherically-symmetric cluster model on the data.
In this analysis, we fitted the model to the data to obtain the best fitting, maximum-likelihood parameters.
We then conducted an MCMC analysis to obtain chains of parameters.
As we described in Section \ref{sect:params}, these chains were processed to calculate a random subset of density profiles.
From these profiles, the distributions of the central density, inner density slope (cuspiness) and concentration were calculated.
The resulting distributions were then used to compute the median values and $1\sigma$ percentiles.
The median cluster position in the chain is used as the `best fitting' cluster position and used to derive $F$.

For the shape parameters, $\epsilon$, $H$, $M_1$, $M_2$, $M_3$ we repeated the fits to the data using \texttt{MBProj2D}, but allowing the shape parameters to vary.
In these analyses, we took the best fitting model from the above symmetric fit, added new free parameters which vary the morphology, then refit the data.
Rather than give the median parameter values from the chain for the shape parameters or angles (which tend to 0.5 for the shape parameters for low count rates), we provide the maximum-likelihood values, but compute the uncertainties using the MCMC distribution about the best fit.

We also repeated the first \texttt{MBProj2D} analysis with a fixed cluster peak position.
The peak was computed from a Gaussian smoothed image of the cluster (see Section \ref{sect:fit_peak_offset}).
The fit started from the previous best-fitting parameters and the analysis was identical, except for freezing the main cluster position to the location of its peak.

\subsection{Image-derived parameters}
\label{sect:image_params}
Several parameters are derived directly from the X-ray images rather than fitting models.
Power ratios, the Gini coefficient and centroid shift calculations require input images which do not contain point sources or neighbouring clusters.
Therefore for these measurements we need to `fill' in the regions occupied by these neighbouring sources.
The photon asymmetry parameter, however, is derived using a mask of these structures.

All these images initially require a mask.
Firstly we computed a model of the sky from the best fit radially symmetric \texttt{MBProj2D} model, including the effect of the PSF, but not background.
We masked out regions where neighbouring clusters or point sources were at most half the contribution of the main cluster.
In addition we masked out regions where these objects were more than $1/10$ of the background level.

To make the filled input images where these masked regions were replaced, we computed the best fitting model of the main cluster and background from the \texttt{MBProj2D} analysis, including PSF, and computed a Poisson realisation of this.
The regions which were masked out in the input image were replaced by this simulated image.

\subsection{SPT cluster data analysis}
\label{sect:spt_data}
For comparison with the eROSITA clusters, we study a subset of SPT-detected clusters \citep{Bleem15}, observed in X-rays using \emph{Chandra} and previously analysed in \cite{Sanders18}.
For the \emph{Chandra} analysis we use the same X-ray data and X-ray bands as described in \cite{Sanders18}.
However, we use \texttt{MBProj2D}, rather than MBProj2 as described in that paper and therefore fit images rather than profiles.

The results for these clusters are compared with both the standard eRASS1 processing and those obtained from deeper eRASS:4 data.
To make a better comparison for the eRASS:4 data we use very similar analysis procedures for both \emph{Chandra} and the deeper eROSITA data.
These data and procedures are close to those described in B24 for the eROSITA-\emph{Chandra} flux comparison.
We use a very similar mask for both datasets, where the cluster is masked beyond a radius of 5~Mpc and outer point sources outside a radius of 3 arcmin from the cluster core are excluded with a mask size with a minimum of 0.5 arcmin radius.
The eRASS:4 eROSITA data are masked where they are outside of the field of view of the \emph{Chandra} data.
In the inner part of the cluster, due to the high spatial resolution, the point sources are masked for the \emph{Chandra} analysis, while the sources are modelled within \texttt{MBProj2D} for the eROSITA data with the positions fixed from \emph{Chandra} values.
The \emph{Chandra} data are binned to a spatial resolution of 4 pixels (2 arcsec), while we use the standard eROSITA spatial binning.

For both analyses, the clusters are assumed to be isothermal.
When analysing the density profile for the symmetric cluster profile case for \emph{Chandra}, we fit multiple energy bands in order to constrain the cluster temperature.
The \emph{Chandra} background in each energy band is assumed to have the shape of the standard blank sky background, but with a variable log normalisation factor.

For eROSITA we fit monochromatic 0.2 to 2.3 keV TM8 images of the clusters, but fix the temperature to the median value obtained by \emph{Chandra}.
The centre of the cluster is fixed to the \emph{Chandra} peak.
When studying the morphological parameters which affect the shape of the cluster (see Section \ref{sect:chandra_descr}), we fit a single band \emph{Chandra} image in the band 0.5 to 4.0 keV, again fixing the temperature at the value obtained from the multi-band fit.

For both eRASS:4 and \emph{Chandra} we assume the same density parametrization as our eRASS1 analysis, with the same priors on the parameters.
The metallicity is fixed to be 0.3 solar, like for eRASS1.

\section{Cluster simulations}
\label{sect:sim}
To better understand our results we undertook simplified simulations of clusters on the eROSITA sky.
The main aim of these simulations is to investigate the parameter biases due to the measurement process and the quality of data.
In addition we wanted to check whether cluster morphology can have a significant affect on the eROSITA selection function.

Rather than making a full eROSITA simulation of the whole sky \citep{Comparat20,Seppi22} we made simulations of individual clusters at random positions on the eROSITA sky and attempted to detect and measure their properties.
The clusters were assigned luminosities and redshifts from a grid, with 9 logarithmic-spaced values of $L_{500}$ between $10^{43}$ and $10^{45}$~erg~s$^{-1}$ and redshift values of $0.02$, $0.035$, $0.05$, $0.07$, $0.1$, $0.15$, $0.2$, $0.3$, $0.4$, $0.5$, $0.6$, $0.7$, $0.8$, $1.0$ and $1.2$.
Using the covariance matrix method described in \cite{Comparat20}, we generated random cluster masses, temperatures and profiles.
Clusters were assigned to a grid point by generating a mass function for the relevant redshift and assigning randomly generated clusters nearby in mass and redshift.
Those clusters with luminosities close to the grid point value were assigned to it.

To make the simulated image of a randomly selected cluster for the grid point, we used \texttt{MBProj2D} to generate a model image for the cluster profile, temperature and Galactic absorption.
The images are created in the band 0.2 to 2.3 keV, with a dimension of 70~arcmin for $z=0.02$, 60~arcmin for $z=0.035$, 50~arcmin for $z=0.05$ and 40~arcmin otherwise.
The cluster location was chosen randomly to lie within the region of sky that the cluster sample is obtained from B24.
The image was convolved with the eROSITA PSF and multiplied by the eROSITA eRASS1 exposure map at that sky position.
Wavelet-filtered maps of the eROSITA sky were used as a background, where structures below scales of around 30 arcmin were removed.
No AGN were included in the simulations.
We split the eROSITA TMs into two sets (TMs 5 and 7, and the rest), which were simulated separately.
Poisson realisations of the two sets of model images were created.
The summed image of the two sets of TMs was used for detection, while only the one excluding TMs 5 and 7 was used for characterisation.

To tell whether a cluster would be detected, we applied the \texttt{eSASS} pipeline to the simulated image.
Initially, \texttt{erbox} is used on the simulated image to make a list of sources (with \texttt{likemin=6}, \texttt{nruns=3}, \texttt{boxsize=4} and \texttt{bkima\_flag=N}).
This source list was supplied to \texttt{erbackmap} to make an initial background map (using \texttt{scut=0.00005}, \texttt{mlmin=6}, \texttt{maxcut=0.5}, \texttt{smoothval=15}, \texttt{snr=40}, \texttt{smoothmax=360}).
\texttt{erbox} is run a second time using the background map (with \texttt{likemin=4} and \texttt{bkima\_flag=Y}).
We run \texttt{erbackmap} a second time to produce a new background map based on the generated source list.
\texttt{erbox} is run again using the previous background map and a new background map is generated using \texttt{erbackmap}.
We now run the maximum-likelihood source detection based on the previous source list, using \texttt{ermldet} with parameters \texttt{likemin=5}, \texttt{extlikemin=3}, \texttt{cutrad=15}, \texttt{multrad=15}, \texttt{extmin=2}, \texttt{extmax=15}, \texttt{nmaxfit=4}, \texttt{nmulsou=2}, \texttt{shapelet\_flag=no} and \texttt{photon\_flag=no}.
A cluster is detected if there is an object in the output catalogue with non-zero extent within 2 arcmin of the input position.
These parameters are the same as used for the main eROSITA catalogue.

One of the major differences between this simplified detection pipeline and the standard one is that we use image-based rather than photon-based detection.
Photon-based detection uses a separate PSF for each photon, making use of smaller PSF at the centre of the telescopes relative to the outskirts as sources scan across the field of view.
Image-based detection uses an average survey PSF and is significantly faster, but should be less sensitive close to the detection threshold.

The detected clusters are analysed in a very similar way to the real clusters.
The images of the clusters, excluding TMs 5 and 7, are analysed with the \texttt{MBProj2D} and image-based routines.
The background component is assumed to have a flat count rate over the 40 arcmin region.
To obtain the density-based parameters, such as concentration, we fit a Vikhlinin model using \texttt{MBProj2D} with the same priors as used for the real data here.
Rather than run a full MCMC analysis for each simulated cluster, we just find the maximum-likelihood parameters.
The forward-modelled shape parameters ($\epsilon$, $H$, $M_1$, $M_2$, $M_3$ and $M_4$) are obtained by fitting in \texttt{MBProj2D} using the same density profile priors as for the density analysis.
We note that the centre of the cluster is allowed to be free during these fits.
The power ratios, photon asymmetry, centroid shift and Gini parameters are calculated from the simulated X-ray image, however assuming the input $R_{500}$, rather than determining it from the data.

In our input models we simulate a variety of cluster shapes including spherical, elliptical (with $\epsilon=0.7$), sloshed (with $H=0.3$), or with $M_1$, $M_2$, $M_3$ or $M_4$ equal to 0.3.
This allows us to test our ability to recover the input parameter shape values or examine how cluster shape affects the obtained image-based parameters (such as power ratios or photon asymmetry).

\begin{figure*}
  \includegraphics[width=\textwidth]{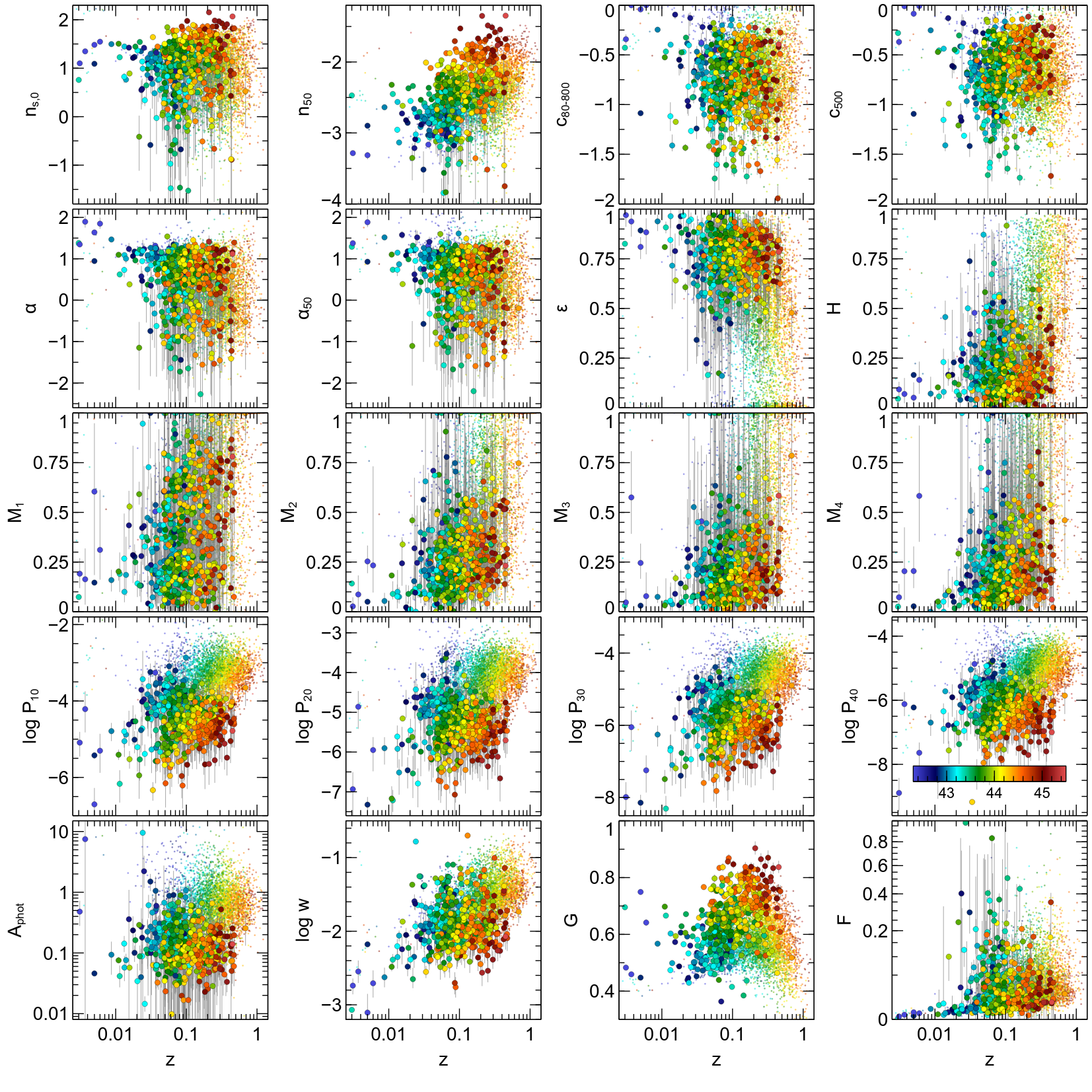}
  \caption{
    Morphological parameters against redshift.
    The parameters are plotted for the bright cluster subset ($\geq 300$ counts) with larger markers.
    The small markers show clusters with at least 25 counts, without plotting uncertainties.
    The colour scale shows the X-ray luminosity (log erg~s$^{-1}$ in the 0.2-2.3 keV band) inside an aperture of radius 800 kpc.
    The parameters are the central scaled gas density ($n_{s,0}$), central electron density ($n_{50}$), concentration from 80-800 kpc ($c_{80-800}$), concentration from $0.1R_{500}$ to $R_{500}$ ($c_{500}$), central density slope ($\alpha$), slope at fixed radius ($\alpha_{50}$), ellipticity ($\epsilon$), slosh ($H$), multiple magnitudes ($M_1$ to $M_4$), power ratios ($P_{10}$ to $P_{40}$), photon asymmetry ($A_\mathrm{phot}$), centroid shift ($w$), Gini coefficient ($G$) and fit-peak offset ($F$).
  }
  \label{fig:param_z}
\end{figure*}

\begin{figure*}
  \includegraphics[width=\textwidth]{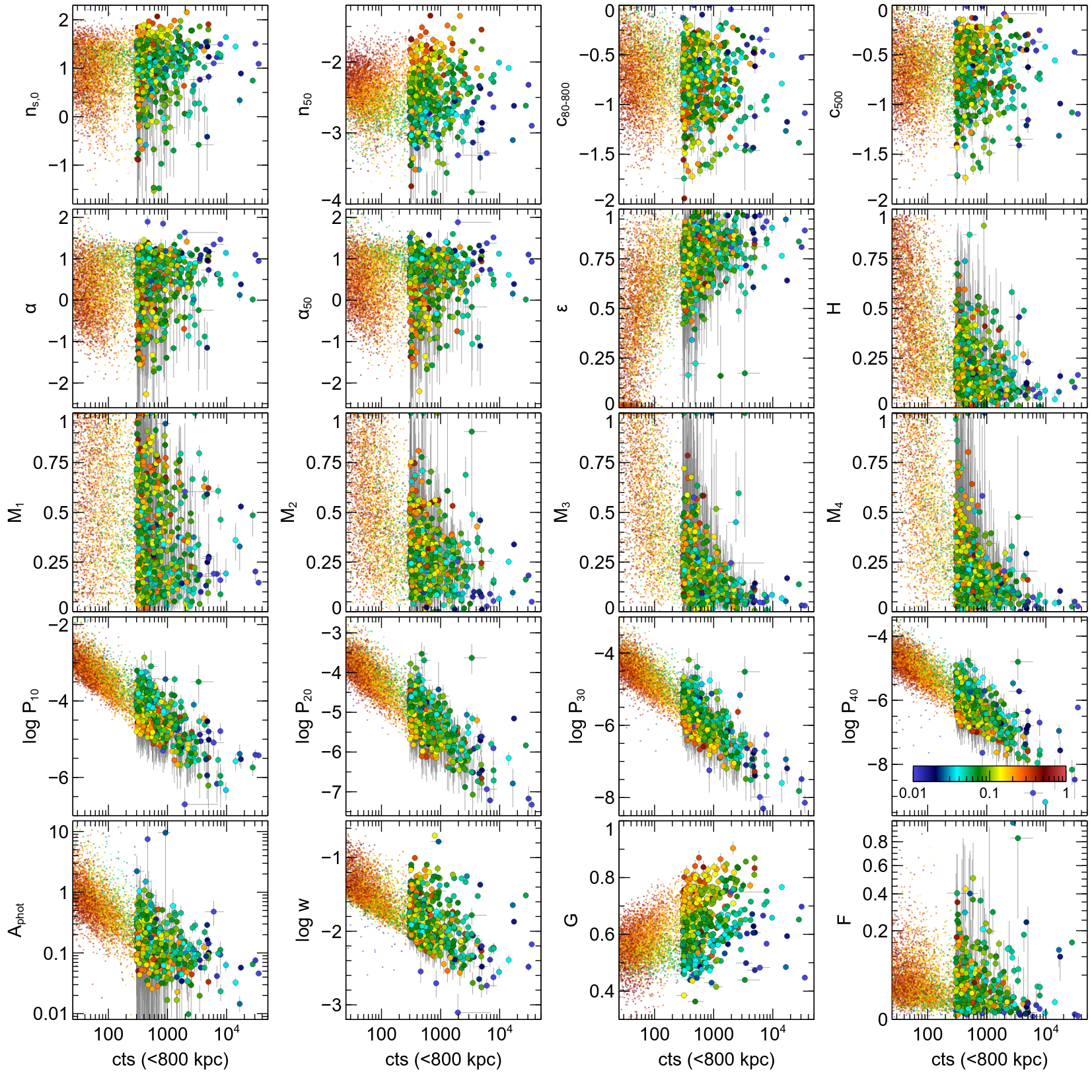}
  \caption{
    Parameters against number of counts in an 800 kpc aperture.
    These parameters are plotted for the bright cluster subset ($\geq 300$ counts) with larger markers.
    The small markers show clusters with at least 25 counts, without plotting uncertainties.
    The colour scale shows the cluster redshift.
  }
  \label{fig:param_cts}
\end{figure*}

\begin{figure*}
  \includegraphics[width=\textwidth]{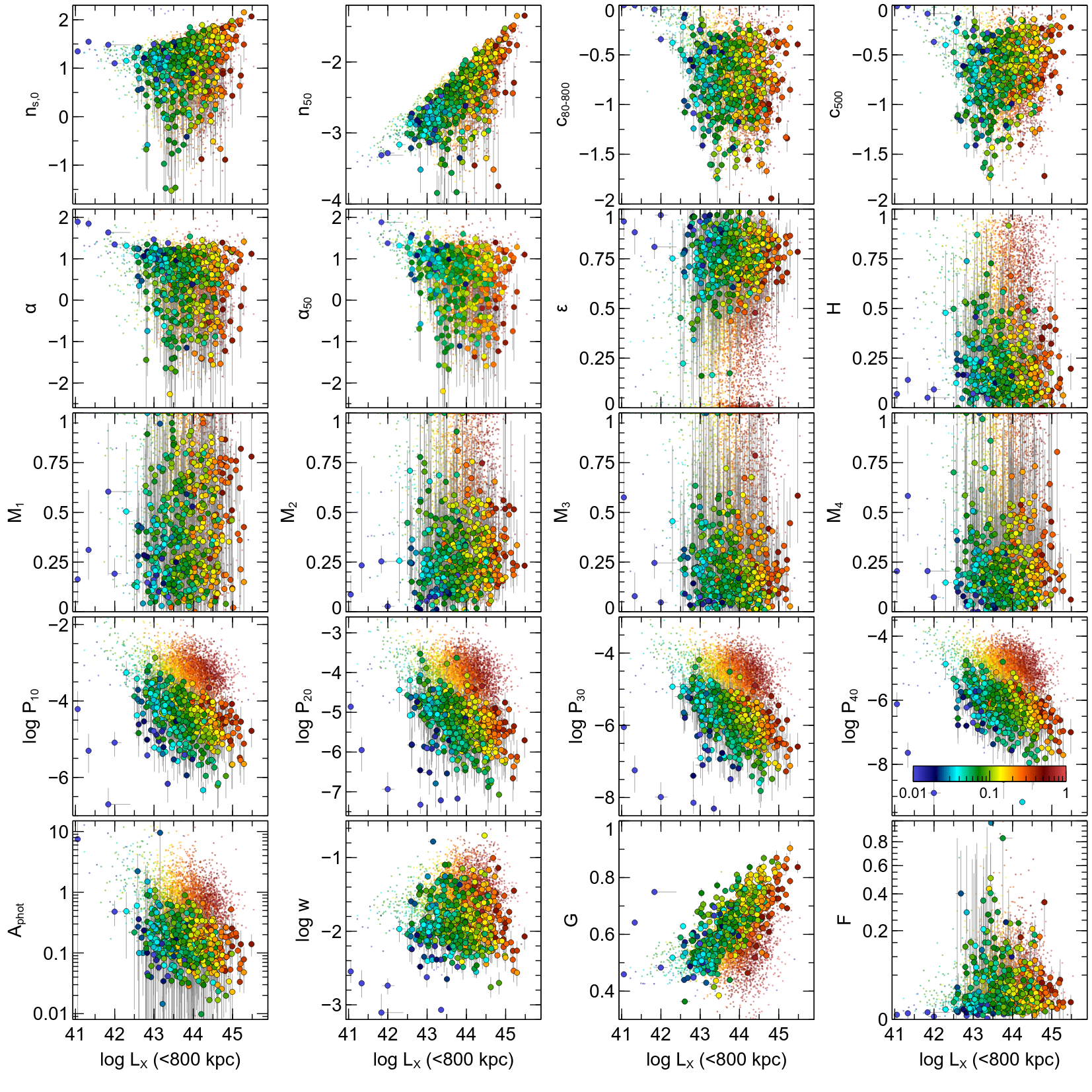}
  \caption{
    Parameters against X-ray luminosity.
    These parameters are plotted for the bright cluster subset ($\geq 300$ counts) with larger markers.
    The small markers show clusters with at least 25 counts, without plotting uncertainties.
    The X-ray luminosity is inside an 800~kpc radius aperture in the 0.2-2.3 keV band.
    The colour scale shows the cluster redshift.
  }
  \label{fig:param_L}
\end{figure*}

\section{Results}
\label{sect:results}

\begin{table}
  \caption{eRASS1 cluster subsamples.}
  \centering
  \begin{tabular}{lr}
    \hline
    Sample & Number \\ \hline
    All clusters & $12075$ \\
    $\geq 25$ counts, $\mathcal{L}_\mathrm{ext}>6$ & $5387$ \\
    $\geq 300$ counts, $\mathcal{L}_\mathrm{ext}>6$ & $541$ \\
    $\geq 400$ counts, $\mathcal{L}_\mathrm{ext}>6$ & $395$ \\
    $\geq 800$ counts, $\mathcal{L}_\mathrm{ext}>6$ & $175$ \\
    $\mathcal{L}_\mathrm{ext}>6$, $\mathcal{L}_\mathrm{det}>20$, cosmology & $4160$ \\
    $\mathcal{L}_\mathrm{ext}>6$, $\mathcal{L}_\mathrm{det}>40$, cosmology & $2484$ \\
    $\mathcal{L}_\mathrm{ext}>6$, $\mathcal{L}_\mathrm{det}>80$, cosmology & $1193$ \\
    $\mathcal{L}_\mathrm{ext}>6$, $\mathcal{L}_\mathrm{det}>320$, cosmology & $360$ \\
    eRASS1/\emph{Planck} comparison & $66$ \\
    eRASS1/SPT comparison & $66$ \\
    eRASS4/SPT comparison & $83$ \\
    eRASS1/eFEDS comparison & $69$ \\
    \hline
  \end{tabular}
  \tablefoot{
    The subsamples marked cosmology were additionally selected to be part of the cosmology subsample from B24.
    Count selection is done within 800 kpc in the 0.2-2.3 keV band.
  }
  \label{tab:subsamples}
\end{table}

In this section we present the morphological parameters obtained from the real eROSITA cluster sample.
Detailed information about the catalogue is given in Appendix \ref{appen:cat}.
In Fig.~\ref{fig:param_z} we plot each of the parameters against the redshift of the source.
For this plot we split the clusters into two subsamples (Table \ref{tab:subsamples}).
The first is a bright subsample of clusters, where the number of counts in a 800~kpc aperture is more than 300 and the number of counts is greater than three times the lower uncertainty on the number of counts.
For these points we colour markers using the X-ray luminosity within the 800~kpc aperture and show the uncertainties on the values.
A fainter subsample with at least 25 significant counts are plotted with smaller markers.

Figure~\ref{fig:param_cts} shows the parameters plotted as a function of the number of counts in the cluster, computed within an aperture of 800~kpc.
As some of our measurements are sensitive to the quality of the data, this plot is ideal for identifying biases due to this.
The colour scale indicates  the redshift of the cluster.
The same results are plotted in Fig.~\ref{fig:param_L}, but showing them as a function of X-ray luminosity within an aperture of 800~kpc.
The colour scale in this figure shows the redshift of the clusters.

When the parameters are plotted as a function of the number of counts, we see that many of the parameters are clearly increasingly biased towards the low count regime.
This is particularly true for those parameters which are not based on the \texttt{MBProj2D} forward modelling method  (e.g. power ratios and photon asymmetry).
In addition we see that when plotting these values as a function of luminosity, the clusters are clearly separated in redshift for these parameters.
The same parameters are also separated as a function of luminosity, if plotted as a function of redshift.

In theory, it could be possible that there are intrinsically strong variations in these parameters as a function of redshift and luminosity.
However, in Section \ref{sect:noise_distance} our simulations show that these strong variations with luminosity and redshift (and therefore counts) are what is expected based on the level of noise in the survey and the PSF size.

The density parameters show increasingly higher values for more luminous or higher redshift clusters.
However, there is no clear evolution with the number of counts.
The scaled central density shows a weaker evolution than the physical density at fixed radius.
The concentration parameters do not show any clear evolution with redshift of luminosity.

For the shape parameters, $\epsilon$, $H$ and $M_1$ to $M_4$, the results become roughly randomly distributed in the lower count regime below 100--200 counts.

\begin{figure*}
  \centering
  \includegraphics[width=0.7\textwidth]{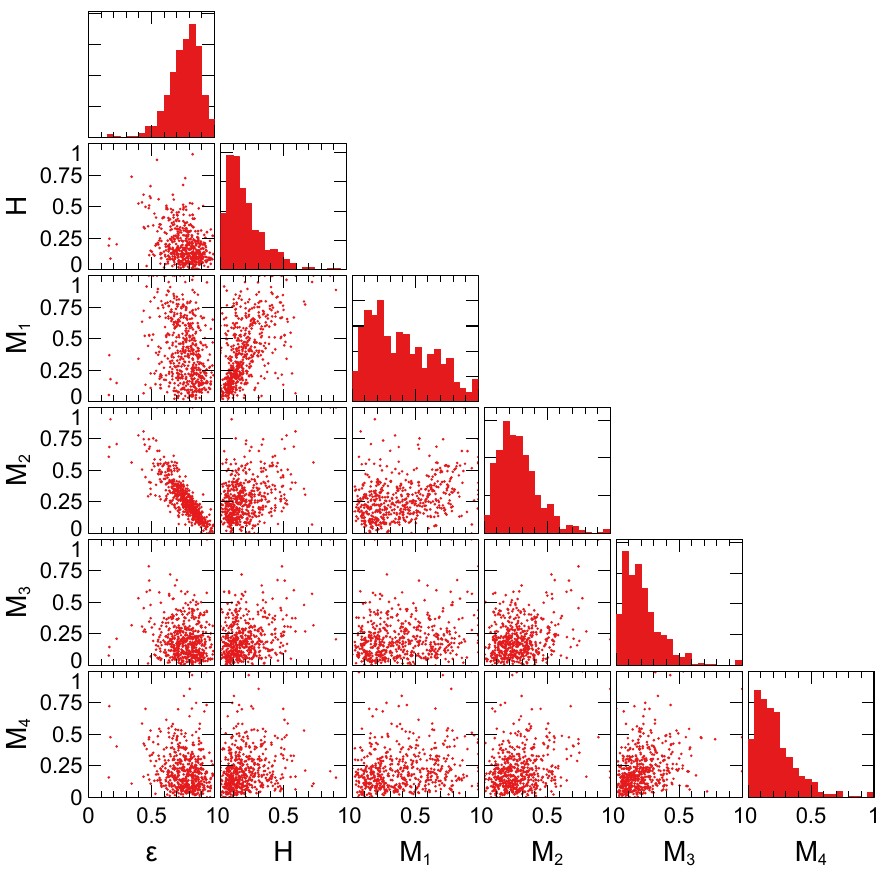}
  \caption{
    Corner plot of the median \texttt{MBProj2D} shape parameters for the bright cluster ($\geq 300$ count) subset.
    The quantities are plotted against each other, while the rightmost panels show the probability density distribution of each value.
  }
  \label{fig:shape_corner}
\end{figure*}

Similarly to the \emph{Chandra}-observed SPT sample (Fig.~\ref{fig:spt_shape}) we can plot a corner plot of the various shape parameters.
Figure \ref{fig:shape_corner} shows the parameters $\epsilon$, $H$, $M_1$, $M_2$, $M_3$ and $M_4$ for the subset of the brightest clusters ($\geq 300$ counts).
The overall distribution of the parameters shows a very similar distribution compared to the SPT sample results, although they are noisier due to the lower number of counts.
We see that some of the parameters are correlated with each other, like for \emph{Chandra}, with $\epsilon$ and $M_2$ being inversely correlated and $H$ and $M_1$ correlated.

\begin{figure*}
  \centering
  \includegraphics[width=0.9\textwidth]{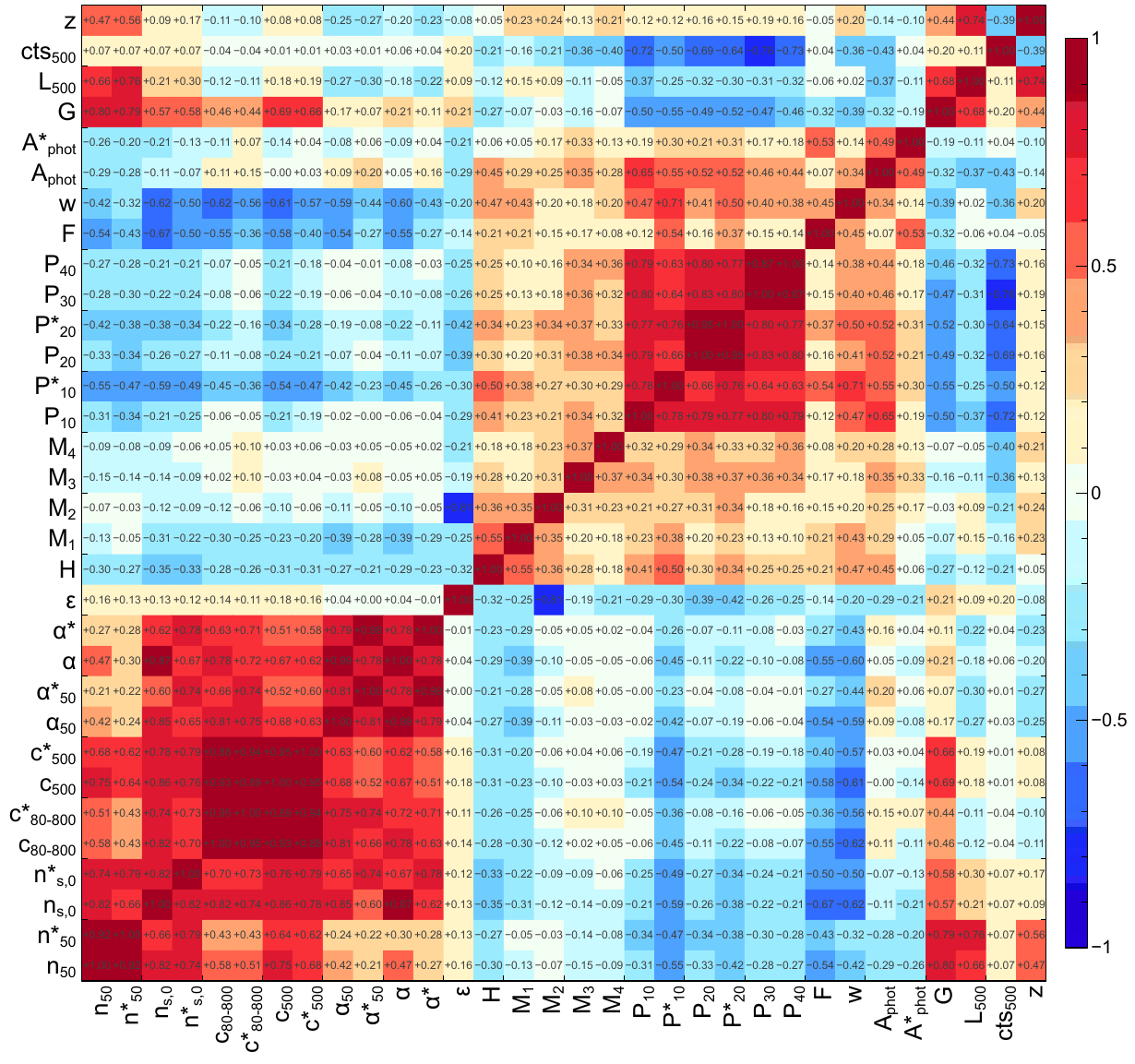}
  \caption{
    Correlation matrix between different parameters for the bright cluster subset.
    Values were measured for clusters with more than 300 counts to reduce the effect of the statistical errors.
    Log values were taken for the photon asymmetry, power ratios and centroid shift.
    We also show the correlation with log cluster luminosity ($L_{500}$), log number of counts in $R_{500}$ ($\mathrm{cts}_{500}$) and log redshift.
  }
  \label{fig:correlation}
\end{figure*}

In Fig.~\ref{fig:correlation} we show a correlation matrix between the various parameters for a subset of bright clusters.
We only include the bright clusters, as the parameters become noise dominated when a cluster is faint, leading to a reduction in any correlation.
In this plot we also include the peak and best fit versions of those parameters when appropriate.
It can be seen that there are generally strong correlations between a high central density, concentration and central slope.
There are some weaker correlations between $n^{*}_{50}$ and the central slope and concentrations, compared to $n_{50}$.
The concentrations measured in physical or scaled apertures, or measured around the peak and best fit positions are all strongly correlated.

There is some mild correlation between the ellipticity (noting that higher ellipticities are rounder clusters) and the central density and concentration.
The parameters $H$, $M_1$ to $M_4$, power ratios, $F$ and $w$ are generally inversely correlated with the central density, concentration and central slope.
The ellipticity and $M_2$ are strongly inversely correlated, while $H$ and $M_1$ show correlation.
The Gini coefficient is correlated with central density, concentration and inner slope.
Several of the parameters measured from images (power ratios, photon asymmetry and centroid shift) are negatively correlated with the number of counts (see Section \ref{sect:noise_distance}).

\begin{figure*}
  \centering
  \includegraphics[width=0.95\textwidth]{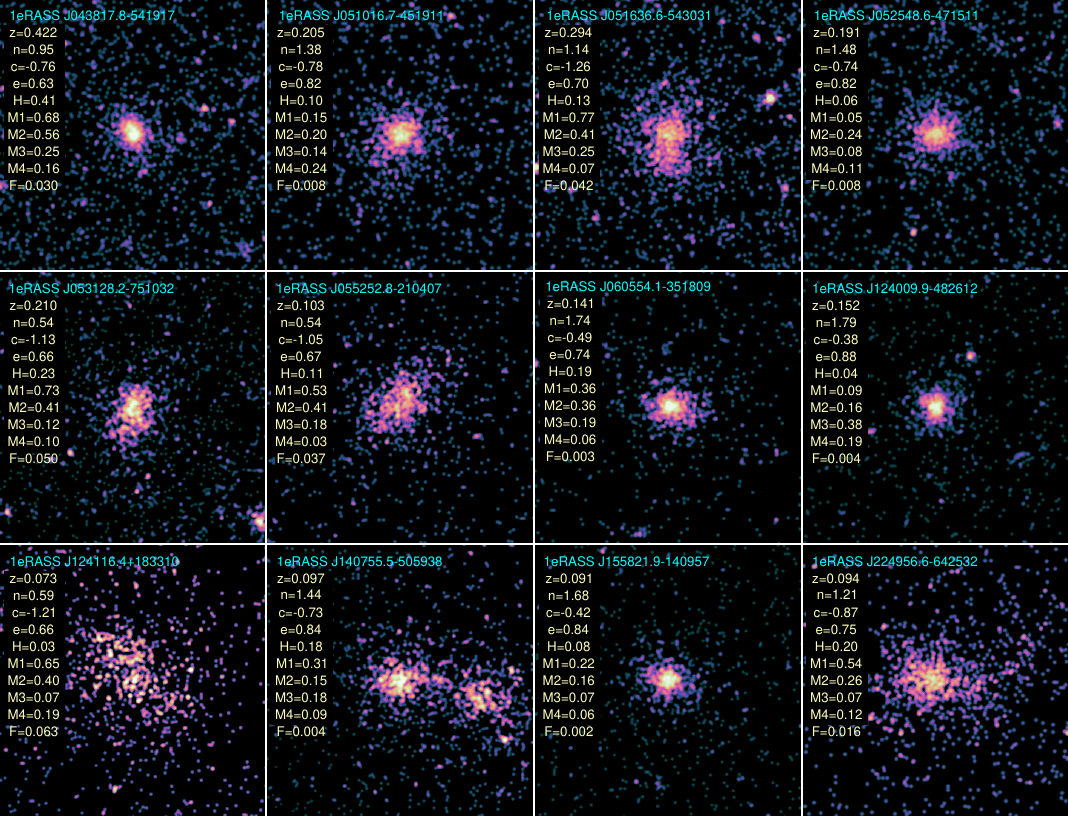}
  \caption{
    Example cluster images and morphological parameters.
    The clusters chosen are those from the catalogue with between 920 and 1080 counts within 800~kpc radius and an uncertainty on the number of counts of less than 5\%.
    The exposure-corrected images are in the 0.2-2.3 keV band with the same angular scale and have been smoothed by a Gaussian with $\sigma=8$~arcsec.
    We show the cluster name, redshift, scaled central density ($n_{\mathrm{s},0}$), concentration ($c_{80-800}$), ellipticity ($\epsilon$), slosh ($H$), multipole magnitudes ($M_1$ to $M_4$) and peak-fit offset ($F$).
  }
  \label{fig:examples}
\end{figure*}

In Fig.~\ref{fig:examples} we show example brighter clusters with a similar number of counts, showing a diversity of morphology.
Examples include a cluster with a high concentration and central density (1eRASS\,J124009.9-482612) and another with low concentration and central density (1eRASS\,J124116.4+183310).
Elliptical clusters (low $\epsilon$) include 1eRASS\,J043817.8-541917, 1eRASS\,J053128.2-751032 and 1eRASS\,J124116.4+183310, while 1eRASS\,J124009.9-482612 has a low ellipticity.
The elliptical clusters generally have larger $M_2$ values.
1eRASS\,J043817.8-541917 has a high slosh ($H$) because of its offset central peak and similarly with 1eRASS\,J053128.2-751032, 1eRASS\,J055252.8-210407, 1eRASS\,J124116.4+183310 and 1eRASS\,J224956.6-642532 has a large $M_1$ value.
1eRASS\,J224956.6-642532 has the largest $M_3$ value because there is bright triangular shape feature around the central peak.
1eRASS\,J051016.7-451911, 1eRASS\,J124009.9-482612 and 1eRASS\,J124116.4+183310 have the largest $M_4$ values in this subsample.
Those clusters with the largest peak-fit offsets are 1eRASS\,J051636.6-543031, 1eRASS\,J053128.2-751032 and 1eRASS\,J124116.4+183310, while 1eRASS\,J124009.9-482612, 1eRASS\,J140755.5-505938 and 1eRASS\,J155821.9-140957 show small offsets.

Although for the most disturbed appearing clusters (e.g. 1eRASS\,J124116.4+183310) several of the parameters indicate their disturbance, there are cases where only one or a few of the parameters indicates some disturbance (e.g. 1eRASS\,J124009.9-482612).
One interesting example is J043817.8-541917, where visual inspection would likely miss the disturbance-indicating values of $\epsilon$, $H$, $M_1$, $M_2$ and $M_3$.


\section{Parameter biases}

\begin{table*}
  \centering
  \caption{List of potential biases.}
  \begin{tabular}{L{0.15\textwidth}p{0.05\textwidth}p{0.35\textwidth}p{0.32\textwidth}}
    \hline
    Name & Section & Description & Effect \\ \hline
    Morphological detection bias & \ref{sect:morph_det_bias} & Morphology of cluster, in particular the peakiness of its profile, affects how easy an object is to be detected. & Produces a biased sample, giving biased distributions of parameters, in particular those associated with detection (e.g. concentration).\\
    Cool core mass and radius bias & \ref{sect:ccmasrad_bias} & Presence of a cool core gives higher luminosity and therefore biased mass or radius, affecting parameter measurements. & Affects parameters measured at scaled radii, see Table \ref{tab:params}, e.g. $c_{500}$, $n_{\mathrm{s},0}$, $\alpha$, $P_{10}$, $A_\mathrm{phot}$, $G$ and $w$.\\
    Position bias & \ref{sect:pos_bias} & Choice of cluster centre affects parameter & Parameters sensitive to core, e.g. $n_{\mathrm{s},0}$, $n_{50}$, $\alpha$, $c_{500}$,  $c_{80-800}$, $P_{10}$ and $A_\mathrm{phot}$.\\
    PSF bias & \ref{sect:psf_bias} & PSF blurs out features affecting parameters which are not model based and increasingly strongly for higher redshifts. & Affects image measured parameters, including power ratios, $G$, $A_\mathrm{phot}$, $w$.\\
    Fixed parameter range bias & \ref{sect:range_bias} & Parameter range is small compared to uncertainties, leading to biases in median values from MCMC. & Affects $\epsilon$, $H$, $M_1$ to $M_4$, and to a lesser extent, concentrations.\\
    Cluster shape selection & \ref{sect:shape_seln} & Related to the Morphological detection bias, a non-circular cluster shape can affect how easy a cluster is to be detected & A sample biased towards more relaxed clusters, affecting the distribution of all parameters \\
    Noise and distance bias & \ref{sect:noise_distance} & More distant clusters produce fewer counts and more blurred images, giving rise to biased parameters. & Affects many image-derived parameters, including power ratios, $G$, $w$, $A_\mathrm{phot}$.\\
    Resolution bias & \ref{sect:res_bias} & Ability to resolve inner parts of cluster depends on its profile, distance and brightness & Not important, if parametric profile is realistic.\\
    Temperature and metallicity bias & \ref{sect:t_bias} & Non-isothermal temperature profile or varying metallicity could affect density estimates. & For groups could affect $n_{50}$, $n_{\mathrm{s},0}$, $\alpha$. \\
    Unresolved point sources & \ref{sect:ptsrc_bias} & Point sources may be unresolved within the cluster. & Could affect all parameters and also lead to biases in the sample selection.\\
    \hline
  \end{tabular}
  \label{tab:biases}
\end{table*}

\label{sect:biases}
\subsection{Introduction}
There are several biases which can affect the obtained parameters.
It is important to understand these biases before using the parameters blindly, as they can be a function of other cluster properties.
For example, some parameters show strong evolution with redshift or luminosity with no change in cluster morphology.
The morphology of the cluster can also affect how likely it is to be selected, presenting a biased subset of clusters with respect to certain characteristics.
We summarise the various biases in Table \ref{tab:biases}.

\subsection{Morphological detection bias}
\label{sect:morph_det_bias}
The morphology of a cluster may affect how easy it is for a cluster to be detected in the X-ray waveband.
For example, cool core clusters are known to have higher luminosities for the same mass compared to non-cool core clusters \citep[e.g.][]{EdgeStewart91,Pratt09,Hudson10,Mittal11b}.
Clusters with disturbed morphologies are known to have lower luminosities.
Therefore for a fixed detection threshold it may be easier to detect cool core clusters and less easy to detect disturbed clusters \citep{Eckert11}, although there are methods which can reduce this bias \citep[e.g.][]{Kaefer20}.
However, as a result of the PSF fitting maximum likelihood detection approach used to make the eROSITA catalogues, there may be objects in which the cool core makes the cluster more likely to be detected as a point source and missed in the cluster catalogue \citep[e.g.][]{Bulbul22}, particularly towards higher redshift where the PSF is more important.
At low redshifts the PSF fitting detection method may miss objects which have flat cores in surface brightness, particularly if the core size becomes larger than the fitting radius \citep{Xu22}.
Large groups or clusters with flat cores may be confused with background variations.

This bias will not change the parameters for an individual object, but as cool core clusters are more regular, the parameters obtained for the population near the flux detection limit or towards higher redshift may be biased by this effect.

\begin{figure*}
  \centering
  \includegraphics[width=0.95\textwidth]{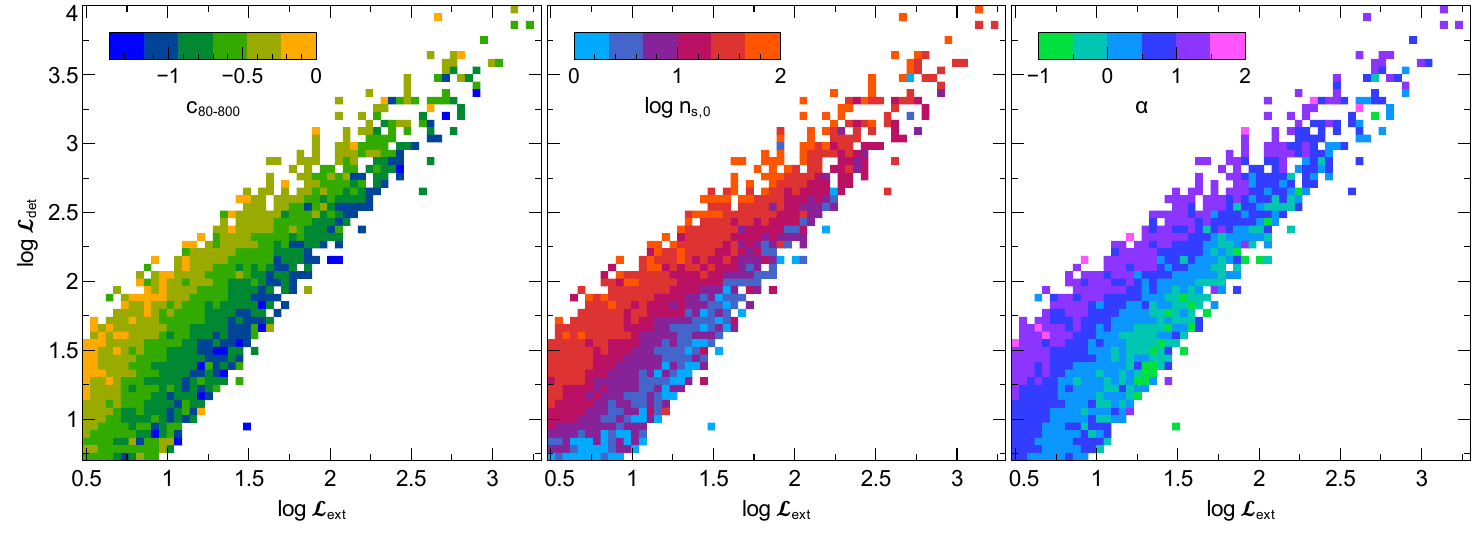}\\
  \includegraphics[width=0.95\textwidth]{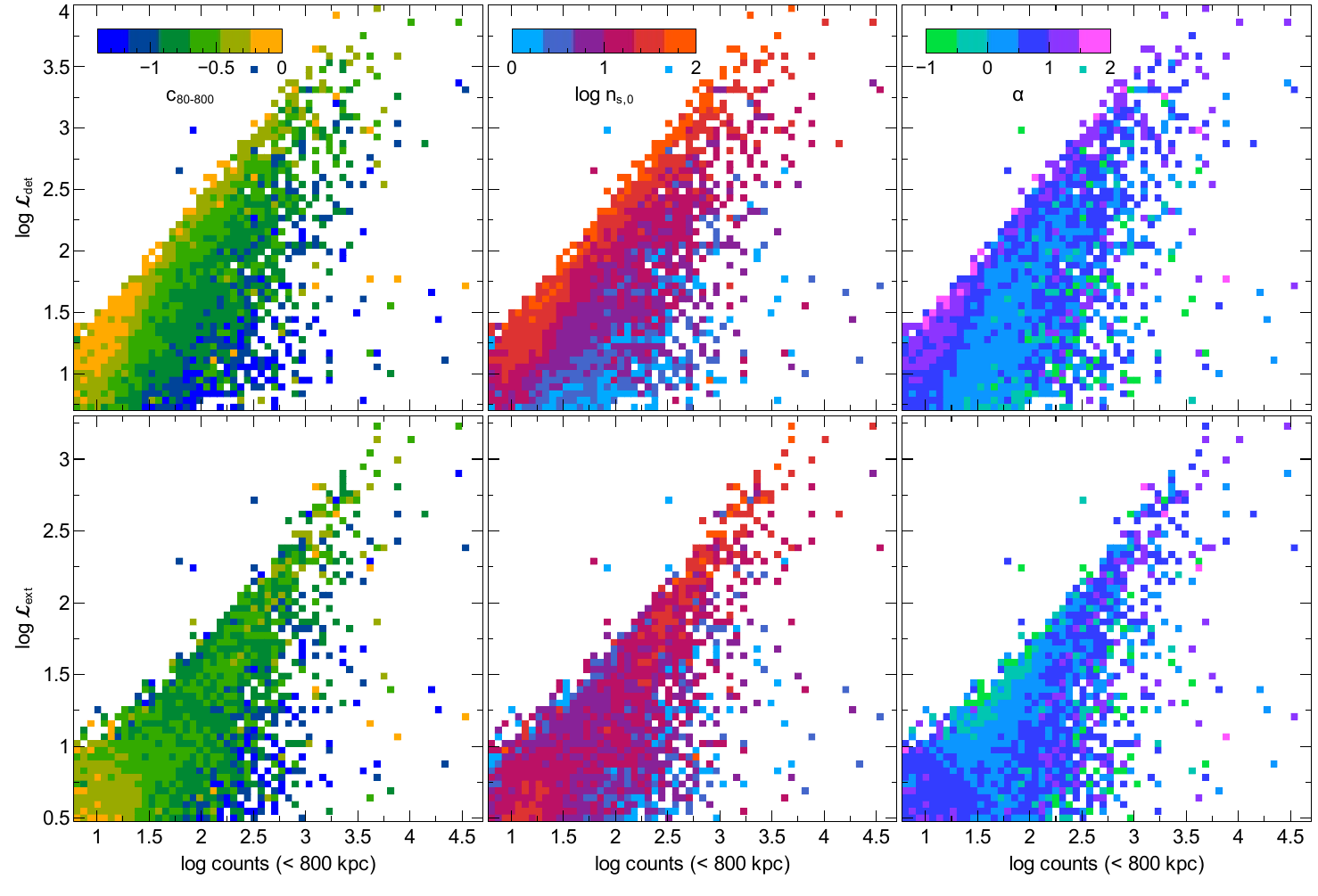}
  \caption{
    Relationship between detection likelihood ($\mathcal{L}_\mathrm{det}$), extension likelihood ($\mathcal{L}_\mathrm{ext}$), counts and morphological parameters.
    (Top panels) 
    The median value of the concentration ($c_{80-800}$; left), central scaled density ($n_\mathrm{s,0}$; centre) and central slope ($\alpha$; right), in grid points of $\mathcal{L}_\mathrm{ext}$ and $\mathcal{L}_\mathrm{det}$.
    (Bottom panels)
    The median morphological parameter values for grid points of counts in an 800~kpc aperture and $\mathcal{L}_\mathrm{det}$ or $\mathcal{L}_\mathrm{ext}$.
  }
  \label{fig:detlike}
\end{figure*}

The easiest to measure indicator of a cool core is the surface brightness concentration of a cluster.
Cool cores also have higher central densities than non cool core clusters and steeper inner surface brightness profiles.
In the eROSITA detection pipeline \citep{Merloni24}, the significance of sources is measured by two parameters, firstly the detection likelihood ($\mathcal{L}_\mathrm{det}$; \texttt{DET\_LIKE\_0} in the catalogue) and secondly the likelihood for the source to be extended ($\mathcal{L}_\mathrm{ext}$; \texttt{EXT\_LIKE} in the catalogue).
When computing the extension likelihood, the source is parametrized as a $\beta$-model with $\beta=2/3$ and a variable extent.

Figure \ref{fig:detlike} shows these parameters for the whole cluster sample, as a function of the cluster detection likelihood or extension likelihood, and the total number of counts in the cluster in an 800~kpc aperture.
It can be seen that for a fixed number of counts in a cluster, those with a higher detection likelihood have a higher concentration, a more dense core and a steeper density slope.
For a fixed detection likelihood, those with fewest counts are more concentrated, more dense and have steeper profiles.
The pattern of extension likelihood is less clear.
A selection in count space is less affected by the cluster morphology, particular above the threshold of $\sim 40$ counts.

\begin{figure*}
  \includegraphics[width=\textwidth]{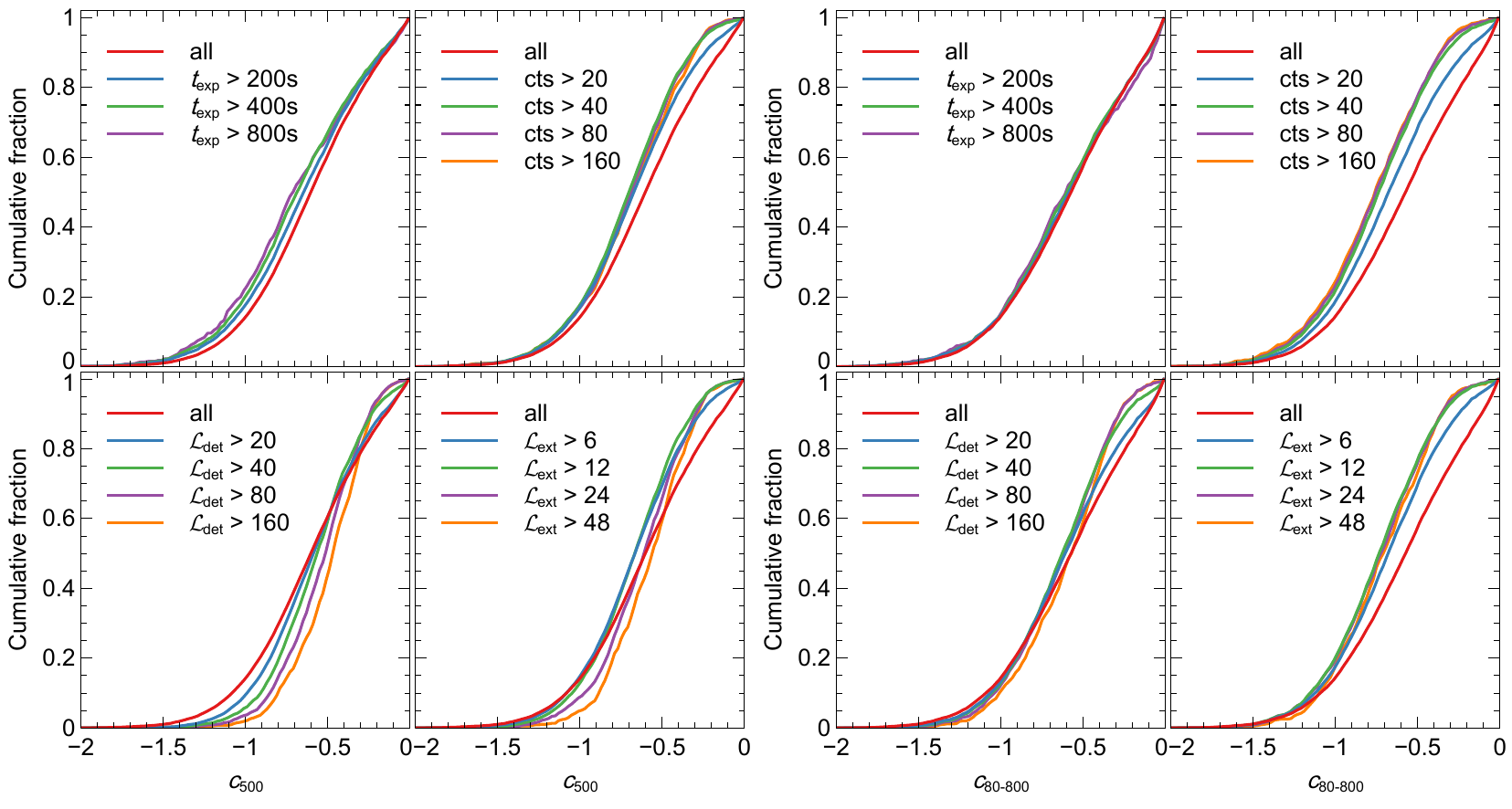}
  \caption{
    Cumulative distribution of concentration for different subsamples of the eRASS1 catalogue.
    The cumulative distributions are shown as a function of exposure time ($t_\mathrm{exp}$), number of counts in a $800$~kpc aperture, detection likelihood ($\mathcal{L}_\mathrm{det}$) and extension likelihood ($\mathcal{L}_\mathrm{ext}$).
    The left panels show the distribution of $c_{500}$, while the right panels show the distribution of $c_{80-800}$.
  }
  \label{fig:cuml_conc}
\end{figure*}

The effect of concentration on selection is shown in further detail in Fig.~\ref{fig:cuml_conc}, where the cumulative distribution of concentration is shown for subsamples with different cuts on exposure, counts, detection likelihood and extension likelihood.
The effect of increased exposure time is to reduce the peak of the distribution to lower values of $c_{500}$, but the distribution of $c_{80-800}$ is relatively unchanged.
Lower numbers of counts is associated with larger values of both concentration values, although above 80 counts, the distributions do not change further.
Increasing the detection likelihood threshold leads to distributions of $c_{500}$ with a peak at higher values, but a narrower width.
The effect of $\mathcal{L}_\mathrm{det}$ on $c_{80-800}$ is mostly to narrow the distribution peak.
The effect of extension likelihood cuts on $c_{500}$ is less clear, with low and high extension likelihoods showing a peak at higher values.
For $c_{80-800}$, higher $\mathcal{L}_\mathrm{ext}$ thresholds lead typically to lower concentration values, although this effect seems to saturate near values of around 12.
We also note that for the low count regime, that the distributions will be broadened out due to statistical noise.

We measure concentration with apertures with angular sizes that vary with redshift, and for $c_{500}$ also vary with mass.
Selecting different cluster susbsamples, for example according to exposure time, will choose clusters with different redshift and mass distributions.
Therefore, the interaction between how likely a cluster is detected, concentration, exposure time, number of counts and detection threshold cuts is rather complex.
However, it might be noted that the maximum shift for the median concentration is 0.2 dex for $c_{80-800}$ between the high and low count cluster subsamples.

\begin{figure}
  \includegraphics[width=\columnwidth]{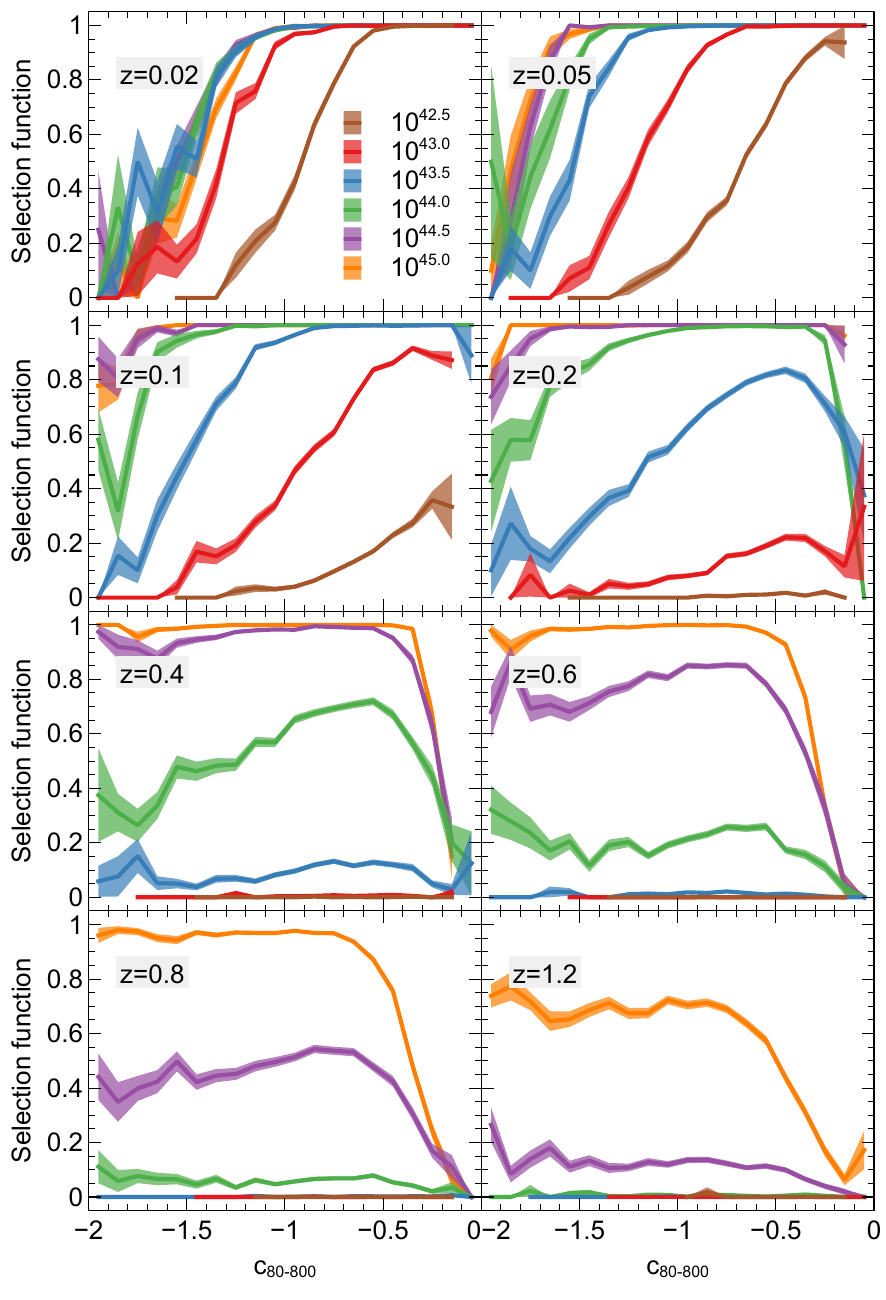}
  \caption{
    Selection function of clusters with different redshifts and luminosities, as a function of concentration, calculated from simulated observations of spherical clusters.
    The panels show the detected fraction of clusters at different redshift.
    For each redshift, we show the detection fraction as a function of concentration for different luminosities (erg~s$^{-1}$).
    The log concentration ($c_{80-800}$) is computed from the input model profile and is binned into bins of 0.1 width.
  }
  \label{fig:sim_conc_selfn}
\end{figure}

We can examine the effect of concentration on selection in more detail with the aid of simulations.
Figure \ref{fig:sim_conc_selfn} shows the fraction of clusters detected as a function of concentration at different redshifts and luminosities.
However, we note that the concentration in this plot is that computed from the input and not from the reconstructed profiles.
It can be seen at the lowest redshifts bins we would not detect groups or clusters if they have a flat surface brightness profiles ($c_{80-800} \lesssim -1$).
The effect is stronger for the less luminous (fainter) objects.
This is likely due to the fixed size of the detection aperture and the background map generation process.

At higher redshifts we would not detect clusters if they have very concentrated surface brightnesses, for example due to a cool core.
At a redshift of 0.4, clusters with $c_{80-800} \gtrsim -0.2$ (or around 60\% of their flux within 80 kpc) are lost, even if they have luminosities as high as $10^{45}$~erg~s$^{-1}$ (roughly a $10^{15}$~M$_{\odot}$ cluster).
This threshold reduces to $-0.4$ at $z=0.6$, $-0.5$ at $z=0.8$ and $-0.6$ at $z=1.2$ (with 25\% of their flux within 80 kpc).
For the SPT-selected clusters observed by \emph{Chandra} (Section \ref{sect:spt_data}), we find 4\% of objects have $c_{80-800} \gtrsim -0.4$, 9\% are at value greater than $-0.5$ and 13\% greater than $-0.6$.
This selection effect is due to these objects becoming more point like and the \texttt{eSASS} detection pipeline not detecting the extension of the object, due to a low extension likelihood.
Therefore extreme cool core objects at high redshifts may be missing from the cluster sample.
One example is the extreme cool core Phoenix cluster \citep{McDonald12}, which is not detected as an extended source in the eRASS1 catalogue (also see \citealt{Bulbul22}).

\begin{figure}
  \includegraphics[width=\columnwidth]{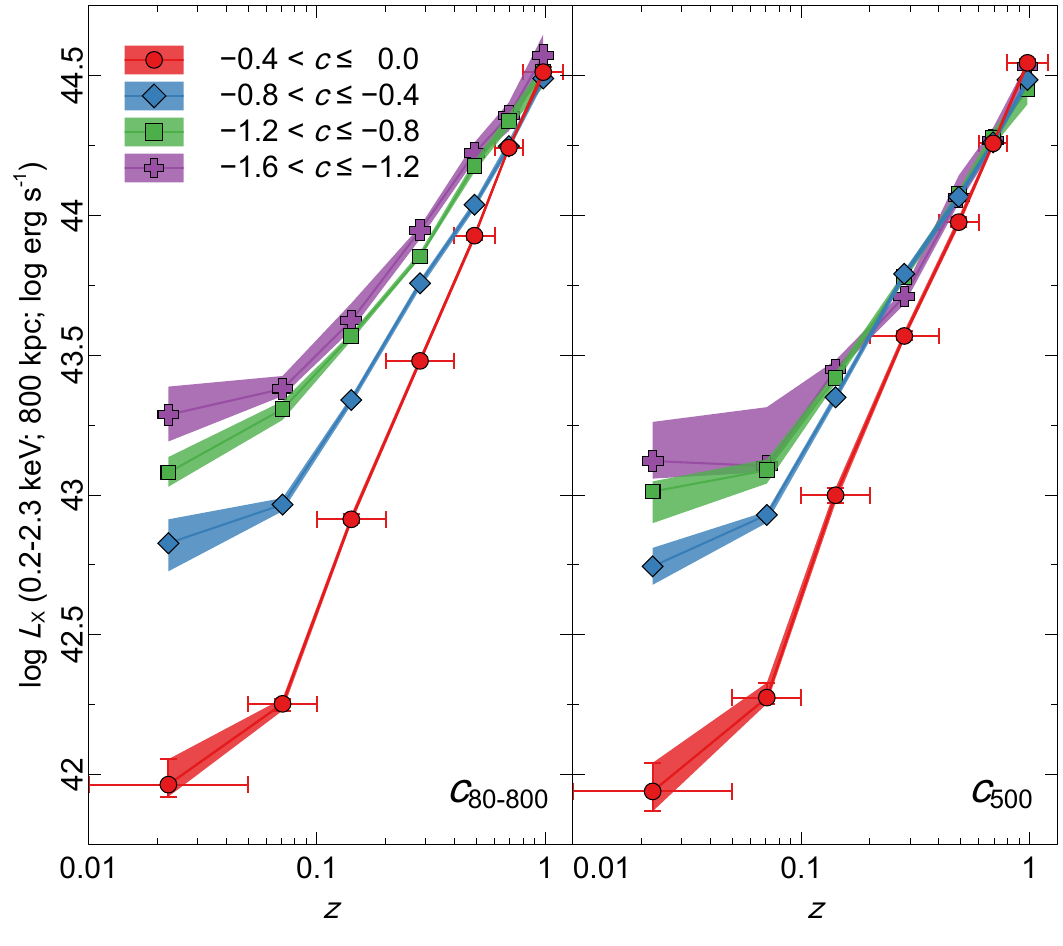}
  \caption{
    The median X-ray luminosity of clusters in bins of concentration.
    The left panel uses bins of $c_{80-800}$, while the right panel shows $c_{500}$.
    The uncertainties shown are calculated with bootstrap resampling.
    All eRASS1 clusters are included in this analysis.
  }
  \label{fig:lx_z_conc}
\end{figure}

The selection effect seen above can also be seen in the observed clusters if we plot the average cluster luminosity as a function of redshift in bins of concentration (Fig.~\ref{fig:lx_z_conc}).
At lower redshift the least luminous clusters are the most concentrated.
Clusters with similar luminosities but flatter cores would not be detected by the source detection.
The opposite effect is present at high redshifts for $c_{500}$, where the least luminous clusters have the flattest cores, as peaked clusters which look like point sources are not detected as extended clusters.

\subsection{Cool core mass and radius bias}
\label{sect:ccmasrad_bias}
Parameters measured at or within radii of a radius scaled with $R_{500}$ are sensitive to the accuracy of the cluster radius measurement.
Cool core clusters have high central gas densities and therefore X-ray bright cores.
As the cluster mass and radius is obtained using the cluster luminosity in the eRASS1 cluster catalogue, the presence of a cool core can bias the cluster radius and parameters which are measured at a scaled radius.
A cool core will increase the cluster luminosity and $R_{500}$, leading to a measurement at fixed scale radius being measured at a higher physical radius than expected.
However, the cluster radius is only scales weakly with the mass with a $1/3$ power.
$n_{s,0}$, will be reduced in cool core clusters by this effect.
Similarly, non-cool core clusters may have boosted central densities, although the effect will be smaller as they have a flat core in their density profile.

Many of the parameters are affected by this bias in radius determination including $n_{s,0}$, $\alpha$, $c_{500}$, power ratios, $w$ and $A_{phot}$.
Parameters unaffected include $n_{50}$, $c_{80-800}$, $\epsilon$, $H$ and multipole magnitudes.
Those parameters which are affected by this issue are indicated in Table \ref{tab:params}.
For example, if we artificially increase $R_{500}$ by 10\%, the median $c_{500}$ increases by $0.035$, $c_{500}^*$ by $0.026$, $n_{\mathrm{s},0}$ by $-0.022$, $n_{\mathrm{s},0}^*$ by $-0.036$, $\alpha$ by $0.038$ and $\alpha^*$ by $0.033$.
Similar changes in the opposite direction are seen if $R_{500}$ is decreased by 10\%.

\subsection{Position bias}
\label{sect:pos_bias}
The central positions of our clusters are determined by the fitting of a model for the surface brightness to an image.
For those parameters obtained using \texttt{MBProj2D}, the central cluster position is allowed to vary during the analysis, while other clusters and modelled point sources in the field are frozen at their best fit location.
For those parameters not measured using \texttt{MBProj2D}, we use the median chain value from the \texttt{MBProj2D} symmetric modelling to use as the cluster centre.
Some of these positions can differ from the standard cluster catalogue location, due to the slightly different modelling procedure.
The cluster positions obtained from the \texttt{MBProj2D} analyses are the locations of the centre of the model which fit the cluster best overall, in a statistical manner.

\begin{figure}
  \includegraphics[width=\columnwidth]{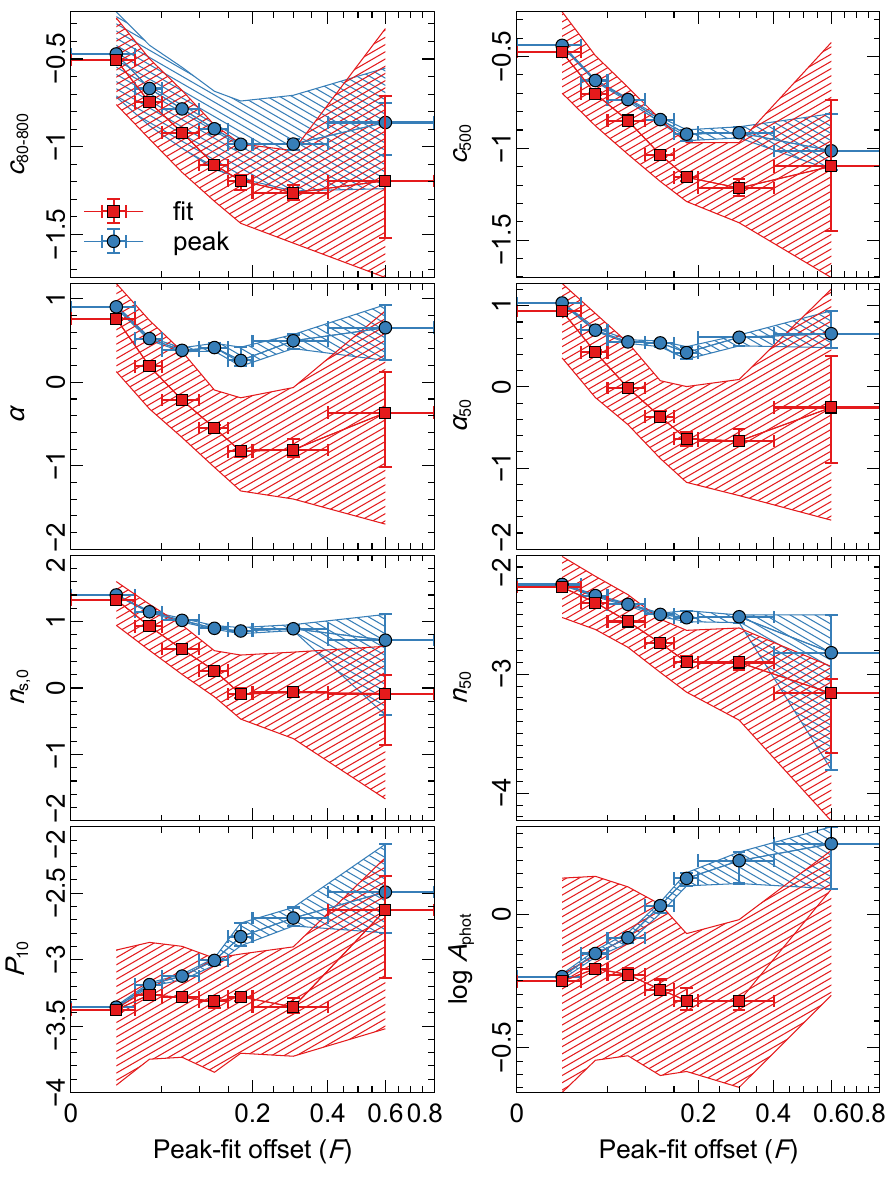}
  \caption{
    Fit-centred and peak-centred ($*$) quantities as a function of the distance between the cluster peak and the best fit position, for a cluster subsample with $\mathcal{L}_\mathrm{ext}>6$ and $\mathcal{L}_\mathrm{det}>20$.
    The median quantities in bins of $F$ are shown with bootstrap resampling uncertainties.
    The shaded regions show the the 1$\sigma$ percentile width of the data points within each bin.
  }
  \label{fig:peak_vs_quantity}
\end{figure}

As stated previously, for some parameters where the choice of centre makes a difference, we provide fitted centre and peak centre values.
Fig.~\ref{fig:peak_vs_quantity} shows average quantities in bins of offset between the peak and best-fitting positions ($F$).
The central density parameters, concentration, central slope, photon asymmetry and power ratio $P_{10}$ are all significantly different if there is an offset between the peak and best fitting parameters.
We also see lower dispersions for some quantities when using the peak position.

For the \texttt{MBProj2D}-derived shape parameters, the cluster centre is also allowed to be free in the analysis.
For parameters $\epsilon$, $H$, and $M_1$ the size of the offset between the standard fit position and the one with varying 2D shape is clearly correlated with the shape parameter.
At more extreme parameter values, this shift can be $0.1$ to $0.2R_{500}$.

\subsection{PSF bias}
\label{sect:psf_bias}
As we look at objects towards higher redshifts (or more precisely, angular diameter distance) or lower masses, the finite size of the PSF of eROSITA becomes increasingly important and makes it harder to measure morphological parameters.
Those quantities measured using the forward-modelling procedure of \texttt{MBProj2D} already take account of the PSF of eROSITA.
If we look at more distant objects, the uncertainties on the measured parameters are increased due to the increased importance of the PSF size, but there should be no bias on the measured parameters if the functional form of the fitted profile is realistic (however, see Section \ref{sect:range_bias}).

There are several parameters, however, which do not consider the PSF in their determination.
These include power ratios, $w$, $A_\mathrm{phot}$ and $G$.
Clusters at increasing distances with the same degree of disturbance will look less disturbed with these parameters, as the structure is smoothed out.
In Section \ref{sect:noise_distance} we present the intrinsic variations of these parameters due to both the effects of noise and the PSF, as a function of redshift and luminosity.

\subsection{Fixed parameter range bias}
\label{sect:range_bias}
Some parameters, including $\epsilon$, $H$, the multipole magnitudes and $G$, are only defined for a range of $0$ to $1$.
The definition of concentration restricts values to the range $0$ to $-2$.
In addition, for the very faintest clusters or most distant clusters, model priors may implicitly restrict the range on some of the forward-modelling obtained parameters.

The finite range means that as the uncertainty on a value increases, the obtained median values from an MCMC chain tend towards the centre of the allowed parameter range.
If there is no constraint on a value, its $1\sigma$ error ranges, obtained from the $1\sigma$ percentiles of the chain will also tend towards encompassing $68\%$ of the parameter range.

For those parameters obtained using MCMC, the full posterior probability distribution can still be used as this effect increases, although model priors could become increasingly important.

To reduce this effect for the shape parameters, $\epsilon$, $H$, $M_1$ to $M_4$, rather than quoting the median chain value, we instead give the maximum likelihood value.
The best fitting values are more evenly distributed over the parameter range when the number of counts is low, rather then being clustered around 0.5.
This also makes comparisons with simulations easier, as we cannot run a full MCMC analysis on each simulation.
Fig. \ref{fig:param_cts} shows that for objects with less than 200-300 counts, it is difficult to trust these individual shape parameters for single objects.

\begin{figure}
  \includegraphics[width=\columnwidth]{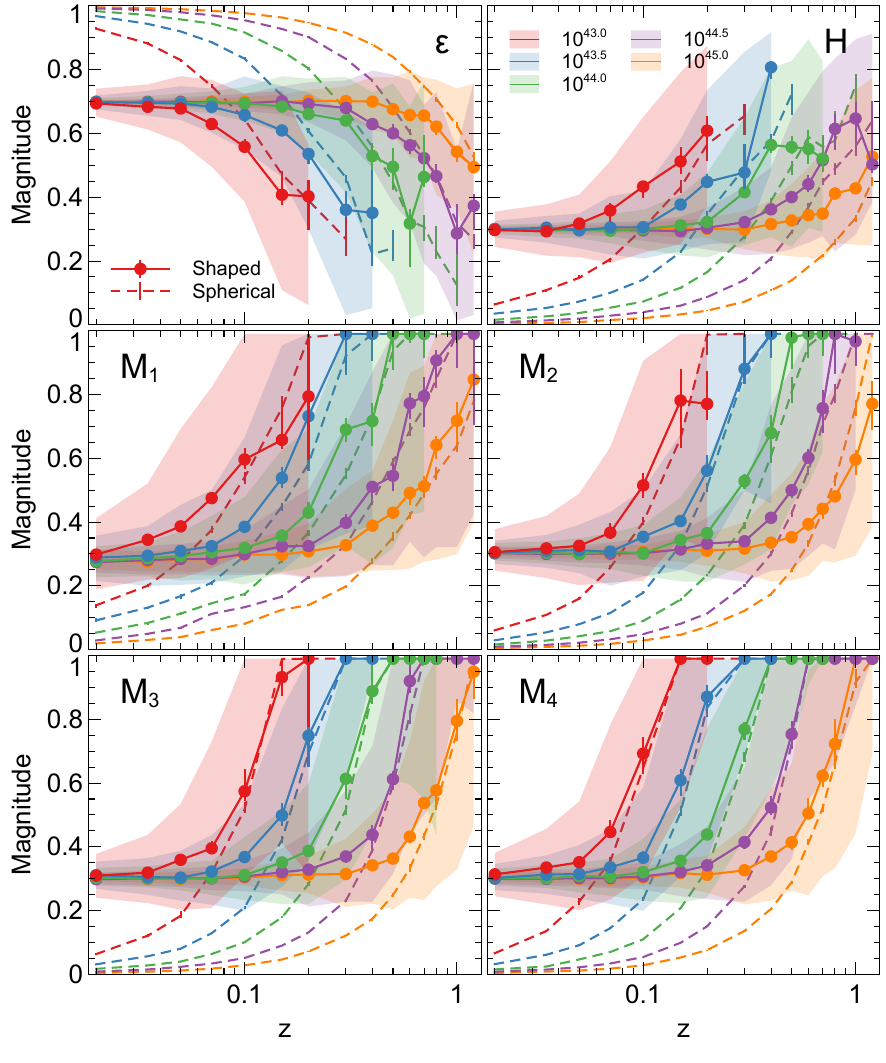}
  \caption{
    Recovery of forward modelled parameters as a function of redshift and luminosity, using maximum likelihood.
    The plots show the recovered parameter values as a function of redshift, for clusters with an input shape compared to an input spherical cluster.
    The error bar shows the average best fitting parameter and the uncertainty on the mean.
    The shaded region shows the $1\sigma$ range of the recovered best fitting parameters.
    The panels show shaped models with $\epsilon=0.7$, $H=0.3$, $M_{1}=0.3$, $M_{2}=0.3$, $M_{3}=0.3$ and $M_{4}=0.3$, respectively, compared to the spherical models.
  }
  \label{fig:shape_recovery}
\end{figure}

With the simulations we also checked how well the fixed-range shape parameters can be recovered from the maximum likelihood.
Figure \ref{fig:shape_recovery} shows the range of parameter values obtained for the input cluster shapes given.
At high luminosities and low redshifts, the input parameter can be recovered.
Surprisingly, we find that the average best fit parameter does not tend towards the input parameter in the low count regime.
The mean values tend towards an extreme case (e.g. $0$ for $\epsilon$ and $1$ for $M_1$ to $M_4$).
The $H$ parameter does not seem so badly affected by this bias.
If we examine extreme cases, which are for low photon count clusters, we find often the maximum likelihood model is a very extreme ellipse which passes through the brightest pixels.
Therefore, one should take care if using the maximum likelihood values for these parameters from our catalogue of parameters in the very low count regime.
The maximum likelihood parameters chosen for objects with low counts are obviously unphysical.
In this case, an informative prior on these parameters could be useful to prevent these poor results, although this could make interpretation more difficult.
More detail about the count range where this issue occurs is discussed in Section \ref{sect:recovered_biases}.

\subsection{Cluster shape affecting selection}
\label{sect:shape_seln}
In addition to how concentrated a cluster is, its 2D shape may affect whether it is selected as an extended source or not.
The \texttt{ermldet} task assumes that a cluster has a symmetric shape with a $\beta$ model profile.

\begin{figure}
  \includegraphics[width=\columnwidth]{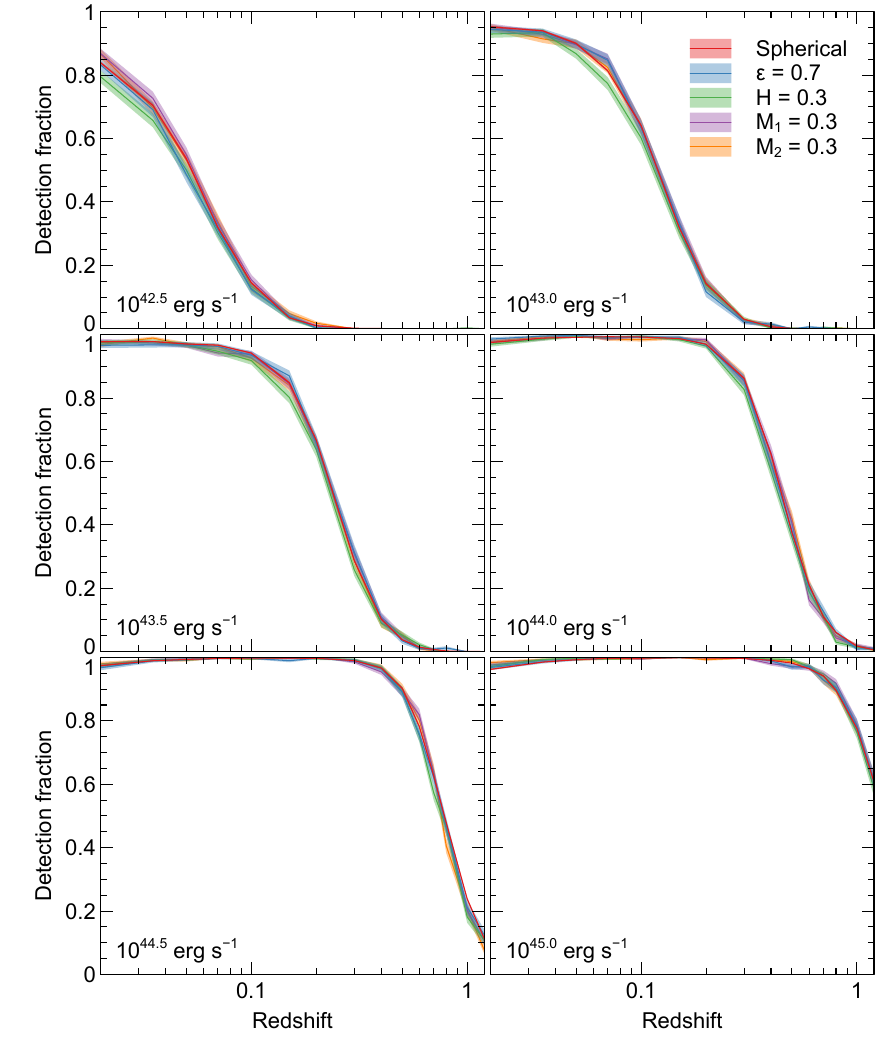}
  \caption{
    Detection fraction as a function of redshift for different luminosities and shapes.
  }
  \label{fig:detfrac_shape}
\end{figure}

In our simulations we tried to assess how much the cluster shape affects the likelihood that a cluster is selected as a function of redshift and luminosity.
We computed this for a limited number of cluster shapes, including spherical, elliptical ($\epsilon=0.7$), sloshed ($H=0.3$), first multipole ($M_1=0.3$) and second multipole ($M_2=0.3$).

Figure \ref{fig:detfrac_shape} shows the fraction detected for these different cluster shapes for different luminosities as a function of redshift.
The results show that for these degrees of non-spherical morphology there are no significant changes in the detection efficiency of clusters.
The largest effect seems to be for low luminosity objects at lower redshifts, where a slosh reduces the detection efficiency by $\sim 5$\%.

\subsection{Noise and distance}
\label{sect:noise_distance}

\begin{figure*}
  \includegraphics[width=\textwidth]{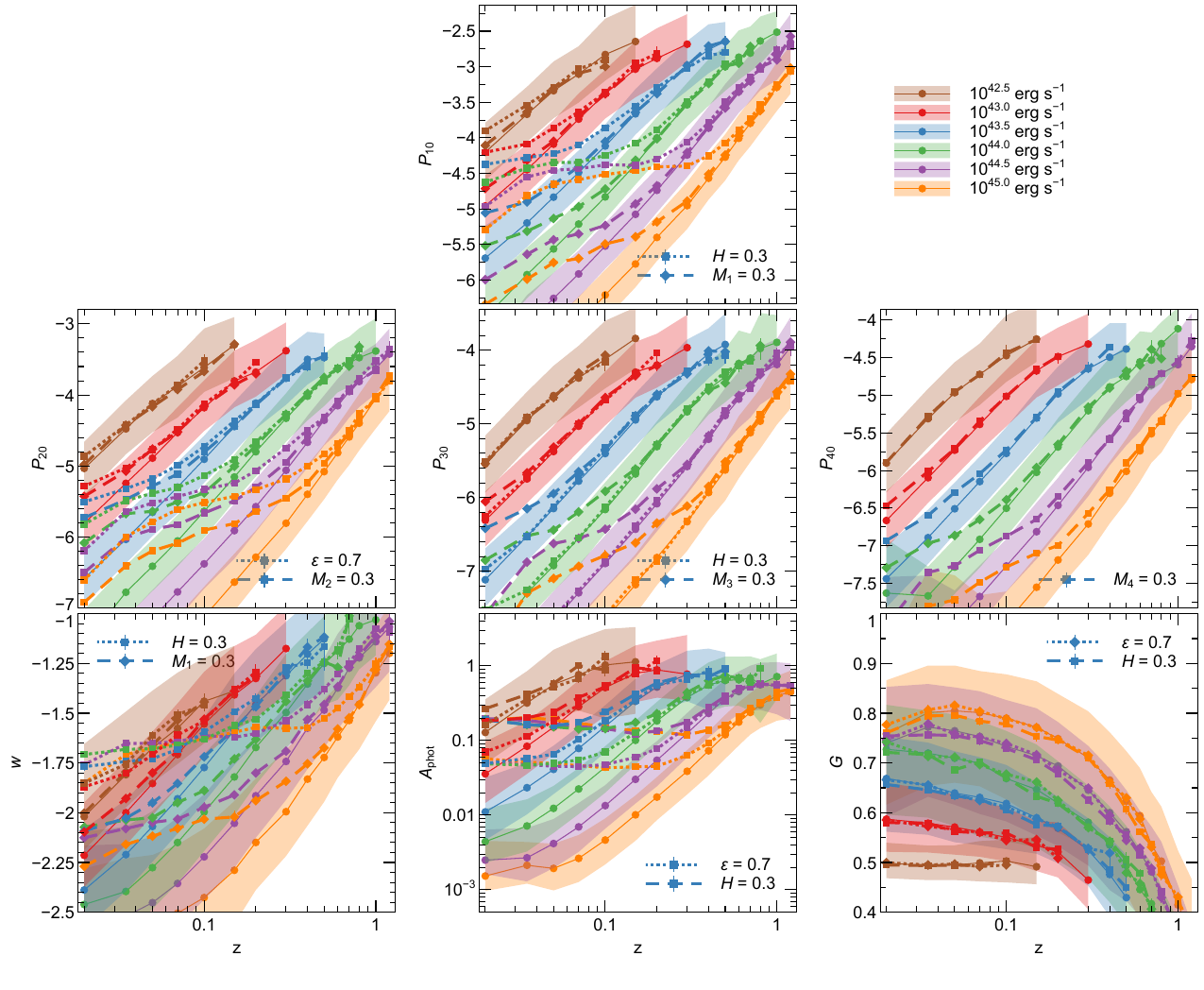}
  \caption{
    Measured non-forward-modelled morphological parameters for simulated model clusters.
    Clusters are simulated with different luminosities and redshifts, after sampling from the mass function.
    The shaded region shows the $1\sigma$ range of recovered parameters of detected clusters simulated from spherical cluster models.
    The dotted and dashed lines show the median obtained parameters for non-spherical detected clusters with the parameters shown.
  }
  \label{fig:recovered_params}
\end{figure*}

Several parameters do not take account of the PSF of the telescope.
In addition, they may be affected by noise in the low count regime.
Figure \ref{fig:recovered_params} shows measured non-forward-modelled morphological parameters for simulated clusters.
For these plots we simulate spherical and shaped clusters with given luminosities and redshifts.
The bands show the $1\sigma$ range of recovered properties as a function of redshift for the spherical clusters of different luminosity.

It can be seen that for these parameters there are strong intrinsic variations with luminosity and redshift.
As these clusters are completely symmetric, this is due to the increase in noise for fainter clusters and increased blurring for distant or less massive clusters.
On the plots we also include the relations of the median for clusters with a shape given (one of $H=0.3$, $\epsilon=0.7$, $M_1=0.3$, $M_2=0.3$, $M_3=0.3$ and $M_4=0.3$).
For the brightest and nearest clusters we see a difference in the recovered parameters between the clusters with shape and the spherical clusters.
However, we note that the Gini coefficient is primarily a measure of concentration, so we do not see a difference for the spherical and clusters with shape.
Some parameters show difference between spherical and non-spherical clusters, even for nearby luminous objects, such as $P_{30}$ and $P_{40}$.

\subsection{Resolution bias}
\label{sect:res_bias}
The more counts detected from a cluster in an annulus, the easier it is to measure properties at that radius.
For example, if want to measure the central entropy of a cluster, then we might, for example, create radial bins to measure a cluster profile and fit a model.
The bins would need to be larger for poorer quality data, leading to a bias in the innermost regions.
This could lead to a floor in the profile caused by the analysis method \citep[see discussion in][]{Panagoulia14}.
Less bias would be present for clusters with more counts in the central region.

Our analysis method does not bin the data, so this should not be important above the pixel scale (although 4 arcsec pixels are 32~kpc at a redshift of 1).
The error bars on a derived quantity should encompass the real value, unless the prior is unrealistic, or the parametric form is different from the real profile, leading to discrepancies in the outer parts of the profile driving the inner part.

\subsection{Temperature and metallicity bias}
\label{sect:t_bias}
We assume that clusters are isothermal and have a $0.3$ Solar metallicity.
If this is not the case, then density measurements or inner density slopes could be inaccurate.
However, variations in temperature and metallicity would need to be strong to produce inaccurate densities, as the X-ray surface brightness is proportional to density-squared.
One case in which they could become important is for groups, where the X-ray emission lines become dominant over bremsstrahlung emission for X-ray temperatures below around 1~keV.
A substantial fraction of eROSITA clusters are lower mass objects with lower temperatures.
Of those with temperature measurements, there are 550 objects with temperatures less than 1 keV and 2\,900 with temperatures less than 3 keV.

\subsection{Unresolved point sources}
\label{sect:ptsrc_bias}
Detected point sources are masked out in our analysis or simultaneously fitted.
Due to the $\sim 30$ arcsec survey PSF, its limited sensitivity and the source detection algorithm, we may miss point sources within clusters, or resolve parts of clusters as point sources.
In addition, some fraction of clusters in the sample are contaminants, where single or multiple point sources could be confused to be extended sources.
The purity of the main cluster catalogue is 86\% (B24).

Unresolved point sources could give rise to artificial substructure within an object and so affect the morphological parameters.
In addition, very peaked substructure could be detected as a point source and therefore removed.
The strength of this effect could be redshift and luminosity-dependent.
To better assess this in a future work would likely require the analysis of more realistic simulations incorporating a full range of cluster morphologies and source populations.
Our existing simulations do not include point sources.
If our simulations were to include a point source population, it would likely give rise to additional bias in the parameters, particularly in low luminosity and higher redshift objects.

\subsection{Recovered parameter biases and uncertainties}
\label{sect:recovered_biases}

\begin{figure*}
    \includegraphics[width=\textwidth]{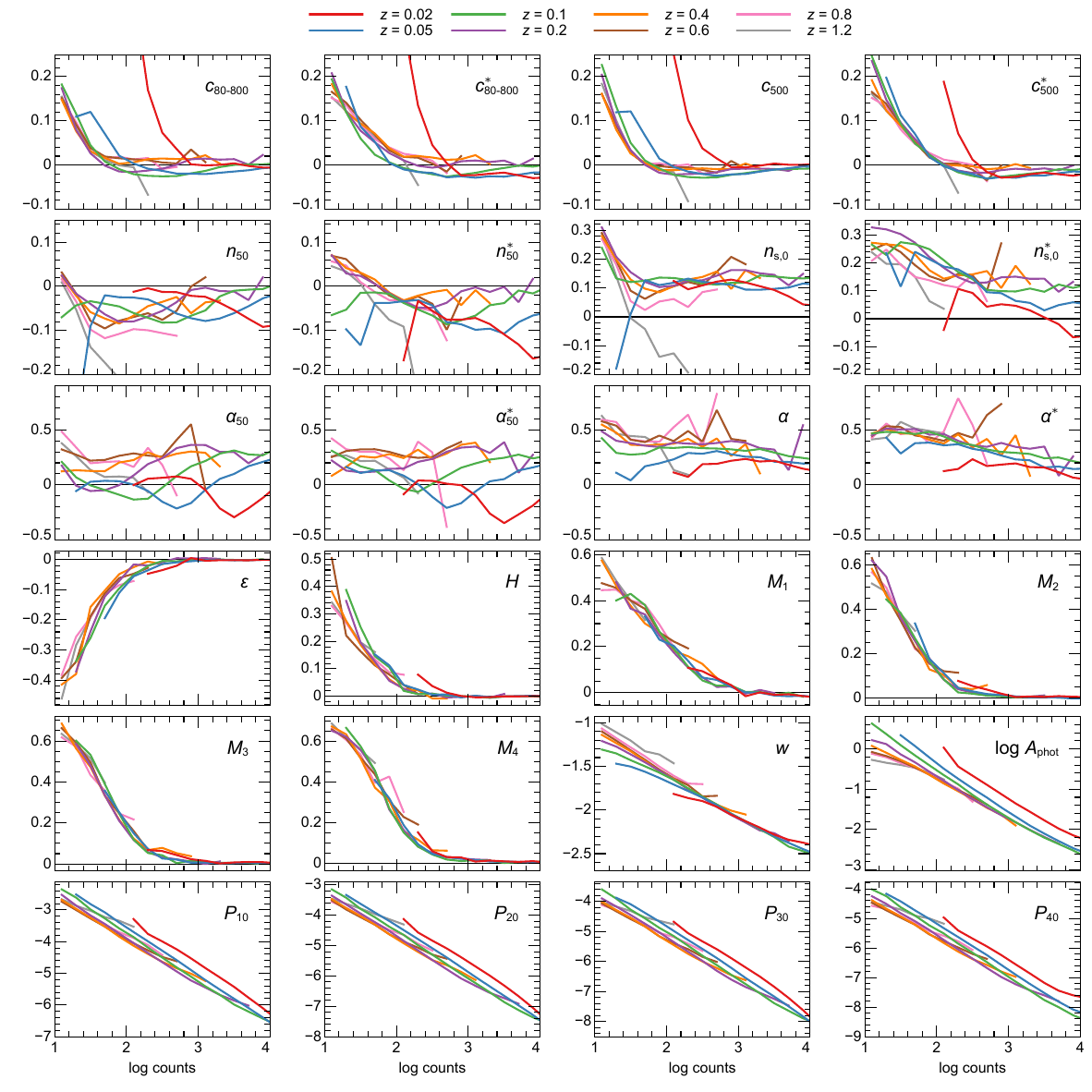}
    \caption{
    Recovered parameter bias from simulated clusters as a function of the redshift and number of counts inside 800 kpc radius.
    For the concentration, central density and cuspiness, shown is the difference between the recovered values and the input profile, after deprojecting the emissivity profile.
    For the shape parameters, $\epsilon$, $H$ and $M_1$ to $M_4$, we show the difference between the recovered and input values (0.7 for $\epsilon$ and 0.3 for the others), for simulations with non-spherical clusters.
    For the parameters $w$, $A_\mathrm{phot}$ and $P_{10}$ to $P_{40}$, we show the value for a spherical cluster.
    }
    \label{fig:sim_bias}
\end{figure*}

\begin{figure*}
    \includegraphics[width=\textwidth]{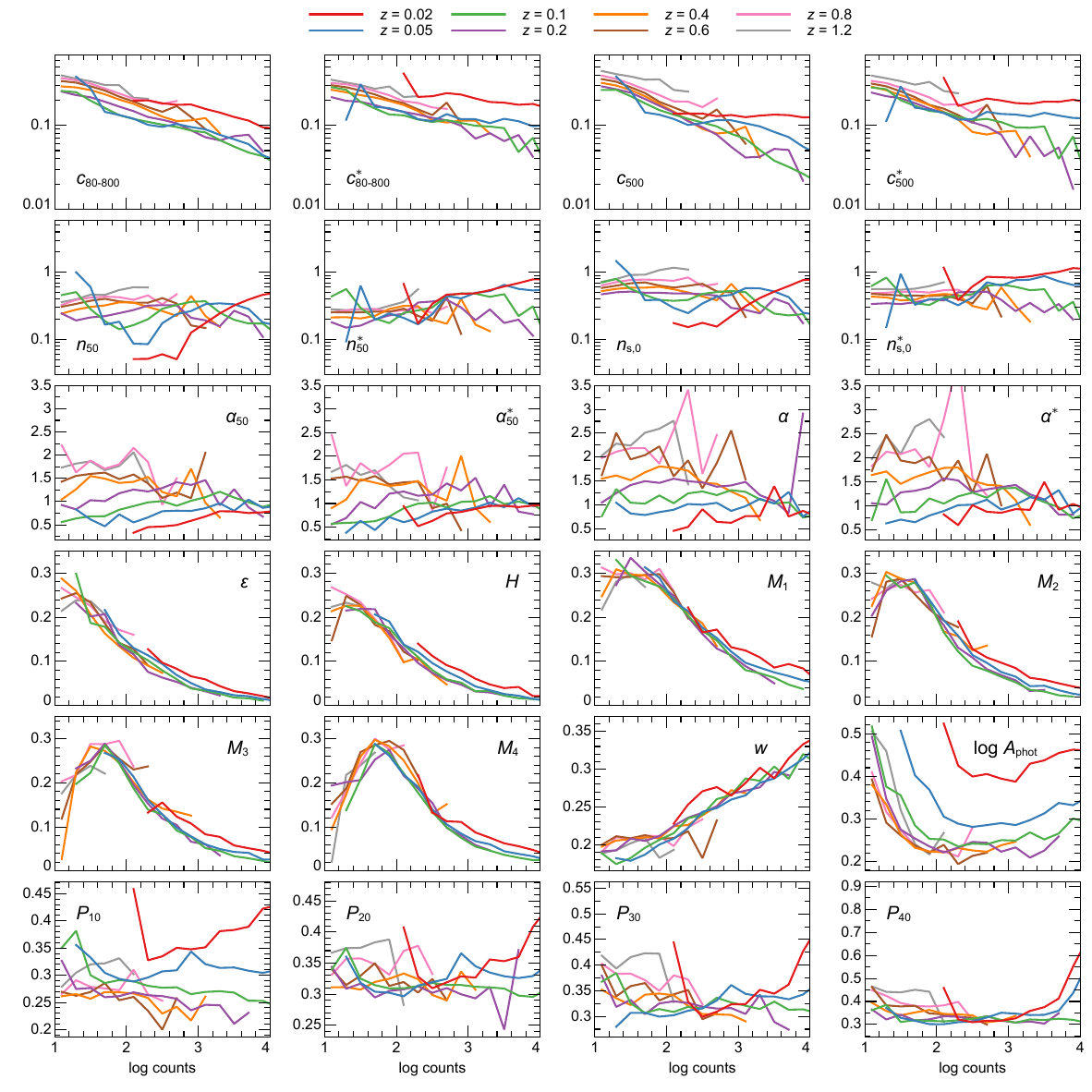}
    \caption{
    Recovered statistical uncertainty in measured parameters from simulated clusters.
    This is the standard deviation on the difference between the input and recovered values, as in Fig.~\ref{fig:sim_bias}.
    }
    \label{fig:sim_err}
\end{figure*}

Some of the listed biases affect whether a cluster is detected and others affect the measured parameters.
Here we examine the combined effect of biases on the measured parameters using our simulations, ignoring selection effects.
These simulations take account of the PSF, the data quality, the fixed parameter range and the choice of centre (fitted or peak), although point sources are not included.
We also compute the statistical uncertainty of the parameters.
These biases and uncertainties are studied as a function of redshift and number of cluster counts.
Counts are more reliable than luminosity in predicting the data quality because it takes account of survey depth and absorption.

Fig.~\ref{fig:sim_bias} shows the difference between the recovered parameter value and its input.
We emphasise that we are just using the median obtained parameter (or maximum likelihood in the case of the shape parameters) and do not include the probability density function for the parameter.
For the density parameters, the input value is obtained by deprojecting the emissivity surface brightness profile (noting that this is not valid in all cases).
For the shape parameters ($\epsilon$, $H$ and $M_1$ to $M_4$), we simulate clusters with known fixed parameter values and look at the difference between the input and output.
For the image based parameters ($w$, $A_\mathrm{phot}$ and $P_{10}$ to $P_{40}$), we simulate spherical clusters and just plot the average value, noting that given perfect signal to noise the value should be $-\infty$ in the log space used.

The largest biases are seen for fainter objects in the very lowest redshift bin, where the extent of the source means that the background level varies significantly over the region of the object.
At these redshifts, the background level means that for the same number of counts the signal to noise ratio is lower.
For higher redshifts, we see fairly moderate evolution in bias with redshift.
The concentration measurements show some positive bias for low count clusters, although it is rather flat and small when there are more than 50-100 counts, except at very low redshifts.
The peak centred concentrations show larger bias, although these simulated clusters are unrealistic by being perfectly spherical.
The central density shows some bias of the order of 0.05-0.15 dex, although the sign and magnitude depends on whether the density at fixed radius is measured.
Some positive bias in the central slope is seen, of between 0 and 0.5.
For the shape parameters, there is some bias below 100 counts or so, due to the finite range of the parameter (Section \ref{sect:range_bias}).
For the image based parameters, we see the same biases as seen in Fig.~\ref{fig:recovered_params} and described in Section \ref{sect:noise_distance}.

The uncertainties on the density profile parameters generally decrease with the number of counts and increase with redshift.
There is not such a strong decrease in uncertainty on the central density with increasing counts if the peak position is used, which is likely because the central density is very sensitive to the central position and we are only using a subset of the data to determine this position.
The statistical uncertainty on the cuspiness is generally large and redshift dependent.
The forward-modelled shape parameters show declining uncertainties with increasing number of counts, and relatively little redshift evolution.

\section{Comparisons with previous results}
\label{sect:other_samp}

\begin{figure*}
  \centering
  \includegraphics[width=\textwidth]{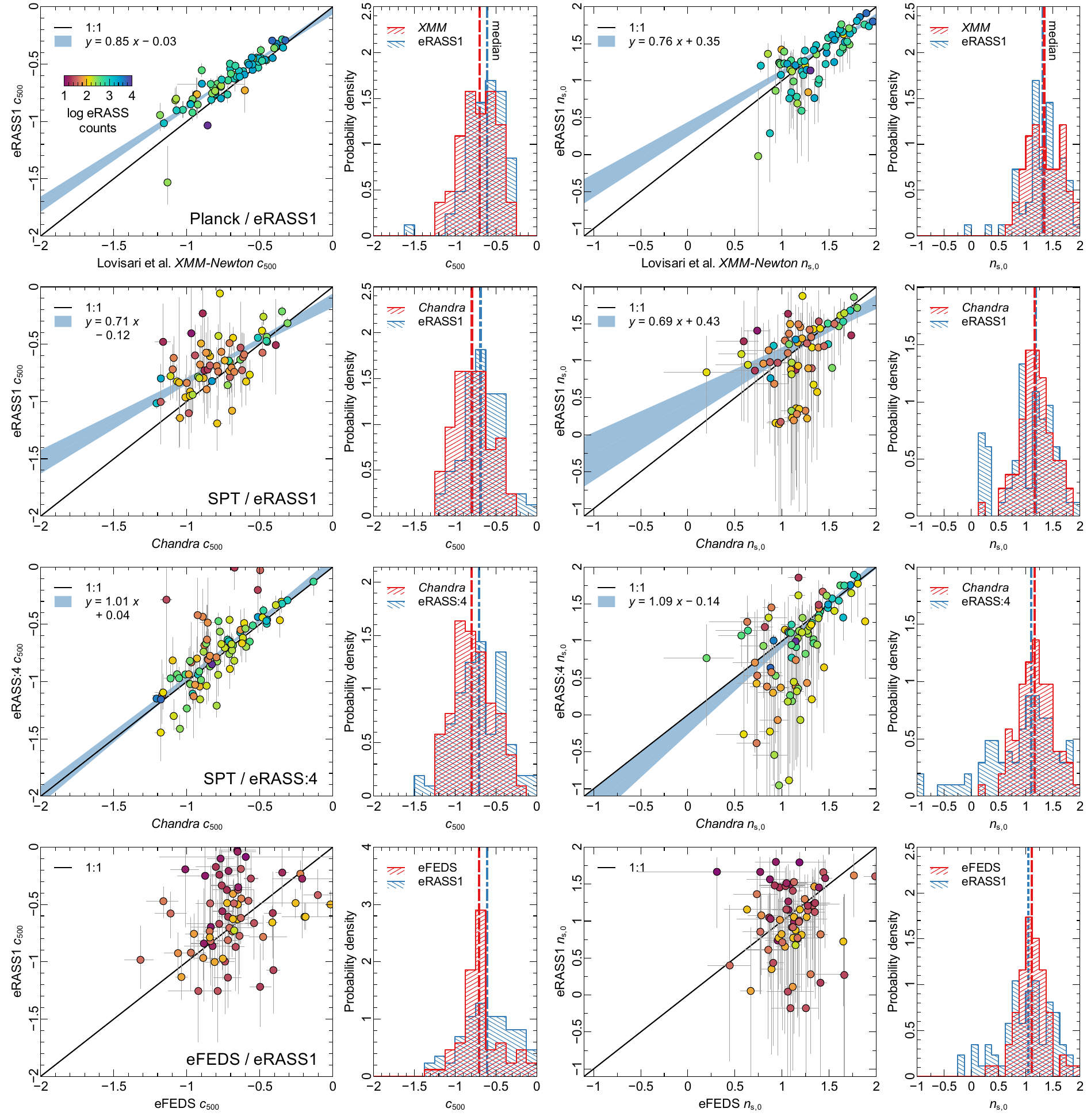}
  \caption{Comparison of concentration and scaled central density measurements for the same cluster samples between eRASS and other measurements.
    The two left columns compare $c_{500}$ and its distributions, while the two right columns show $n_{\mathrm{s},0}$.
    Median values are plotted as vertical lines.
    The number of counts for the cluster in the eRASS1 or eRASS:4 data is indicated by the colour.
    (Top row) 66 \emph{Planck} ESZ and eRASS1-detected clusters, with \emph{XMM-Newton} results from \cite{Lovisari17} and using the eRASS1 peak values.
    (Second row) 66 SPT and eRASS1-detected clusters, analysed using \emph{Chandra} data and taking the peak values from eRASS1.
    (Third row) 83 SPT-detected clusters, extracting the values from \emph{Chandra} eRASS:4 data using similar analysis procedures.
    (Bottom row) 69 clusters detected in both eFEDS and eRASS1.
    The eFEDS survey results are from \cite{Ghirardini22}, while the eRASS1 results are the best-fit position values.
    For the top three rows, we compute the best fitting linear relation between the results, assuming a normal distribution around the relation.
  }
  \label{fig:conc_compar}
\end{figure*}

\subsection{Matched subsamples}
\label{sect:matched}
To check for consistency, we compare the $c_{500}$ concentrations and $n_{\mathrm{s},0}$ central densities for the same clusters with different analyses in Fig. \ref{fig:conc_compar}.
This comparison will be unaffected by selection effects.
For three of these comparisons we compute using MCMC a median linear relation between eROSITA and the other values, assuming that the distribution is normal about the relation.
We note that B24 previously reported 15\% lower fluxes in eROSITA first data release (DR1) processing compared to \emph{Chandra}, which likely will give rise to a difference in central density of roughly half this value.

The first comparison is an \emph{XMM-Newton} analysis of the \emph{Planck}-detected early SZ (ESZ) sample \citep{Lovisari17}.
These clusters were matched to counterparts in the eRASS1 catalogue within a radius of 4 arcmin to make a jointly detected sample of 66 clusters, with a median eROSITA mass of $M_{500}=10^{14.92}$~M$_\odot$.
We note that the values of $R_{500}$ used in the two analyses are different, which could lead to systematic differences, although $R_{500}$ scales fairly weakly with mass.
The cluster centres will also be different, although both use X-ray peak centres.
Despite these possible differences, there is a reasonable agreement between the sets of concentrations, although there appears to be an offset increasing with lower concentration.
The median cluster has a 0.1 dex higher concentration in eRASS1.
If we examine the central density, there is scatter above the relation, which may be due to the assumed density profile form being too simple for these very bright clusters.
There is evidence for a non-unity slope or offset between the relations, with low central densities for eROSITA found higher than \emph{XMM-Newton}, and high central densities lower, although the small uncertainties on the bright objects are likely driving this fit.
The median value, however, is exactly the same between the two sets of measurements for the same clusters.
The differences between the two analyses could be the result of the $R_{500}$ differences, background modelling, particularly due to the large field of view of eROSITA, point sources, and the differences in the assumed functional form of the density or surface brightness.

We can also compare with a sample of clusters detected by SPT \citep{Bleem15}, in particular those previously studied in X-rays using \emph{Chandra} by \cite{Sanders18}.
We match an eROSITA to SPT cluster if its distance is less than 4 arcmin, to make a sample of 66 objects, with a median eROSITA mass of $M_{500}=10^{14.14}$~M$_\odot$.
Section \ref{sect:spt_data} describes the \emph{Chandra} data analysis, which uses the same parametrization for the density profile as for the eROSITA data.
For both these analyses, we use the concentration as defined by the peak of the cluster.
However, $R_{500}$ for the \emph{Chandra} analysis is from the SPT catalogue, while the eROSITA $R_{500}$ is the value in the eRASS1 catalogue.
The best fitting relation between eROSITA and \emph{Chandra} is similar to the \emph{Planck} clusters, with an increasing offset towards lower concentrations.
The median cluster again shows a 0.1 dex higher concentration for eRASS1, although the median concentration is lower than for the \emph{Planck} sample.
For the central densities, again the median values are exactly the same between the two different sets of analyses.
However, there is a similar non one-to-one relation found as with the \emph{Planck} data, if fitting the datapoints.

Rather than match the eRASS1 and \emph{Chandra} SPT objects, we can do a detailed analysis of eRASS:4 data for the whole \emph{Chandra} SPT sample, using as similar a methodology as possible (see Section \ref{sect:spt_data}).
In this case, we use the SPT $R_{500}$ for both analyses and also study objects which were not detected as clusters in eRASS1, producing a sample of 83 objects.
The best fitting relation between the two sets of values is consistent with 1:1.
The increased agreement compared to the eRASS1 sample is likely a combination of using exactly the same extraction regions and cluster centre and masking or modelling the same point source regions.
However, the median cluster again shows a $\sim 0.1$ dex higher concentration for eROSITA.
The central densities show a good agreement between the two analyses, with a relation which is consistent with $1:1$.
The median density values are also very close to each other.

Finally, we compare two different eROSITA surveys, eRASS1 and the deeper eROSITA Final Equatorial-Depth Survey \citep[eFEDS; ][]{Ghirardini22}, for clusters which match with a counterpart within 2 arcmin of the other survey, finding a sample of 69 objects.
The median eROSITA DR1 mass of this sample is $M_{500}=10^{14.41}$~M$_\odot$.
eFEDS shows a rather narrower concentration distribution for the same clusters, with a median around 0.1 dex lower than eRASS1.
The width of the concentration distribution is also narrower than the \emph{XMM} and \emph{Chandra} results, which may be due to the prior on the inner slope of the density parametrization, $\alpha>0$, used by \cite{Ghirardini22}.
The median central densities are similar for the two different analyses and datasets.

Overall, there is a reasonable agreement between the different analyses for the same cluster samples, considering that the datasets are different and some of the analyses are independent.
However, there are suggestions for the analyses using the eRASS1 clusters combined with \emph{Planck} ESZ or SPT, that there might be an offset for the lowest concentration or lowest density objects.
There is good agreement between the matched analysis procedures for the eRASS:4 and \emph{Chandra} data.
The scatter seen in the eRASS data is affected by the number of counts.
\emph{Planck} clusters are X-ray bright (mean in log space of 740 counts in eRASS1) compared to the SPT sample (log-space mean of 100 counts).
The eFEDS sample is fainter than the SPT objects, with a log-space mean of 35 eRASS1 counts.
The reasons for the differences in the other comparisons could be due to different analysis procedures, the locations of the cluster centres, unresolved point sources in eRASS1, the spatial masks for the datasets, background modelling and the 15\% offset in flux found in the DR1 processing (B24).

\begin{figure*}
  \centering
  \includegraphics[width=0.9\textwidth]{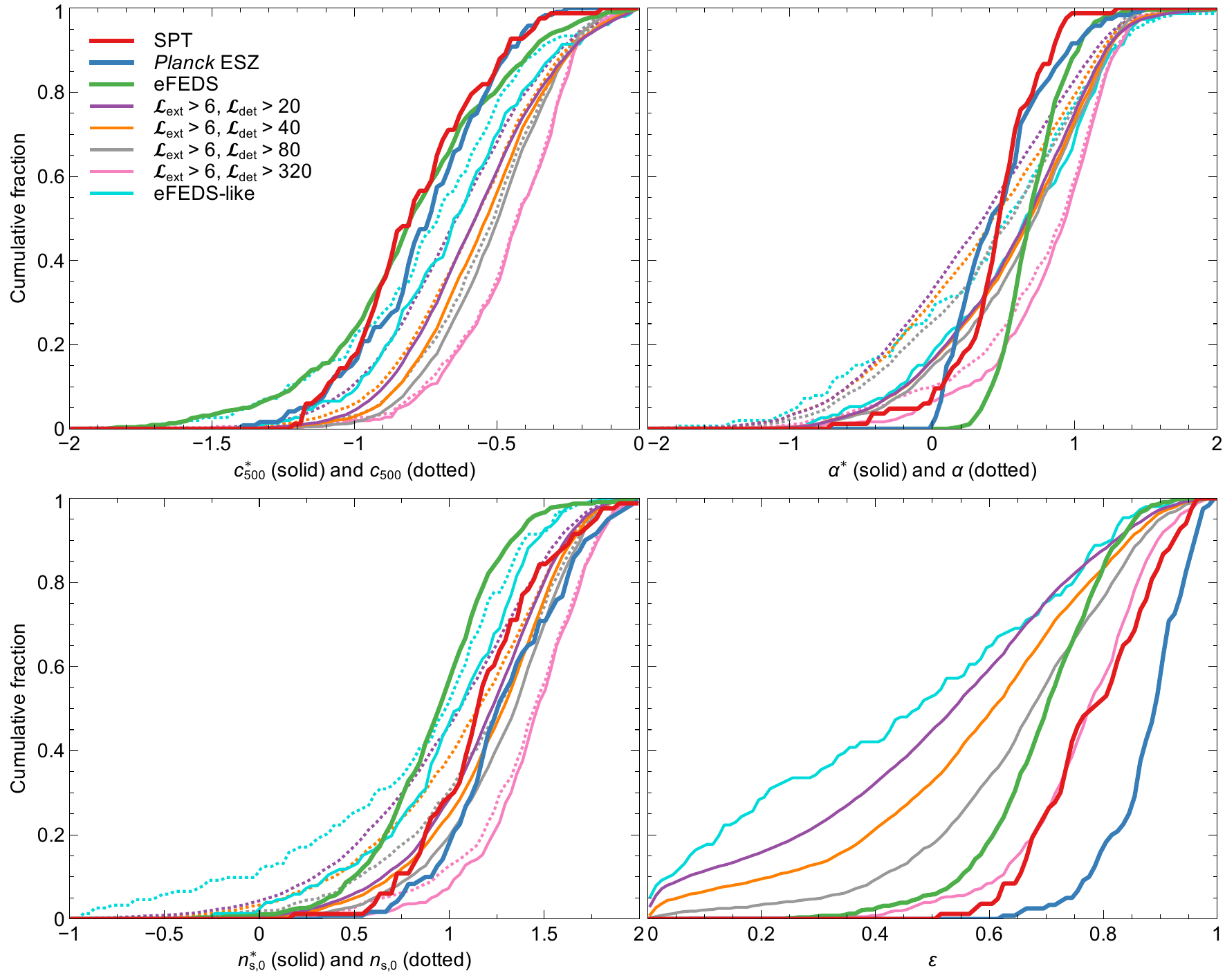}
  \caption{
    Cumulative distribution of parameters for SPT, \emph{Planck} ESZ and eFEDS clusters compared to eRASS1 for different detection thresholds.
    For eROSITA the solid lines show the values fixing the cluster position at the peak, while the dotted lines show the distribution for when the cluster position varies during the fit.
    Shown are the concentration (top left), cuspiness (top right), central scaled density (bottom left) and ellipticity (bottom right).
    \emph{Planck} values are taken from \cite{Lovisari17} and eFEDS values are obtained by \cite{Ghirardini22}.
    The eFEDS-like curves are for eRASS1 clusters with $\mathcal{L}_\mathrm{det}>5$, $\mathcal{L}_\mathrm{ext}>6$ and an exposure time between 1000 and 1400s, to match eFEDS.
    The median published log masses of the SPT, \emph{Planck} and eFEDS samples are $14.7$, $14.8$ and $14.0$, respectively.
    The eROSITA sample median log M$_\odot$ masses are $14.4$, $14.5$, $14.6$ and $14.8$, for detection likelihood thresholds of $20$, $40$, $80$ and $320$, respectively.
  }
  \label{fig:sample_dist_cuml}
\end{figure*}

\subsection{Whole samples}

We can also compare whole samples, although it must be noted that selection effects may dominate any differences between them.
In Paper II we investigate the underlying distributions, taking account of selection effects.
Figure \ref{fig:sample_dist_cuml} shows cumulative distributions of parameters for the \emph{Chandra} observed SPT sample, the \emph{XMM-Newton} observed \emph{Planck} ESZ sample \citep{Lovisari17} and the eFEDS sample \citep{Ghirardini22}.
For eFEDS1, we show distributions of parameters for subsamples with increasing cuts on the detection likelihood.
High detection likelihood objects are the most significant in the survey and have the smallest statistical uncertainties on the parameter values.
For comparison with eFEDS, we also show an eRASS1 subsample with the same cuts in detection and extension likelihood and with a similar observation depth, although we do not match the background levels.

Where appropriate, we also show the distributions for both the parameters measured with a free cluster centre and those measured where the peak position is used.

We see that the SPT and \emph{Planck} samples show similar distributions of concentration.
eFEDS shows a median concentration close to the SPT value, although the wings are broader.
The eRASS1 clusters have systematically higher concentrations than the other samples.
The $c_{500}^*$ values are larger than $c_{500}$, as expected.
The eRASS1 median concentration is around 0.3 dex larger than SPT or eFEDS medians.
If we examine subsamples with larger $\mathcal{L}_\mathrm{det}$ thresholds, the median shifts to higher concentration values, as previously seen in Fig.~\ref{fig:cuml_conc}.
However, as we discussed for Fig.~\ref{fig:detlike} and Fig.~\ref{fig:cuml_conc}, cuts in count space above $\sim 40$ do not substantially change the concentration distribution.

The reason for the different concentration distribution for eRASS1 compared to the other surveys could be due to its X-ray selection and relatively shallow depth, although we note that there are differences in the measured parameters for the same clusters which could contribute (Section \ref{sect:matched}).
X-ray selection is more sensitive to concentrated clusters (Section \ref{sect:morph_det_bias}).
Fig.~\ref{fig:cuml_conc} also shows that increasing $\mathcal{L}_\mathrm{ext}$ would not reduce the median concentration to bring it in agreement with the SZ-selected samples.
Increasing the depth of the survey decreases the median concentration, although the effect is relatively modest up to 800s.
\cite{Lovisari17} give a median concentration for the REXCESS sample of representative X-ray selected clusters \citep{Bohringer07} as $c_{500} \sim -0.6$, which is consistent with what we find for eRASS1 (Fig.~\ref{fig:sample_dist_cuml}).

eFEDS contains less concentrated clusters, despite its X-ray selection, which may be due to the survey depth.
We check this using the `eFEDS-like' eRASS1 subsample, noting that the eFEDS cluster positions are fitted for.
Comparing the eFEDS-like concentrations when the central position is fitted for, we see a similar distribution to eFEDS.
The cumulative fraction at the most and least concentrated peaks are very similar to eFEDS.
However, eFEDS-like is more concentrated by around 0.1 dex at the median.
This may be due to the difference in the fitting method, including the functional form used, or the average background level.

The median $\alpha^*$ is similar to eFEDS, although eFEDS does not use peak cluster positions.
The difference in slope between eFEDS and eFEDS-like is likely due to the  positive central density slope prior used by \cite{Ghirardini22} compared to the wide prior used here.
The SPT and \emph{Planck} clusters do not show such steep inner profiles, with median values around 0.25 smaller than eRASS1.
Larger $\mathcal{L}_\mathrm{det}$ thresholds increase $\alpha^*$, although the effect is small for modest increases.

The median central scaled densities in eRASS1 clusters with a lower detection threshold are similar to those found in the SPT and \emph{Planck} clusters, although the central density for eRASS1 strongly increases with $\mathcal{L}_\mathrm{det}$.
eFEDS clusters show a density distribution with a median reduced by around 0.3 dex from eRASS1 with $\mathcal{L}_\mathrm{det}>20$.
The differences between eFEDS and eFEDS-like is again likely affected by the different priors on the central density slope assumed in the analyses.

The raw ellipticity distribution for eRASS1 is dominated by measurement errors for lower detection thresholds.
As the values can only lie between 0 and 1, and clusters are more circular than elliptical, values are biased low for low numbers of counts
As the detection threshold increases, the distribution approaches the SPT distribution, with a median of $\epsilon=0.77$.
eFEDS values are lower, but they also have significant measurement errors.
It is unclear why the eFEDS and eFEDS-like samples show different ellipticity distributions.
\emph{Planck} clusters, show significantly larger ellipticity values, indicating they are more circular.
However, we can compare directly the eRASS1 and \emph{XMM-Newton} measured \emph{Planck} values for the matched sample.
The \emph{XMM-Newton} values are significantly larger, despite the low bias in the values expected in the eROSITA high count regime (Fig.~\ref{fig:sim_bias}).

Fitting a Gaussian linear model, we find a relation between the two of $\epsilon_{\mathrm{eRASS1}} = 0.51 \epsilon_{XMM} + 0.32$.
It is unclear what the origin of these differences is, as both should be the ratio of the minor to major axis, although it may be because the two measurements might be more sensitive to different parts of the cluster due to the field of view and PSF.
In addition, there may be statistical differences due to the parameter range and the larger number of counts in the \emph{XMM-Newton} data.

\section{Combined disturbance indicators}
\label{sect:disturb}
Different morphological parameters can be combined together to produce a common disturbance parameter, as proposed in \cite{Rasia13} and also demonstrated in \cite{Ghirardini22} and \cite{Campitiello22}.
We use a different method to those papers to characterise the disturbance of a cluster.
Many of the parameters measure similar aspects of the morphology of a cluster and they can have strong biases.
Therefore, it is probably not a good idea to combine all possible parameters.
Our method is to fit a Gaussian mixture model (GMM) to the parameters of brightest galaxy clusters.
In such a model the data values are fitted by a mixture of multidimensional Gaussian distributions.
In our analysis we decide to use two such components, to separate the clusters into regular and disturbed populations.
The underlying assumption is that two components are sufficient for distinguishing these populations.
The model can then be applied to the whole cluster sample to say whether a set of parameters for a cluster is better represented by the distribution given by the regular component or the disturbed one.

The choice of fitted parameters is important for how well this technique works.
To avoid the issues of strong bias with redshift or luminosity, we restrict ourselves to the forward-modelled parameters.
In addition, the shape of the cluster may vary independently from the presence of a bright core.
We therefore create two disturbance indicators, one of which is only dependent on shape, and the other also includes the concentration.
$D_\mathrm{shape}$ is the disturbance score for shape-based measurements, fitted to $\epsilon$, $H$, $M_1$, $M_2$, $M_3$ and $M_4$, and the peak-fit offset $F$.
$D_\mathrm{comb}$ is a combined disturbance indicator, fitted to the same parameters as $D_\mathrm{shape}$ but adding $c_{80-800}$ to include the effects of bright cores.

\begin{table*}
  \centering
  \caption{Best fitting parameters of the GMM for the morphological parameters.}
  \begin{tabular}{l|cccc|cccc}
\hline
Model & \multicolumn{4}{c|}{$D_\mathrm{comb}$} &  \multicolumn{4}{c}{$D_\mathrm{shape}$} \\ \hline
Parameter & Mean R & Width R & Mean D & Width D & Mean R & Width R & Mean D & Width D \\
\hline
Amplitude  & \multicolumn{2}{c}{$0.739$}  & \multicolumn{2}{c|}{$0.261$} & \multicolumn{2}{c}{$0.769$} & \multicolumn{2}{c}{$0.231$}  \\
$\epsilon$ & $0.813$ & $0.079$ & $0.746$ & $0.097$ & $0.818$ & $0.075$ & $0.722$ & $0.093$ \\
$H$        & $0.113$ & $0.066$ & $0.241$ & $0.140$ & $0.109$ & $0.059$ & $0.259$ & $0.141$ \\
$M_1$      & $0.238$ & $0.123$ & $0.522$ & $0.238$ & $0.227$ & $0.107$ & $0.588$ & $0.209$ \\
$M_2$      & $0.204$ & $0.104$ & $0.286$ & $0.129$ & $0.192$ & $0.096$ & $0.327$ & $0.114$ \\
$M_3$      & $0.076$ & $0.024$ & $0.117$ & $0.062$ & $0.071$ & $0.024$ & $0.136$ & $0.062$ \\
$M_4$      & $0.074$ & $0.029$ & $0.114$ & $0.063$ & $0.070$ & $0.024$ & $0.130$ & $0.063$ \\
$F$        & $0.007$ & $0.004$ & $0.107$ & $0.153$ & $0.004$ & $0.003$ & $0.106$ & $0.159$ \\
$c_{80-800}$&$-0.656$ & $0.244$ & $-1.175$ & $0.179$ & -- & -- & -- & -- \\
\hline
  \end{tabular}
  \tablefoot{
    Parameters for the two GMM for the disturbance based on shape and concentration ($D_\mathrm{comb}$) and purely based on the shape ($D_\mathrm{shape}$).
    Shown are the amplitudes for the two Gaussian components (relaxed, R, and disturbed, D) in each of the models, and the mean and width of these Gaussians in the dimension of the parameter given.
    The width of the Gaussian is the square-root of the diagonal components of the covariance matrix.
    We do not list the non-diagonal entries in the matrix.
  }
  \label{tab:gmm}
\end{table*}

These parameters were fitted using the \texttt{pyGMMis} package \citep{MelchiorGoulding18}, which has the advantage of being able to account for measurement errors by allowing a covariance matrix for each input data points.
As we measure our parameters independently, we do not know the covariance between the parameters for each cluster, but only their uncertainties.
We therefore use a diagonal covariance matrix for each cluster, which does not take account of the likely covariance between parameters (for example $\epsilon$ and $M_2$).
This may have the effect of double counting some aspects of the morphology in the model.
The parameter error bars are symmetrized by computing the root mean squares to make the matrix (noting that is not strictly valid as fixed range parameters may be close to a limit).
The two-component GMM is fitted to the data and covariances from clusters, only using those 175 with more than 800 counts to not be dominated by measurement errors.
We assume that one of these Gaussians is closer to disturbed clusters and one is closer to relaxed clusters.
Table \ref{tab:gmm} shows the parameters of the two GMMs, one for $D_\mathrm{comb}$ and one for $D_\mathrm{shape}$, fitted to the bright clusters.

To classify the clusters, we take this model and compute the likelihood of each of the Gaussians given the cluster parameters and their uncertainties.
To better take account of the model uncertainties, we repeat this for a random selection of model parameters taken from the MCMC chains (or bootstrap values for $F$).
The score for a cluster ($D$) is the fraction of times the Gaussian corresponding to disturbed clusters has a higher likelihood compared to the one for relaxed clusters.

\begin{figure*}
  \includegraphics[width=\textwidth]{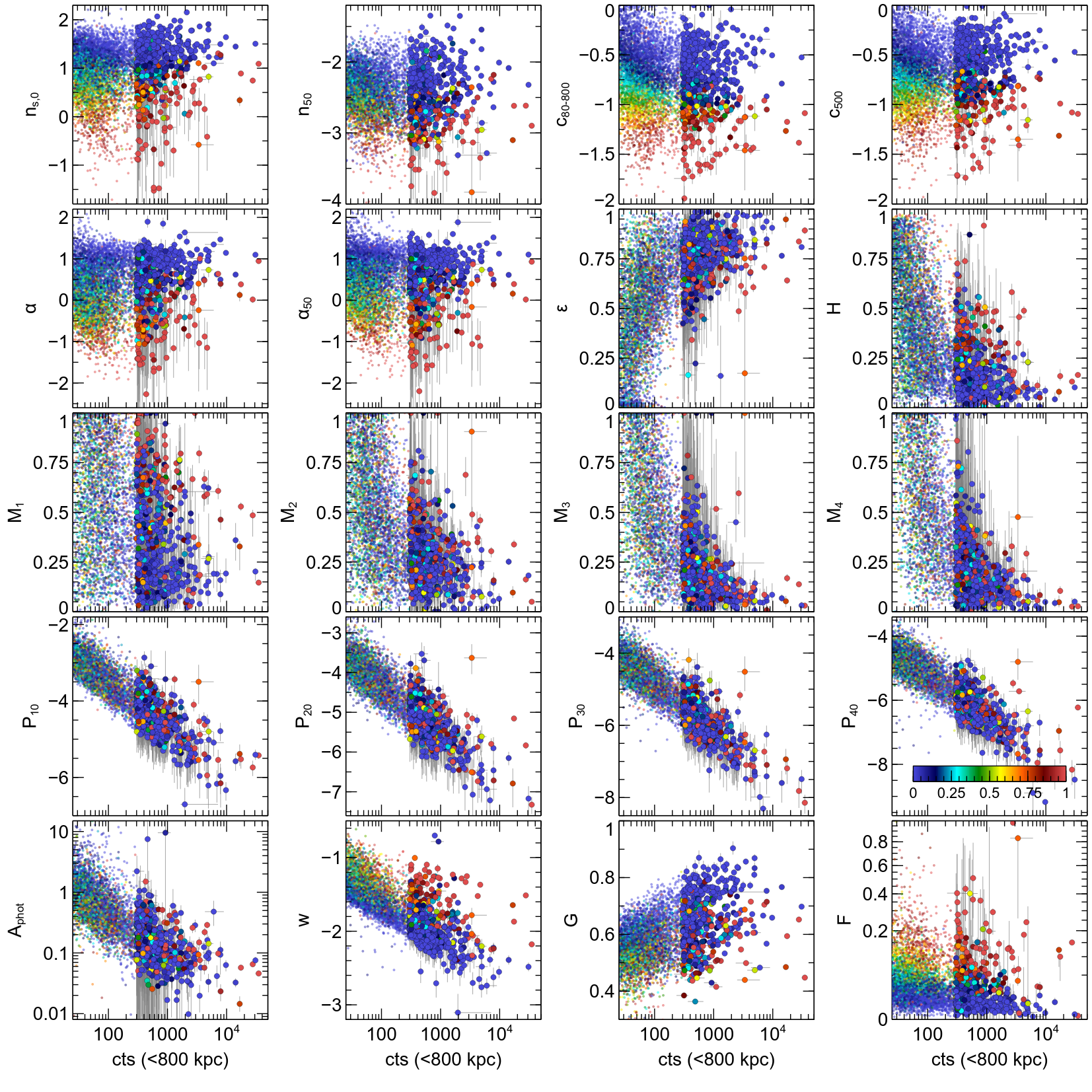}
  \caption{
    Parameter values as a function of counts, coloured according to the combined shape and concentration disturbance score, $D_\mathrm{comb}$.
  }
  \label{fig:param_score}
\end{figure*}

In Fig.~\ref{fig:param_score} we show the parameters plotted as a function of number of counts, as in Fig.~\ref{fig:param_cts}, but coloured by the disturbance score.
The plot shows that the method is preferentially detecting relaxed clusters as having a high central density, high concentration, steep central density profile, higher $\epsilon$, lower slosh, lower $M_1$, lower $M_2$, low $w$, higher $G$ and low $F$.

\begin{figure*}
  \includegraphics[width=\columnwidth]{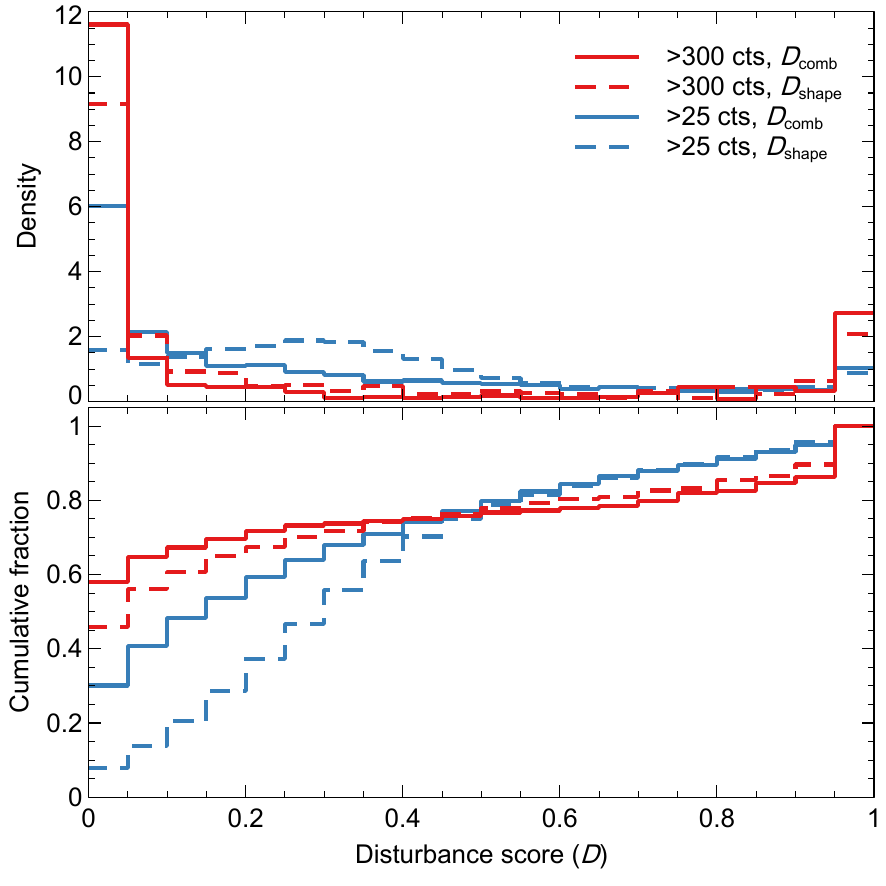}
  \includegraphics[width=\columnwidth]{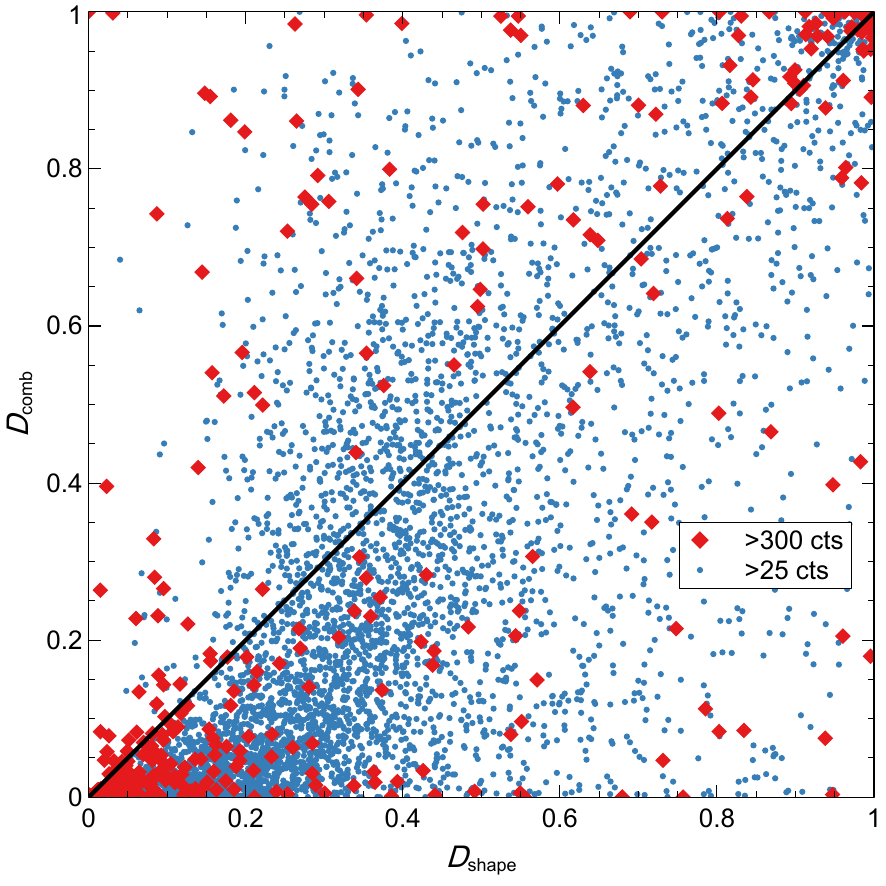}
  \caption{
    Disturbance scores for the sample, for clusters with greater than 25 and greater than 300 counts.
    (Left) Histograms of the distributions of $D_\mathrm{comb}$ (including concentration) and the purely 2D shape based $D_\mathrm{shape}$.
    (Right) $D_\mathrm{comb}$ plotted against $D_\mathrm{shape}$ for the two count thresholds.
  }
  \label{fig:param_disturb_hist}
\end{figure*}

The fraction of clusters which have a high disturbance score $D_\mathrm{comb}$ is relatively constant with the number of counts in the object.
Using thresholds of 0, 25, 100 and 300 counts there are 27, 31, 27 and 25\% of objects with scores greater than 0.5, respectively (see Fig.~\ref{fig:param_disturb_hist}).
20\% of clusters with more than 300 counts reach the higher disturbance threshold of $D_\mathrm{comb}>0.8$, while 66\% have a $D_\mathrm{comb}<0.2$.
Looking at clusters with more than 25 counts, there are 14\% with $D_\mathrm{comb}>0.8$ and 46\% with $D_\mathrm{comb}<0.2$.
Fainter clusters have $R_\mathrm{shape}$ which lie more toward the centre of the range.
This is likely because the shape parameters have large error bars when the number of counts is low, compared to the concentration and therefore we know less about the cluster shape.
In contrast, $D_\mathrm{comb}$ is driven largely by concentration in the fainter objects.

As mentioned above, it matters which parameters are used to make the mixture model.
We have created scores including ($D_\mathrm{comb}$) and non including ($D_\mathrm{shape}$) concentration.
In Fig.~\ref{fig:param_disturb_hist} we also plot two scores against each other for a sample of bright and fainter clusters.
Although the two scores are close for a substantial number of clusters, there are cases where they lie at opposite sides of the scale.
The statistical uncertainty on concentration is typically much smaller than the other parameters, so it can drive the disturbance score when included.
If $D_\mathrm{shape}$ is large, but $D_\mathrm{comb}$ is small, this may indicate a cluster which has a disturbed shape but also a bright core.
If $D_\mathrm{comb}$ is large, but $D_\mathrm{shape}$ is small, this may indicate a symmetric cluster with a very flat surface brightness profile.

There is also some choice of how many clusters are initially fitted by the GMMs.
We chose a threshold of $800$ counts to ensure that statistical uncertainties on the input parameters were relatively small.
We also reran the analysis using a threshold of $400$ counts.
The results were fairly similar, except the $D$ scores were generally lower, with around 20\% of objects classified as disturbed.
If we increase the number of GGM components to three and fit the combined parameters, we find that the third component has a low 4\% amplitude compared to the others (64 and 32\% for 800 counts) and lies at a more extreme position in the irregular cluster parameter space.
Fitting the 400 count clusters yields a similar result.
If we add a fourth component, the modelling also prefers this to have a low amplitude, with around 13\% in the smallest two components.
Therefore, the GGM model does not show strong evidence for more than two components.

\begin{figure*}
  \includegraphics[width=\textwidth]{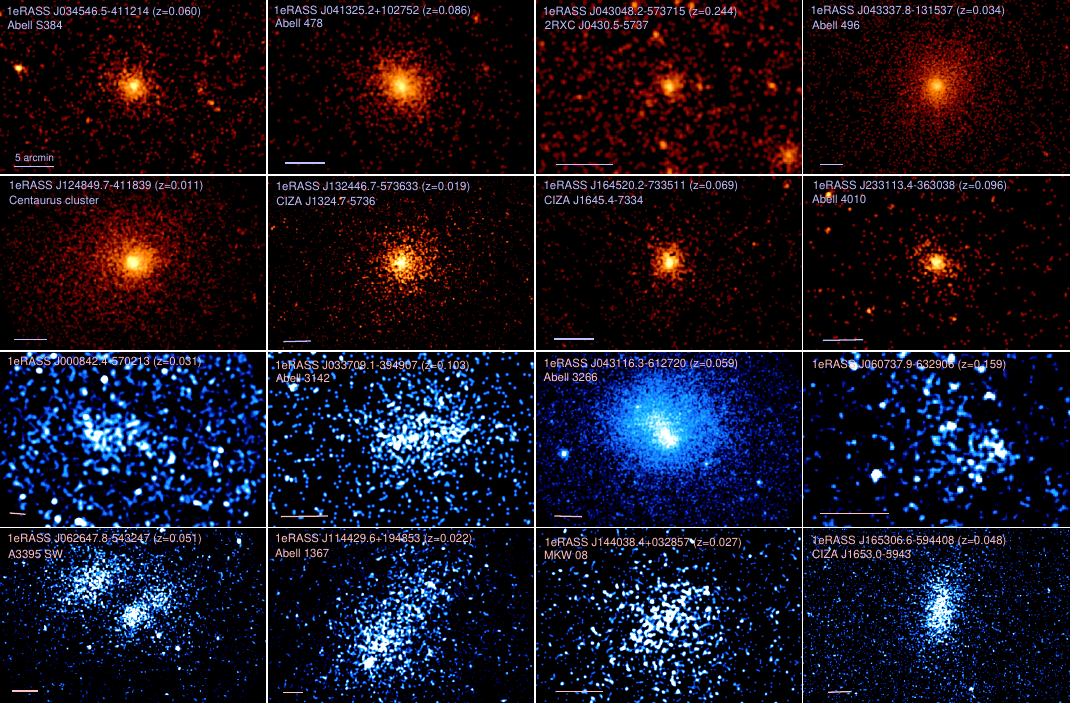}
  \caption{
    Examples of eight relaxed and eight disturbed clusters.
    Clusters are randomly selected from the sample with more than 300 counts, where $D_\mathrm{comb}=0$ (top panels) or $D_\mathrm{comb}=1$ (bottom panels).
    The exposure corrected images in the 0.2--2.3 keV band have been smoothed with a Gaussian of $\sigma=8$ arcsec or 64~arcsec in the case of 1eRASS J000842.4-570213.
    The bars show a fixed angular scale of 5 arcmin.
  }
  \label{fig:relax_disturb}
\end{figure*}

Some examples of clusters with extreme low and high disturbance scores are shown in Fig.~\ref{fig:relax_disturb}.
In this brighter cluster regime, the parameter appears to be able to easily separate relaxed and disturbed clusters.
The clusters include well known relaxed and disturbed objects.

We find a large fraction of relaxed objects.
X-ray selection is better at finding clusters with dense central cores, likely associated with relaxed cool core clusters, while SZ selection finds a larger number of disturbed systems.
Like some previous X-ray surveys, we find a large fraction of relaxed objects.

\cite{Rossetti17} studied the morphology of clusters selected using the SZ effect by \emph{Planck}, using \emph{Chandra} and some \emph{XMM-Newton} data.
Based on the concentration in 40 and 400~kpc apertures, they found that $29 \pm 4$\% of their clusters were cool-core objects.
They conducted the same measurement on the \emph{ROSAT} all-sky X-ray selected MACS sample \citep{Mann12}, finding $59 \pm 5$\% could be classified as cool-core.

\cite{Lovisari17} studied the morphology of the \emph{Planck} ESZ clusters using \emph{XMM-Newton}.
By visual inspection, they classified 32\% of objects as relaxed, 39\% disturbed and 29\% were of an intermediate state.
All of their clusters which are visually identified as relaxed have a concentration of more than $0.15$ or $c_{500} > -0.82$ in our log units.
Ignoring the statistical uncertainties on our values, this threshold would imply 80\% of our eRASS1 objects are relaxed, similar to the fraction identified by our GMM.

\cite{Campitiello22} studied the morphology of 188 clusters observed by \emph{XMM-Newton} and selected using \emph{Planck}.
Using a combination of visual inspection and combining morphological parameters, they found that 13\% were robustly classified as relaxed, while 21\% were disturbed.

\cite{Ghirardini22} found that the X-ray selected eROSITA eFEDS cluster sample had a similar fraction of cool cores to SZ selected surveys.
They construct a relaxation score based on a combination of several morphological parameters, finding that concentration probes the inner part of the cluster but is not overall a good measure of relaxation.
Around 70\% of their clusters at the lowest redshifts were found to be relaxed, similar to what is found by us.
However, this decreases rapidly to 35-50\% at $z>0.2$, becoming closer to the SZ-selected samples.
This difference in relaxation over redshift could be evolution or selection effects.

\section{Conclusions}
We measure 29 morphological parameters for over 12 thousand clusters found in the eROSITA survey by B24.
The parameters we measure include concentration, central density, inner density slope, peak offset, power ratios, Gini coefficient, photon asymmetry, centroid shift, ellipticity, slosh and multipole magnitudes.
In our work we introduce new slosh and multipole magnitude forward modelled parameters.
This paper provides the largest X-ray derived set of cluster morphological parameters to date.
We provide a catalogue of these parameters and their uncertainties.

Several of the measured quantities (e.g. power ratios, centroid shift, Gini coefficient, photon asymmetry) show values which correlate strongly with the number of counts, luminosity or redshift.
We conduct simulations of clusters observed in the survey finding similar relations.
These correlations are biases due to noise or PSF and therefore these parameters are unreliable for eROSITA clusters.
We discuss the various biases which can be present in the parameters and provide the typical biases and uncertainties as a function of redshift and the number of counts.

We see using both simulations and data that the concentration of clusters strongly affects whether they are detected in the eROSITA survey.
At low redshifts ($z \lesssim 0.2$) we preferentially detect more highly concentrated objects and do not detect low luminosity groups and clusters with flat surface brightness profiles.
At high redshifts ($z \gtrsim 0.4$), we will miss the most concentrated clusters.
Alternative detection algorithms (e.g. wavelets) may find more flat surface brightness objects at low redshifts.

We make comparisons of results for matched cluster subsamples for eRASS and other analyses.
There is reasonable agreement in the concentration and density with matched \emph{Planck} ESZ and SPT subsamples.
The median values are close to those obtained using eROSITA, although there may be mild offsets or differences in slopes between the analyses.
These differences may be due to different analysis procedures, data quality, choice of cluster centre and parametrization, or unresolved AGN in the datasets.

Comparing whole cluster samples, we find that eRASS1 clusters have higher concentrations than SPT, \emph{Planck} ESZ and eFEDS.
The difference is around 0.3 dex in our log units, implying that the ratio of flux inside $0.1 R_{500}$ to $R_{500}$ is twice what is found in these other samples.

We construct combined parameters based in a Gaussian mixture model for measuring disturbance of clusters ($D_\mathrm{shape}$ and $D_\mathrm{comb}$), finding that $1/4$ objects are classified as disturbed.
The high relaxation fraction is likely due to X-ray selection selecting highly concentrated clusters which are more likely to be relaxed.

\begin{acknowledgements}
This work is based on data from eROSITA, the soft X-ray instrument aboard \emph{SRG}, a joint Russian-German science mission supported by the Russian Space Agency (Roskosmos), in the interests of the Russian Academy of Sciences represented by its Space Research Institute (IKI), and the Deutsches Zentrum für Luft- und Raumfahrt (DLR). The \emph{SRG} spacecraft was built by Lavochkin Association (NPOL) and its subcontractors, and is operated by NPOL with support from the Max Planck Institute for Extraterrestrial Physics (MPE).

The development and construction of the eROSITA X-ray instrument was led by MPE, with contributions from the Dr. Karl Remeis Observatory Bamberg \& ECAP (FAU Erlangen-Nuernberg), the University of Hamburg Observatory, the Leibniz Institute for Astrophysics Potsdam (AIP), and the Institute for Astronomy and Astrophysics of the University of Tübingen, with the support of DLR and the Max Planck Society. The Argelander Institute for Astronomy of the University of Bonn and the Ludwig Maximilians Universität Munich also participated in the science preparation for eROSITA.

The eROSITA data shown here were processed using the \texttt{eSASS/NRTA} software system developed by the German eROSITA consortium.

E.B., V.G., A.L. and X.Z. acknowledge financial support from the European Research Council (ERC) Consolidator Grant under the European Union’s Horizon 2020 research and innovation program (grant agreement CoG DarkQuest No 101002585).

Funded in part by the Deutsche Forschungsgemeinschaft (DFG, German Research Foundation) - 450861021.
\end{acknowledgements}

\bibliographystyle{aa}
\bibliography{refs}

\begin{appendix}
\onecolumn
\section{Description of catalogue}
\label{appen:cat}
Table \ref{tab:cat} describes the columns in the morphology catalogue, giving the column name and respective morphological parameter.
Most of the parameters are dimensionless, except for the inner physical gas density.
Data will be uploaded to CDS for the final paper version and are currently available at \url{https://erosita.mpe.mpg.de/dr1/AllSkySurveyData_dr1/Catalogues_dr1/}.

\begin{table*}[h!]
  \caption{Column names in morphology catalogue.}
  \centering
  \begin{tabular}{lll}
    \hline
    Column       & Value & Unit \\ \hline
    \texttt{NAME}         & IAU Cluster name \\
    \texttt{DETUID}       & eROSITA detection ID \\
    \texttt{CTS\_800}     & Number of counts \\
    \texttt{NS\_0}        & $n_{\mathrm{s},0}$ \\
    \texttt{NS\_0\_P}     & $n_{\mathrm{s},0}^*$ \\
    \texttt{NE\_50}       & $n_{50}$ & log~cm$^{-3}$ \\
    \texttt{NE\_50\_P}    & $n_{50}^*$ & log~cm$^{-3}$ \\
    \texttt{ALPHA}        & $\alpha$ \\
    \texttt{ALPHA\_P}     & $\alpha^*$ \\
    \texttt{ALPHA\_50}    & $\alpha_{50}$ \\
    \texttt{ALPHA\_50\_P} & $\alpha_{50}^*$ \\
    \texttt{C\_R500}             & $c_{500}$ \\
    \texttt{C\_R500\_P}          & $c_{500}^*$ \\
    \texttt{C\_80\_800}          & $c_{80-800}$ \\
    \texttt{C\_80\_800\_P}       & $c_{80-800}^*$ \\
    \texttt{F}                   & $F$ \\
    \texttt{P10}                 & $P_{10}$ \\
    \texttt{P10\_P}              & $P_{10}^*$ \\
    \texttt{P20}                 & $P_{20}$ \\
    \texttt{P20\_P}              & $P_{20}^*$ \\
    \texttt{P30}                 & $P_{30}$ \\
    \texttt{P40}                 & $P_{40}$ \\
    \texttt{GINI}                & $G$ \\
    \texttt{APHOT}               & $A_\mathrm{phot}$ \\
    \texttt{APHOT\_P}            & $A_\mathrm{phot}^*$ \\
    \texttt{CENTROID}            & $w$ \\
    \texttt{ELLIPTICITY}         & $\epsilon$ \\
    \texttt{ELLIPTICITY\_THETA}  & $\theta_0$ for $\epsilon$ fit & deg \\
    \texttt{SLOSH}               & $H$ \\
    \texttt{SLOSH\_THETA}        & $\theta_0$ for $H$ fit & deg \\
    \texttt{M1}                  & $M_1$ \\
    \texttt{M1\_THETA}           & $\theta_0$ for $M_1$ fit & deg \\
    \texttt{M2}                  & $M_2$ \\
    \texttt{M2\_THETA}           & $\theta_0$ for $M_2$ fit & deg \\
    \texttt{M3}                  & $M_3$ \\
    \texttt{M3\_THETA}           & $\theta_0$ for $M_3$ fit & deg \\
    \texttt{M4}                  & $M_4$ \\
    \texttt{M4\_THETA}           & $\theta_0$ for $M_4$ fit & deg \\
    \texttt{D\_SHAPE}            & $D_\mathrm{shape}$ \\
    \texttt{D\_COMB}             & $D_\mathrm{comb}$ \\
    \hline
  \end{tabular}
  \tablefoot{
    For each numerical column there are corresponding columns with the suffixes \texttt{\_POS} and \texttt{\_NEG} which denote the $1\sigma$ uncertainties.
    The number of counts is that within 800 kpc radius in the 0.2-2.3 keV band.
  }
  \label{tab:cat}
\end{table*}
\end{appendix}

\end{document}